\documentclass[journal,onecolumn,11pt]{IEEEtran}
\usepackage{epsfig,rotating,setspace,latexsym,amsmath,epsf,amssymb,amsfonts,bm,epstopdf,cite,authblk,bbm,mathtools,amsthm,algorithm,color,systeme,mathtools,amsmath,derivative,bigints,mathrsfs,tabularx}
\usepackage[short,c2]{optidef}
\usepackage[utf8]{inputenc}
\usepackage[T1]{fontenc}
\usepackage[thinc]{esdiff}
\usepackage[noend]{algpseudocode}
\usepackage[caption=false,font=footnotesize]{subfig}

\theoremstyle{plain}
\newtheorem{theorem}{Theorem}
\newtheorem{lemma}{Lemma}
\newtheorem{corollary}{Corollary}

\newtheorem{remark}{Remark}
\newenvironment{Proof}[1]{\medskip\par\noindent{\bf Proof:\,}\,#1}{{\mbox{\,$\blacksquare$}\par}}
\newenvironment{Proof Sketch}[1]{\medskip\par\noindent{\bf Proof Sketch:\,}\,#1}{{\mbox{\,$\blacksquare$}\par}}

\algrenewcommand\algorithmicforall{\textbf{foreach}}
\algrenewcommand\algorithmicindent{.8em}

\allowdisplaybreaks

\begin{document}

\title{Fundamental Tradeoffs in Sampling, Transmission and Information Freshness: Instantaneous Feedback}

\author{Subhankar Banerjee \qquad Sennur Ulukus\\
\normalsize Department of Electrical and Computer Engineering\\
\normalsize University of Maryland, College Park, MD 20742\\
\normalsize \emph{sbanerje@umd.edu} \qquad \emph{ulukus@umd.edu} }

\maketitle

\begin{abstract}
    We study a time-slotted generate-at-will status-update system over an erasure channel with transmission success probability $q$, subject to average sampling- and transmission-rate constraints $f_s$ and $f_r$, respectively. Due to separate constraints on sampling and transmission rates, a three-action control problem emerges. Depending on the timeliness requirements and the constraints, the sampler-transmitter pair must jointly decide whether to remain idle, retransmit a stale packet, or preempt a stale packet and replace it with a freshly sampled packet. We first derive a universal lower bound on the optimal average age of information for this system. We then analyze a class of stationary randomized policies and propose a policy with a $4$-approximation guarantee. We show that the joint problem reduces to the sampling-constrained problem when $f_s\leq qf_r$, and to the transmission-constrained problem when $f_s\geq f_r$. In the remaining regime $qf_r<f_s<f_r$, we consider the two constraints jointly. We formulate the joint sampling and transmission problem as a constrained Markov decision process (CMDP) and introduce a policy class $\bar{\Pi}$, consisting of $(W,R)$ policies, where $W$ specifies a waiting threshold for transmission and $R$ specifies a packet-replacement threshold. We show that, for every choice of the Lagrange multipliers, the corresponding Lagrangian-relaxed Markov decision process (MDP) admits an optimal policy in $\bar{\Pi}$. We then formulate a linear program (LP) over probability distributions of $(W,R)$ policies and show that choosing a $(W,R)$ policy according to an optimal solution of the LP yields an optimal policy for the CMDP. Moreover, there exists an optimal solution of the LP that assigns positive probability to at most three $(W,R)$ policies. Since $\bar{\Pi}$ is countably infinite, for any $\epsilon>0$, we construct a finite subset $\bar{\Pi}_{\epsilon}\subset \bar{\Pi}$ in a principled way and show that the optimal solution of the LP restricted to $\bar{\Pi}_{\epsilon}$ induces an $\epsilon$-optimal policy.  We further derive a stronger pair of lower bounds by separately considering systems with only the sampling-rate constraint and only the transmission-rate constraint. For the regime $qf_r<f_s<f_r$, we further construct a policy that randomizes among four $(W,R)$ policies and show that it provides a $2$-approximation guarantee. 
\end{abstract}

\section{Introduction}
Many emerging networked systems, such as remote monitoring, networked control, industrial automation, intelligent transportation, and IoT applications, rely on the timely delivery of status updates. In these systems, the usefulness of received data critically depends on its freshness. The age of information has emerged as a fundamental metric for quantifying such information freshness \cite{YatesSurvey}. In this work, we consider a generate-at-will status-update system, in which fresh information about the source can be acquired by a sensor whenever the sensor chooses to generate an update \cite{akar2025age}. 

A generate-at-will status-update system consists of two principal controllable components: a sampler that acquires fresh information and a transmitter that communicates the acquired information to a remote monitor. These operations generally consume different resources. Sampling may require sensing energy, access to a measurement device, or computational resources for processing and encoding an observation, whereas transmission may consume communication energy, spectrum, channel uses, or access opportunities on a shared network. Thus, it is natural to impose separate average-rate constraints on sampling and transmission. These constraints provide a simple and broadly applicable abstraction of the resource limitations associated with the two essential stages of delivering a fresh update: acquiring information at the source and communicating it to the monitor.

\begin{figure}[t]
    \centerline{\includegraphics[width = 0.5\columnwidth]{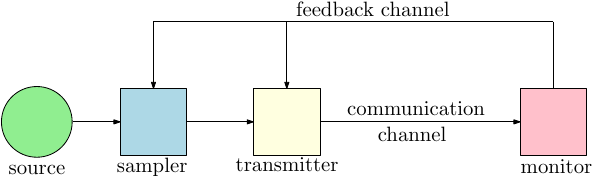}}
    \caption{System model and its components.}
    \label{fig:1}
\end{figure}

We study a time-slotted system over an erasure channel with transmission success probability $0< q \leq 1$. The maximum allowable average sampling and transmission rates are denoted by $0<f_s\leq 1$ and $0<f_r \leq 1$, respectively. In this work, we refer to the age of information at the monitor as the age of the monitor. The system is illustrated in Fig.~\ref{fig:1}. We study sampling and transmission policies that minimize the age of the monitor under instantaneous feedback from the monitor.

In each slot, the sampler-transmitter pair may remain idle; sample a fresh packet and transmit it if the transmitter has no stored packet or retransmit the packet currently stored at the transmitter; or discard the stored packet, generate a fresh one, i.e., preempt a stale packet with a fresher packet, and transmit it immediately. These actions affect both the system state and resource consumption differently. Idling consumes neither resource but increases both the age of the monitor and the age of the stored packet. Retransmission preserves the generation time of the stored packet and consumes only transmission resources, whereas preempting the stored packet with a freshly sampled packet and transmitting it provides the greatest possible freshness improvement upon successful delivery, and consumes both sampling and transmission resources. Consequently, in general, the two resource constraints cannot be treated independently. The controller must jointly determine when to wait, when to retransmit the stored packet, and when the potential freshness improvement justifies consuming an additional sampling opportunity. This gives rise to a three-action control problem.

There is an extensive literature on constrained age of information minimization. Under a sampling-rate constraint, existing works optimized when a fresh sample should be generated and transmitted through communication systems with random delay or service time \cite{sun2019sampling,sun2019sampling2,pan2023optimal,tang2023age}, or over a slotted erasure channel \cite{banerjee2025equidistant}. In particular, \cite{banerjee2025equidistant} showed that randomization between at most two equidistant-sampling policies is optimal under an average sampling-rate constraint. However, transmission attempts are unconstrained in that model, and the resulting optimal policy need not satisfy the separate transmission-rate constraint considered in this work. The other works derived deterministic or randomized threshold-sampling policies when the channel or service statistics were known \cite{sun2019sampling,sun2019sampling2,pan2023optimal}, and an online asymptotically optimal sampling policy when the delay statistics were unknown \cite{tang2023age}. Apart from the model in \cite{banerjee2025equidistant}, a common feature of these models is that service is non-preemptive: once the transmission of a packet begins, the sampler cannot discard it while it is in service and replace it with a fresher one. Consequently, the controller primarily determines how long to remain idle before generating the next sample, and there is no separate decision between retransmitting a stored packet and replacing it with a fresh one. Thus, these sampling-constrained formulations do not capture the three-action control problem described above, making our problem fundamentally different from the problems considered in this line of work.

Under transmission-side resource constraints, one line of work scheduled fresh updates subject to power, energy, bandwidth, or per-slot transmission limitations \cite{hatami2022demand,tang2020minimizing,chen2023minimizing}, whereas another line of work determined whether to remain idle, transmit a fresh update, or retransmit a previously failed update \cite{wang2019minimizing,ceran2019average}. Both lines of work considered only transmission-side resource constraints. In the former, packets either arrived exogenously or a fresh packet was generated whenever transmission was selected. Hence, the controller essentially chose only between remaining idle and transmitting an available fresh packet, without separately deciding between retransmitting a stored packet and replacing it with a fresh one. In the latter, retransmission and packet replacement were explicitly considered; however, the generation of a fresh packet was unconstrained. Consequently, the choice between retransmitting the stored packet and replacing it with a fresh one was driven by channel reliability and the resulting freshness improvement, without accounting for the consumption of a separately limited sampling resource. The additional sampling constraint fundamentally changes the tradeoff between retransmitting the stored packet and replacing it with a fresh one, leading to an optimal-policy structure that differs from those obtained in transmission-constrained systems with an unconstrained sampler \cite{wang2019minimizing,ceran2019average}. In \cite{gong2022sleep}, sensing and transmission costs were included in the objective together with the average age. The authors also studied a single constraint on average energy consumption, which bounds a weighted sum of the sampling and transmission rates. A closely related work imposing separate sampling and transmission constraints is \cite{zakeri2022query}, which minimizes query age of information, under which freshness is relevant only when a query occurs. This objective is fundamentally different from the conventional time-average age of information considered in this work, which continuously accounts for the age of the monitor.

Several works formulated constrained age minimization problems as constrained Markov decision processes (CMDPs). Some obtained optimal policies by directly applying numerical dynamic-programming or occupation-measure methods to finite-state models \cite{hatami2022demand,zakeri2022query,delfani2025optimizing}, whereas others exploited the properties of their particular CMDPs to establish threshold structures. Sampling-constrained problems commonly yielded a threshold on the waiting time before generating the next sample \cite{sun2019sampling,sun2019sampling2,pan2023optimal}, while transmission-constrained problems yielded thresholds in the age of the monitor, channel state, packet age, or retransmission count \cite{ceran2019average,wang2019minimizing,tang2020minimizing,chen2023minimizing}. These structural results were obtained for problems with a single constrained resource and therefore do not directly characterize the interaction between fresh sampling and retransmission when the two actions consume different combinations of constrained resources. Moreover, our system state must track both the age of the monitor and the age of the packet stored at the transmitter, resulting in a two-dimensional countably infinite state space. Directly solving a numerically truncated CMDP would neither reveal the underlying policy structure nor, without further analysis, quantify the loss caused by truncation. Our objective is therefore to identify a structure-aware policy class and use it to construct a principled optimal policy, or a policy with an explicit optimality guarantee, rather than employing the numerical solution of a bounded-state approximation.

The optimal policies for the corresponding single-constraint problems provide useful insight into the structure of a policy for the joint problem. Under only an average sampling-rate constraint, an optimal policy can be obtained by randomizing between at most two equidistant-sampling policies \cite{banerjee2025equidistant}. Under only an average transmission-rate constraint, an optimal policy follows a waiting-threshold structure: after a successful delivery, the sampler-transmitter pair remains idle until the age of the monitor reaches a threshold and then transmits a fresh packet in every slot \cite{ceran2019average}. Motivated by these observations, we introduce the class of $(W,R)$ policies, which combines these two structures in a principled manner. After each successful delivery, a $(W,R)$ policy remains idle until the age of the monitor reaches the waiting threshold $W$. It then transmits in every slot, retransmitting the stored packet following an unsuccessful transmission and replacing it with a fresh packet whenever its age reaches the packet-replacement threshold $R$. Thus, $W$ determines when an active transmission period begins, while $R$ regulates how frequently sampling resources are consumed during that period.

We consider the system with instantaneous feedback and derive a simple universal lower bound on the minimum achievable average age of the monitor. We analyze a class of stationary randomized policies, optimize the policy parameters under sufficient feasibility conditions, and show that the resulting policy has a $4$-approximation guarantee. Building on the single-constraint policies discussed above, we identify two regimes in which an optimal policy for one of the single-constraint problems remains optimal for the joint problem. When $f_s\leq qf_r$, an optimal policy for the sampling-constrained problem also satisfies the transmission constraint and is therefore optimal for the joint problem. Similarly, when $f_s\geq f_r$, an optimal policy for the transmission-constrained problem also satisfies the sampling constraint and is optimal for the joint problem. Thus, only the intermediate regime $qf_r<f_s<f_r$ requires a joint sampling-and-transmission design. The corresponding problems with only the sampling-rate constraint and only the transmission-rate constraint also provide two additional lower bounds on the optimal average age of the monitor, which jointly strengthen the preceding universal lower bound.

We formulate the problem as a CMDP. We show that, for every choice of the Lagrange multipliers, the corresponding Lagrangian-relaxed Markov decision process (MDP) admits an optimal policy in the proposed $(W,R)$ policy class. We then formulate a linear program (LP) over randomized mixtures of $(W,R)$ policies and show that an optimal solution of this LP induces an optimal policy for the CMDP. Moreover, the LP admits an optimal solution that assigns positive probability to at most three $(W,R)$ policies. Since the LP involves a countably infinite collection of $(W,R)$ policies, for any given $\epsilon>0$, we appropriately restrict the policy class to a finite set and prove that the resulting finite-dimensional LP produces an $\epsilon$-optimal sampling and transmission policy. For the intermediate regime $qf_r<f_s<f_r$, we further construct a policy that randomizes among four carefully selected $(W,R)$ policies. Using the lower bound obtained by considering the system with only the transmission-rate constraint, we show that this policy provides a $2$-approximation guarantee.

\section{System Model and Problem Formulation}
We denote the action of the sampler at time $t$ with $a_{s}(t)\in\{0,1\}$, where $a_{s}(t)=1$ implies that the sampler samples a fresh packet at time $t$ and $a_{s}(t)=0$ implies otherwise. Similarly, we denote the action of the transmitter at time $t$ with $a_{r}(t)\in\{0,1\}$, where $a_{r}(t)=1$ implies that the transmitter transmits the freshest available packet at time $t$ and $a_{r}(t)=0$ implies otherwise. We assume that if the transmitter has no sampled packet available at time $t$, it does not transmit, i.e., $a_{r}(t)=0$. Under a causal policy $\pi$, we denote the action of the sampler-transmitter pair at time $t$, by $(a_{s}^{\pi}(t),a_{r}^{\pi}(t))$. 

\begin{figure}[t]
    \centerline{\includegraphics[width =0.5 \columnwidth]{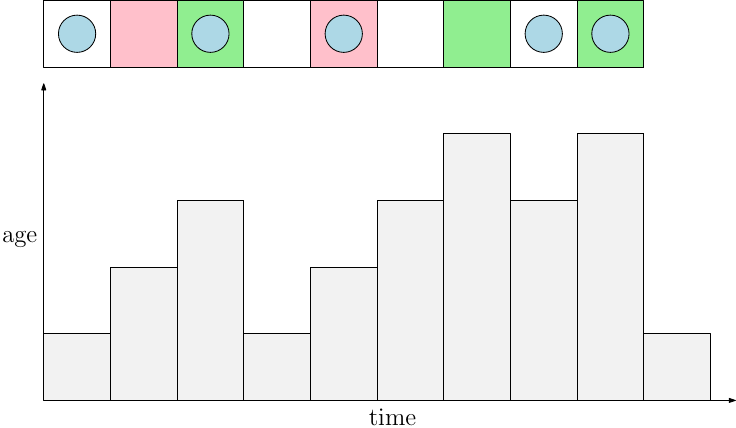}}
    \caption{The cyan circles denote sampling of new update packets, the pink boxes and the green boxes denote unsuccessful transmissions and the successful transmissions of the most recently sampled packet, respectively. }
    \label{fig:age_evo}
\end{figure}

We assume that the sampling of a packet occurs at the beginning of a time slot. In this work, we consider instantaneous feedback. For each transmission, the monitor sends an ACK (NACK) upon a successful (unsuccessful) transmission. The corresponding ACK/NACK is received by the sampler-transmitter pair at the end of the transmission slot. Each transmission attempt of a packet takes one time slot, where the transmission starts at the beginning of a slot and ends at the end of that slot. The channel realization in slot $t$ is denoted by an indicator random variable $Z(t)$, where $Z(t)=1$ denotes a successful transmission. We assume that $\{Z(t)\}_{t\geq1}$ are independent and identically distributed (i.i.d.) Bernoulli random variables with $\mathbb{P}(Z(t)=1)=q$. We denote the age of the monitor at time $t$ under a policy $\pi$ with $v_{1}^{\pi}(t)$. As we are only considering freshness, we assume that if a fresh packet gets sampled, it replaces any stale packet at the transmitter. We denote the age of the packet at the transmitter with $v_{2}^{\pi}(t)$. If there is no packet present at the transmitter, we assume that $v_{2}^{\pi}(t)=\infty$. At time $t$, if the transmitter has no packet and the sampler does not generate a fresh packet, then transmission is infeasible; thus, the transmitter remains idle, i.e., $a_{r}^{\pi}(t)=0$. At time $t$, we denote the state of the system under policy $\pi$ with $s(t)=(v_{1}^{\pi}(t), v_{2}^{\pi}(t))$. Given an arbitrary initial state $s(1)$, the age of the monitor $v_{1}^{\pi}(t)$ evolves for $t\geq 1$ as follows,
\begin{align}
    v_{1}^{\pi}(t+1) = \begin{cases}
        1, & a_{s}^{\pi}(t)=1 \ \textrm{and} \  a_{r}^{\pi}(t)Z(t)=1,\\ 
        v_{2}^{\pi}(t)+1, & a_{s}^{\pi}(t)=0  \ \textrm{and} \  a_{r}^{\pi}(t)Z(t)=1,\\ 
        v_{1}^{\pi}(t)+1, &a_{r}^{\pi}(t)Z(t)=0.
    \end{cases}
\end{align}
We define the evolution of the age of the packet $v_{2}^{\pi}(t)$ for $t\geq 1 $ as follows,
\begin{align}
    v_{2}^{\pi}(t+1) = \begin{cases}
        \infty,  & a_{r}^{\pi}(t)Z(t) = 1, \\ 
        1,  & a_{r}^{\pi}(t)Z(t) = 0 \ \textrm{and} \ a_{s}^{\pi}(t) = 1, \\ 
        v_{2}^{\pi}(t) +1,  & a_{r}^{\pi}(t)Z(t) = 0 \ \textrm{and} \ a_{s}^{\pi}(t) = 0.  
    \end{cases}
\end{align}
In Fig.~\ref{fig:age_evo}, we pictorially show the age evolution under an arbitrary policy $\pi$. Under a policy $\pi$ and an initial state $s(1)$, we denote the average age, sampling rate, and transmission rate over the time horizon $T$ with $\Delta(\pi;T,s(1))$, $\nu_{s}(\pi;T,s(1))$, and $\nu_{r}(\pi;T,s(1))$, respectively,
\begin{align}
    & \Delta(\pi;T,s(1)) = \frac{1}{T}  \sum_{t=1}^{T} \mathbb{E}_{\pi}[v_{1}^{\pi}(t)|s(1)],\label{eq:nn3} \\ 
    &\nu_{s}(\pi;T,s(1)) = \frac{1}{T}\sum_{t=1}^{T}\mathbb{E}_{\pi}[a_{s}^{\pi}(t)|s(1)],\label{eq:nn4} \\
    & \nu_{r}(\pi;T,s(1)) = \frac{1}{T}\sum_{t=1}^{T}\mathbb{E}_{\pi}[a_{r}^{\pi}(t)|s(1)].\label{eq:nn5}
\end{align}
We denote the corresponding limsups of (\ref{eq:nn3})-(\ref{eq:nn5}) with $\Delta(\pi;s(1))$, $\nu_{s}(\pi;s(1))$, and $\nu_{r}(\pi;s(1))$, respectively.  Thus,
\begin{align}
     &\Delta(\pi;s(1)) = \limsup_{T\rightarrow\infty} \Delta(\pi;T,s(1)), \\ 
     &\nu_{s}(\pi;s(1)) = \limsup_{T\rightarrow\infty} \nu_{s}(\pi;T,s(1)), \\ 
     &\nu_{r}(\pi;s(1))= \limsup_{T\rightarrow\infty} \nu_{r}(\pi;T,s(1)).
\end{align}

In this work, we aim to minimize the long-term average age of the monitor subject to the long-term average sampling and transmission resource constraints, i.e., 
\begin{align}\label{eq:opt_prob}
    \Delta^{*}(s(1))= \inf_{\pi\in\Pi} & \quad \Delta(\pi;s(1))\nonumber \\  
    \textrm{s.t.} &  \quad  \nu_{s}(\pi;s(1)) \leq f_{s}, \nonumber\\ 
    & \quad \nu_{r}(\pi;s(1)) \leq f_{r},
\end{align}
where $\Pi$ is the set of all causal policies. Throughout the remainder of this paper, we assume that the initial state is $s(1)=(1,\infty)$ when it is not mentioned explicitly. Thus, we remove the dependence on $s(1)$ in the subsequent expressions. 

\section{Policy Design}\label{sec:insta}
 We first establish a universal lower bound on the optimal average age of the monitor. This bound will be used to obtain a performance guarantee for the stationary randomized policy introduced below. We define,
\begin{align}\label{eq:nn7}
    \bar{\eta}_{\min} = \min\{f_{s},qf_{r}\}.
\end{align}
\begin{theorem}\label{th:LB}
    The optimal average age of the monitor for the problem in (\ref{eq:opt_prob}), $\Delta^{*}$, is lower bounded as
    \begin{align}
        \Delta^* \geq \frac{1}{2}\left(1+\frac{1} {\bar{\eta}_{\min}}\right).
    \end{align}
\end{theorem}
The proof of Theorem~\ref{th:LB} is provided in Appendix~\ref{appen:th1}.

Now, we introduce a stationary randomized policy with non-negative parameters $\omega_{0}$, $\omega_{1}$, and $\omega_{2}$, satisfying  $\omega_{0}+\omega_{1}+\omega_{2}=1$. We denote this policy by $\pi_{\textrm{SR}}(\omega_{0},\omega_{1},\omega_{2})$. Under the policy $\pi_{\textrm{SR}}(\omega_{0},\omega_{1},\omega_{2})$, in every slot, the sampler-transmitter pair independently makes one of the following three selections. With probability $\omega_{0}$, it remains idle. With probability $\omega_{1}$, it retransmits the stored packet if a packet
is available at the transmitter; if no packet is available, it remains idle. With probability $\omega_{2}$, it samples a fresh packet and transmits it immediately. The randomized selection in each slot is independent of the age of the monitor, the age of the stored packet, the selections in all previous slots, the channel realizations, and all available feedback information. We denote this class of stationary randomized policies by $\Pi_{\textrm{SR}}$.

\begin{lemma}\label{lemma:n1}
    For $\omega_{2}>0$, the policy $\pi_{\textrm{SR}}=\pi_{\textrm{SR}} (\omega_{0},\omega_{1},\omega_{2})$ satisfies
    \begin{align}
        \nu_s(\pi_{\textrm{SR}})&=\omega_{2},\label{eq:n11}\\
        \nu_r(\pi_{\textrm{SR}})&\leq\omega_{1}+\omega_{2},\label{eq:n12}\\
        \Delta(\pi_{\textrm{SR}})&=\frac{1}{\omega_{2}}+\frac{1-q}{q(\omega_{1}+\omega_{2})}.\label{eq:n13}
    \end{align}
\end{lemma}
The proof of this lemma is provided in Appendix~\ref{appen:th:sr_four_approx}. 

From Lemma~\ref{lemma:n1}, the conditions
\begin{align}\label{eq:n34}
    \omega_{2}\leq f_s, \qquad \omega_{1}+\omega_{2}\leq f_r
\end{align}
are sufficient for $\pi_{\textrm{SR}}$ to satisfy the sampling-rate and transmission-rate constraints, respectively. Thus, we consider the following optimization problem,
\begin{align}\label{eq:n35}
    \underset{\omega_{0},\omega_{1},\omega_{2}}{\operatorname{minimize}} & \quad \frac{1}{\omega_{2}} +\frac{1-q}{q(\omega_{1}+\omega_{2})} \nonumber\\
    \textrm{s.t.} &\quad \omega_{2}\leq f_s,\nonumber\\
    &\quad \omega_{1}+\omega_{2}\leq f_r,\nonumber\\
    &\quad \omega_{0}+\omega_{1}+\omega_{2}=1,\nonumber\\
    &\quad \omega_{0}\geq 0, \ \omega_{1}\geq0,\ \omega_{2}>0.
\end{align}
The actual transmission rate may be strictly smaller than $\omega_{1}+\omega_{2}$, since a retransmission selection results in idling whenever no packet is available. Thus, (\ref{eq:n35}) finds the best stationary randomized policy
among those whose parameters satisfy the sufficient conditions in (\ref{eq:n34}).

\begin{theorem}\label{th:sr_four_approx}
    The optimization problem in (\ref{eq:n35}) admits the following optimal solution,
    \begin{align}
        \omega_{0}^{*}&=1-f_r,\label{eq:n36}\\
        \omega_{1}^{*}&=f_r-\min\{f_s,f_r\},\label{eq:n37}\\
        \omega_{2}^{*}&=\min\{f_s,f_r\}.\label{eq:n38}
    \end{align}
    The policy $\pi_{\textrm{SR}}^{*}=\pi_{\textrm{SR}}(\omega_{0}^{*},\omega_{1}^{*},\omega_{2}^{*})$ satisfies both constraints in
    (\ref{eq:opt_prob}), and
    \begin{align}\label{eq:n39}
        \Delta(\pi_{\textrm{SR}}^{*})=\frac{1}{\min\{f_s,f_r\}}+\frac{1-q}{qf_r}.
    \end{align}
    Moreover,
    \begin{align}\label{eq:n40}
        \Delta(\pi_{\textrm{SR}}^{*}) \leq \frac{4}{1+\bar{\eta}_{\min}}\Delta^{*}\leq4\Delta^{*}.
    \end{align}
\end{theorem}
The proof of this theorem is provided in Appendix~\ref{appen:th:sr_four_approx}. 

Now, we formulate (\ref{eq:opt_prob}) as a CMDP. First, we define the components of this CMDP. 

\textit{State space:} We denote a state of the system by $s=(v_{1},v_{2})$, where $v_{1}$ is the age of the monitor and $v_{2}$ is the age of the packet at the transmitter. If the transmitter has no update packet, we let $v_{2}=\infty$. For a state $s$, when $v_{2}<\infty$, we assume that $v_{2}\leq v_{1}$. We denote the state space by $\mathcal{S}$. For simplicity of presentation, we denote a state $s=(v_{1}+x,\infty)$ with $s_{np}^{v_{1}+x}$ for $x\geq 0$, where $np$ stands for no packet. Similarly, we denote a state $s=(v_{1}+x,v_{2}+y)$, $x,y\geq 0$, with $s_{ap,v_{2}+y}^{v_{1}+x}$ where $ap$ stands for an available packet.  

\textit{Action space:} We denote the action space with $\mathcal{A}=\{0,1,2,3\}$. Depending on the state, a subset of $\mathcal{A}$ remains valid. For a state of the form $(v_{1},\infty)$, the valid actions are $a=0$, $a=2$, and $a=3$, where $a=0$ implies that the sampler stays idle and does not sample a packet, $a=2$ implies that the sampler samples a new packet and the transmitter transmits it, and $a=3$ implies that the sampler samples a packet, however, the transmitter does not transmit the packet. For a state $(v_{1},v_{2})$, all actions in the set $\mathcal{A}$ are valid. Under action $a=0$, both the sampler and the transmitter stay idle. For action $a=1$, the sampler stays idle, and the transmitter transmits the packet with age $v_{2}$. For action $a=2$, the sampler samples a new packet, and the transmitter drops the stale packet and transmits the newly sampled packet. For action $a=3$, the sampler samples a new packet, and the transmitter replaces the stale packet with the newly sampled packet; however, it does not transmit it. We can show that action $a=3$ can be excluded without loss of optimality. Thus, for the rest of this paper, we do not consider action $a=3$, and we omit policies which employ action $a=3$ for some state $s\in\mathcal{S}$ from the policy set $\Pi$.

\textit{Cost:} For a state-action pair $(s,a)$, we denote the one-step cost by $C(s,a)=v_{1}$, i.e., the age of the monitor. 

\textit{Transition probability:} We denote the transition probability from state $s$ to $s'$ under action $a$ by $P_{a}(s,s')$. Below, we denote all possible non-zero transition probabilities,
\begin{align}
    &P_{0}(s_{np}^{v_{1}},s_{np}^{v_{1}+1}) = 1, 
    && P_{2}(s_{np}^{v_{1}},s_{np}^{1}) = q, 
    && P_{2}(s_{np}^{v_{1}},s_{ap,1}^{v_{1}+1}) = (1-q), 
    && P_{0}(s_{ap,v_{2}}^{v_{1}},s_{ap,v_{2}+1}^{v_{1}+1}) = 1, \nonumber\\
    & P_{2}(s_{ap,v_{2}}^{v_{1}},s_{ap,1}^{v_{1}+1}) = (1-q), 
    && P_{2}(s_{ap,v_{2}}^{v_{1}},s_{np}^{1}) = q, 
    && P_{1}(s_{ap,v_{2}}^{v_{1}}, s_{ap,v_{2}+1}^{v_{1}+1}) =(1-q), 
    && P_{1}(s_{ap,v_{2}}^{v_{1}},s_{np}^{v_{2}+1}) = q. 
\end{align}
We now reformulate (\ref{eq:opt_prob}) as,
\begin{align}\label{eq:opt_prob_cmdp}
    \inf_{\pi\in\Pi} & \quad  \limsup_{T\rightarrow \infty} \frac{1}{T} \sum_{t=1}^{T} \mathbb{E}_{\pi} [C(s(t),a(t))|(1,\infty)] \nonumber \\  
    \textrm{s.t.} &  \quad \limsup_{T\rightarrow\infty} \frac{1}{T}  \sum_{t=1}^{T}\mathbb{E}_{\pi}[\bm{1}_{\{a(t)=1\}\cup\{a(t)=2\}}|(1,\infty)]  \leq  f_{r}  \nonumber, \\ 
    & \quad \limsup_{T\rightarrow\infty}\frac{1}{T}\sum_{t=1}^{T}\mathbb{E}_{\pi}[\bm{1}_{\{a(t)=2\}}|(1,\infty)] \leq f_{s},
\end{align}
where $\bm{1}_{\{a(t)=2\}}$ and $\bm{1}_{\{a(t)=1\} \cup \{a(t)=2\}}$ denote the indicator functions for the events that a sampling and a transmission occur at time $t$, respectively. For fixed Lagrangian multipliers $\lambda\geq 0$ and $\mu\geq 0$, we write the Lagrangian-relaxed MDP of (\ref{eq:opt_prob_cmdp}) as
\begin{align}\label{eq:20}
      \inf_{\pi\in\Pi} \  \limsup_{T\rightarrow \infty} \frac{1}{T} \sum_{t=1}^{T} \mathbb{E}_{\pi} &[C(s(t),a(t)) + \mu \bm{1}_{\{a(t)=2\}} + \lambda\bm{1}_{\{a(t)=1\}\cup\{a(t)=2\}}|(1,\infty)].
\end{align}
The state space, action space, and transition probabilities for the MDP in (\ref{eq:20}) are the same as those of the primal CMDP in (\ref{eq:opt_prob_cmdp}). For a state-action pair $(s, a)$, we denote the cost of the Lagrangian-relaxed MDP as $C_{1}(s,a)$,
\begin{align}
    C_{1}(s,a) = C(s,a) + \mu \bm{1}_{\{a=2\}} + \lambda \bm{1}_{\{a=1\}\cup\{a=2\}}.
\end{align}
The Lagrange multiplier $\mu$ can be interpreted as the unit-cost associated with sampling a fresh packet, while $\lambda$ can be interpreted as the unit-cost associated with transmission of a fresh packet or an older packet. For an initial state $s\in\mathcal{S}$, policy $\pi\in\Pi$, and $0<\alpha<1$, we denote the discounted value function by,
\begin{align}\label{eq:22}
    V_{\alpha}^{\pi}(s) = \sum_{t=1}^{\infty}\alpha^{t-1}\mathbb{E}_{\pi}[ C_{1}(s(t),a(t))|s(1)=s].
\end{align}
In the literature, the discounted value function $V_{\alpha}(s)$, is defined as,
\begin{align}\label{eq:23}
    V_{\alpha}(s) = \inf_{\pi\in\Pi} \ V_{\alpha}^{\pi}(s).
\end{align}
From the literature, \cite{bertsekas2012dynamic,puterman2014markov}, we know that,
\begin{align}
    V_{\alpha}(s) &= \min_{a}\{V_{\alpha}(s;a)\}, \label{eq:24} \\
    V_{\alpha}(s;a) &= C_{1}(s,a) + \alpha \sum_{s'\in\mathcal{S}}P_{a}(s,s') V_{\alpha}(s'),
\end{align}
Now, consider the following iteration for $n\geq 1$, 
\begin{align}
    V_{\alpha,n}(s) &= \min_{a}\{V_{\alpha,n}(s;a)\},\label{eq:n26}\\ 
    V_{\alpha,n}(s;a) &=C_{1}(s,a) + \alpha \sum_{s'\in\mathcal{S}} P_{a}(s,s') V_{\alpha,n-1}(s').
\end{align}
where we initialize $V_{\alpha,0}(s)=0$ for all $s\in\mathcal{S}$. From \cite{sennott1989average}, we know that,
\begin{align}\label{eq:26}
    \lim_{n\rightarrow\infty} V_{\alpha,n}(s) = V_{\alpha}(s).
\end{align}
Note that, the following identities hold for $1\leq v_{2}\leq v_{1}$,
\begin{align}
    V_{\alpha,n+1}(s_{ap,v_{2}}^{v_{1}};0) &= v_{1} + \alpha V_{\alpha,n}(s_{ap,v_{2}+1}^{v_{1}+1}), \label{eq:31}\\ 
    V_{\alpha,n+1}(s_{ap,v_{2}}^{v_{1}};1) &= v_{1} + \lambda + \alpha(q V_{\alpha,n}(s_{np}^{v_{2}+1}) + (1-q) V_{\alpha,n}(s_{ap,v_{2}+1}^{v_{1}+1})), \label{eq:32} \\ 
    V_{\alpha,n+1}(s_{ap,v_{2}}^{v_{1}};2) &= v_{1} + \lambda+\mu + \alpha(q V_{\alpha,n}(s_{np}^{1}) + (1-q) V_{\alpha,n}(s_{ap,1}^{v_{1}+1})), \label{eq:33}\\
    V_{\alpha,n+1}(s_{np}^{v_{1}};0) &= v_{1} + \alpha V_{\alpha,n}(s_{np}^{v_{1}+1}), \label{eq:34} \\ 
    V_{\alpha,n+1}(s_{np}^{v_{1}};2) &=  v_{1} + \lambda+\mu + \alpha(q V_{\alpha,n}(s_{np}^{1}) + (1-q) V_{\alpha,n}(s_{ap,1}^{v_{1}+1})). \label{eq:35}
\end{align}

Next, we present three lemmas that establish several properties of the discounted value function. These properties will be used in the proofs of the subsequent theorems. The proofs of these lemmas are provided in Appendix~\ref{appen:discounted_properties}. 

In the first lemma, we establish lower and upper bounds on $V_{\alpha}(s_{ap,v_{2}}^{v_{1}})$.

\begin{lemma}\label{lemma:2}
    For $0<\alpha<1$, $V_{\alpha}(s_{ap,v_{2}}^{v_{1}})$ is lower bounded by $ \frac{v_{1}}{1-\alpha(1-q)}$ and is upper bounded by $\frac{v_{1}}{1-\alpha (1-q)} + K$, where 
    \begin{align}
        K = \frac{\lambda + \mu + \alpha q V_{\alpha}(s_{np}^{1})}{1 - \alpha(1 - q)} + \frac{\alpha(1 - q)}{(1 - \alpha(1 - q))^{2}}<\infty.
    \end{align}
\end{lemma}

In the next lemma, we show that the transmitter having a packet whose age is equal to the age of the monitor is equivalent to having no packet at the transmitter.

\begin{lemma}\label{lemma:1}
    For $0<\alpha<1$, the following relations hold,
    \begin{align}
         V_{\alpha}(s_{ap,v_{1}}^{v_{1}}) &= V_{\alpha}(s_{np}^{v_{1}}), \label{eq:27}\\
         V_{\alpha}(s_{ap,v_{1}}^{v_{1}};0) &= V_{\alpha}(s_{np}^{v_{1}};0), \label{eq:nnn26} \\  
         V_{\alpha}(s_{ap,v_{1}}^{v_{1}};2) &= V_{\alpha}(s_{np}^{v_{1}};2). \label{eq:nnn27}
    \end{align}
\end{lemma}

Finally, in the next lemma, we establish the monotonicity properties of $V_{\alpha}(s)$.

\begin{lemma}\label{lemma:4}
    For $0<\alpha<1$, $V_{\alpha}(s_{np}^{v_{1}})$ and $V_{\alpha}(s_{ap,v_{2}}^{v_{1}})$ are increasing functions of $v_{1}$ and $V_{\alpha}(s_{ap,v_{2}}^{v_{1}})$ is an increasing function of $v_{2}$. Moreover, for every $x\geq0$, the following hold,
    \begin{align}
        V_{\alpha}(s_{ap,v_{2}}^{v_{1}+x}) &\geq V_{\alpha}(s_{ap,v_{2}}^{v_{1}}) + x, \label{eq:nn25}\\
        V_{\alpha}(s_{np}^{v_{1}+x}) &\geq V_{\alpha}(s_{np}^{v_{1}})+x. \label{eq:nn26}
    \end{align}
\end{lemma}

Now, we show that, for the Lagrangian-relaxed MDP, when the age of the monitor is sufficiently large, there exists an optimal policy that does not choose the idling action. Specifically, when a packet is available, the policy selects either retransmission or sampling and immediately transmitting a fresh packet, while when no packet is available, it selects sampling and immediately transmitting a fresh packet.

\begin{theorem}\label{th:2}
    There exists $\bar{v}_{1}<\infty$ and a stationary policy $\pi_{L}^{*}$ that is optimal for the Lagrangian-relaxed MDP in (\ref{eq:20}), such that, $\pi_{L}^{*}$ does not select action $a=0$ for a state $s_{ap,v_{2}}^{v_{1}}$, where $v_{1}\geq \bar{v}_{1}$. Moreover, for state $s_{np}^{v_{1}}$, where $v_{1}\geq \bar{v}_{1}$, $\pi_{L}^{*}$ chooses action $a=2$ over action $a=0$.      
\end{theorem}

The proof of this theorem is provided in Appendix~\ref{appen:th:2}.

\begin{remark}\label{rem:no_packet}
    The result in Theorem~\ref{th:2} also holds for the discounted-cost problem for every $0<\alpha<1$. In particular, for each $\alpha$, there exists a finite $\bar{v}_{1}(\alpha)$ and a discounted-cost optimal stationary policy having the same structure as in Theorem~\ref{th:2}. Moreover, for every $\alpha\geq\frac{1}{2}$, we can use the same $\bar{v}_{1}$ defined in the proof of Theorem~\ref{th:2}, independent of $\alpha$.
\end{remark}

Now, we introduce the following property of the discounted-cost problem. For every $\mu,\lambda\geq 0$, there exists $0\leq\bar{\alpha}(\mu,\lambda)<1$, such that, for every $\bar{\alpha}(\mu,\lambda)<\alpha<1$, the discounted-cost problem in~(\ref{eq:23}) admits an optimal policy with the following property: if this policy selects $a=2$ in the state $s_{np}^{{v}_{1}}$, then it also selects $a=2$ in $s_{np}^{{v}_{1}+x}$ for every $x\geq0$. We refer to this property as the \emph{no-packet threshold property}. This property is intuitive because, as the age of the monitor increases, sampling and transmitting a fresh packet becomes increasingly desirable. Remark~\ref{rem:no_packet} shows that this property holds for all sufficiently large values of the age of the monitor. We first derive its structural consequences and subsequently prove that it always holds.

The next result establishes a less intuitive structural property of an optimal policy for (\ref{eq:20}). It shows that, the intuitive no-packet threshold property has a nontrivial implication when a packet is already available. In particular, there exists an optimal policy for (\ref{eq:20}) such that, if it samples and transmits a fresh packet in a no-packet state $s_{np}^{v_{1}}$, then, for the same monitor age $v_{1}$, it never idles when a packet is available at the transmitter, irrespective of the age of the stored packet. In such a state, the policy either retransmits the stored packet or samples a fresh packet and transmits it immediately.

\begin{theorem}\label{th:3}
    Under the no-packet threshold property, there exists an optimal stationary policy $\pi_{L}^{*}$ for the Lagrangian-relaxed MDP in (\ref{eq:20}) such that, if  $\pi_{L}^{*}(s_{np}^{v_{1}})=2$, then  $\pi_{L}^{*}(s_{ap,v_{2}}^{v_{1}})\in\{1,2\}$, for all $1\leq v_{2}\leq v_{1}$.
\end{theorem}

The proof of this theorem is provided in Appendix~\ref{appen:th:3}.

\begin{remark}\label{rem:2}
    Under the no-packet threshold property, the structural result in Theorem~\ref{th:3} also holds for the discounted-cost problem. In particular, for every discount factor $\alpha$ satisfying $\bar{\alpha}(\mu,\lambda)<\alpha<1$, and
    \begin{align}\label{eq:n73}
        \lambda\leq\frac{\alpha q}{1-\alpha},
    \end{align}
    there exists an optimal stationary policy for the discounted-cost problem such that, if it selects action $a=2$ at a no-packet state $s_{np}^{v_{1}}$, then it does not select action $a=0$ at any state $s_{ap,v_{2}}^{v_{1}}$, for all $1\leq v_{2}\leq v_{1}$.
\end{remark}

\begin{remark}\label{rem:3}
    For $\alpha>0$ satisfying $\frac{\lambda}{\lambda+q}\leq \alpha <1 $, we define 
    \begin{align}\label{eq:npt_eventual_threshold}
        \hat v_1(\alpha)= \min\{ k\in\mathbb{N} \ | \ V_\alpha(s_{np}^{v_{1}};2) \leq V_\alpha(s_{np}^{v_{1}};0)  \text{ for every }v_{1}\geq k \}.
    \end{align}
    Remark~\ref{rem:no_packet} guarantees that the set in (\ref{eq:npt_eventual_threshold}) is nonempty and, consequently, that $\hat v_{1}(\alpha)<\infty$.
    
    The definition of $\hat{v}_{1}(\alpha)$ guarantees that action $a=2$ is optimal at every no-packet state whose monitor age is at least $\hat {v}_{1}(\alpha)$. However, it does not state the optimality of action $a=0$ at monitor ages below $\hat {v}_{1}(\alpha)$. In particular, action $a=2$ may be optimal at some no-packet states with monitor age smaller than $\hat{v}_{1}(\alpha)$. Thus, at this point $\hat v_1(\alpha)$ is a different quantity than the no-packet threshold $\bar{v}_{1}(\alpha)$. In Theorem~\ref{th:no_packet_threshold}, we show that action $a=0$ is strictly optimal at every no-packet state whose monitor age is smaller than $\hat{v}_{1}(\alpha)$, thereby establishing that $\hat{v}_{1}(\alpha)=\bar{v}_{1}(\alpha)$.
    
    The discounted-cost part of the proof of Theorem~\ref{th:3} uses the no-packet threshold property only through the fact that action $2$ is optimal at every no-packet state above the threshold $\bar{v}_{1}(\alpha)$. It does not use the optimal actions at the no-packet states below that threshold. Therefore, the same backward-induction argument applies with $\hat v_1(\alpha)$ in place of $\bar v_1(\alpha)$. Moreover, the condition $\alpha \geq \frac{\lambda}{\lambda+q}$ is the same as the condition of (\ref{eq:n73}). Thus,  for $v_{1}\geq \hat{v}_{1}$ and $1\leq v_{2}\leq v_{1}$, we have,
    \begin{align}
        V_\alpha(s_{ap,v_2}^{v_1})= \min\left\{V_\alpha(s_{ap,v_2}^{v_1};1), V_\alpha(s_{ap,v_2}^{v_1};2)\right\}.
    \end{align}
\end{remark}

The next result shows that, whenever the age of the monitor exceeds the threshold specified by the no-packet threshold property, the discounted value function is affine in the age of the monitor, with a slope independent of the age of the stored packet. The remaining term depends only on the age of the stored packet. 

\begin{theorem}\label{th:4}
    Under the no-packet threshold property with threshold $\bar{v}_{1}$, for $\bar{\alpha}(\mu,\lambda)<\alpha<1$ and $\lambda \leq \frac{\alpha q}{1-\alpha}$, the following relation holds
    \begin{align}
        V_{\alpha}(s_{ap,v_{2}}^{v_{1}}) = \frac{v_{1}}{1-\alpha(1-q)} + \hat{M}(v_{2}),
    \end{align}
    where $v_{1}\geq \bar{v}_{1}$, $1\leq v_{2}\leq v_{1}$, and $\hat{M}(v_{2})$ is independent of $v_{1}$ and is only a function of $v_{2}$.
\end{theorem}

The proof of this theorem is provided in Appendix~\ref{appen:th_4}.

\begin{theorem}\label{th:5}
    Under the no-packet threshold property with threshold $\bar{v}_{1}$, for $\bar{\alpha}(\mu,\lambda)<\alpha<1$, $\lambda \leq \frac{\alpha q}{1-\alpha}$, and for a state $s_{ap,v_{2}}^{v_{1}}$, where $v_{1}\geq \bar{v}_{1}$ and $v_{2}\leq v_{1}$, there exists an optimal action corresponding to the discounted-cost problem in (\ref{eq:23}), that does not depend on $v_{1}$. Specifically, there exists a finite $\bar{v}_{2}\in\mathbb{N}$, such that, irrespective of the age of the monitor $v_{1}$, whenever the age of the packet is greater than or equal to $\bar{v}_{2}$, then action $a=2$ is optimal, otherwise action $a=1$ is optimal.
\end{theorem}

The proof of this theorem is provided in Appendix~\ref{appen:th:5}.

\begin{remark}\label{rem:4}
    Let $\hat{v}_{1}(\alpha)$ be defined as in Remark~\ref{rem:3}, and suppose that $\frac{\lambda}{\lambda+q}<\alpha<1$. Similar to Remark~\ref{rem:3}, the proofs of Theorems~\ref{th:4} and~\ref{th:5} use the no-packet threshold $\bar{v}_{1}(\alpha)$ only to establish that, at every packet-available state $s_{ap,v_{2}}^{v_{1}}$ with $v_{1}\geq\bar{v}_{1}(\alpha)$, an optimal action can be selected from $\{1,2\}$, and hence action $a=0$ can be excluded. Remark~\ref{rem:3} establishes the same property for every packet-available state $s_{ap,v_{2}}^{v_{1}}$ with $v_{1}\geq\hat{v}_{1}(\alpha)$. Therefore, the same argument used in the proof of Theorem~\ref{th:4}, with $\hat{v}_{1}(\alpha)$ in place of $\bar{v}_{1}(\alpha)$, shows that there exists a function $\hat{M}$, such that, for $v_{1}\geq\hat{v}_{1}(\alpha)$ and $1\leq v_{2}\leq v_{1}$, the following holds
    \begin{align}\label{eq:affine_above_hat_v1}
        V_{\alpha}(s_{ap,v_{2}}^{v_{1}}) = \frac{v_{1}}{1-\alpha(1-q)} + \hat{M}(v_{2}),
    \end{align}
    where $\hat{M}(v_{2})$ is independent of $v_{1}$.
    
    Consequently, the same argument used in the proof of Theorem~\ref{th:5} shows that for every $v_{1}\geq\hat{v}_{1}(\alpha)$, an optimal action is independent of $v_{1}$. Specifically, there exists a finite $\bar{v}_{2}\in\mathbb{N}$, such that, action $a=1$ is optimal when $v_{2}<\bar{v}_{2}$, whereas action $a=2$ is optimal when $v_{2}\geq\bar{v}_{2}$.
\end{remark}

Theorems~\ref{th:3}-\ref{th:5} establish the structural consequences of the no-packet threshold property. We next prove that, for every $\mu,\lambda\geq0$, this property holds.

\begin{theorem}\label{th:no_packet_threshold}
    For $\lambda,\mu\geq0$, let
    \begin{align}\label{eq:npt_alpha_bar}
        \bar\alpha(\mu,\lambda)= \frac{\lambda}{\lambda+q}.
    \end{align}
    For every $\bar\alpha(\mu,\lambda)<\alpha<1$, there exists a deterministic stationary optimal policy $\pi_\alpha^{*}$ for the discounted-cost problem in~(\ref{eq:23}), and a finite $\bar v_1\in\mathbb N$, such that
    \begin{align}\label{eq:npt_statement}
        \pi_\alpha^*(s_{np}^{v_1}) =
        \begin{cases}
            0,&1\leq v_1<\bar v_1,\\
            2,&v_1\geq\bar v_1.
        \end{cases}
    \end{align}
    Consequently, if $\pi_\alpha^*(s_{np}^{v_1})=2$, then $\pi_\alpha^*(s_{np}^{v_1+x})=2$ for every integer $x\geq0$.
\end{theorem}

The proof of Theorem~\ref{th:no_packet_threshold} is provided in Appendix~\ref{appen:no_packet_threshold}.

We now define a specific class of policies, denoted by $\bar{\Pi}$. Each policy $\pi\in\bar{\Pi}$ is parameterized by two positive integers, $W$ and $R$, and is referred to as a $(W,R)$ policy. The parameter $W$ specifies a waiting threshold. Following each successful reception, the sampler-transmitter pair remains idle until the age at the monitor reaches $W$. At that time, the sampler generates a fresh packet, which is immediately transmitted. If the transmission is successful, a new cycle begins and the sampler-transmitter pair again waits until the monitor age reaches $W$. Otherwise, the transmitter repeatedly retransmits the same packet until either it is successfully delivered or its age reaches $R$, the packet-replacement threshold. When the age of the packet stored at the transmitter reaches $R$, the sampler discards that packet, generates a fresh packet, and the transmitter immediately transmits the newly sampled packet. The sampler-transmitter pair continues transmitting, retransmitting, and replacing packets according to this rule until a packet is successfully received by the monitor, at which point a new cycle begins. The policy set $\bar{\Pi}$ also includes policies of the form $(W,\infty)$. Under a $(W,\infty)$ policy, once a packet is sampled, the transmitter repeatedly transmits the same packet until it is successfully delivered to the monitor, without replacing it with a fresher packet. Thus, 
\begin{align}
    \bar{\Pi} =\{(W,R)| (W,R)\in\mathbb{N}^{2}\} \cup \{(W,\infty)|W\in\mathbb{N}\}.
\end{align}

Note that the $(W,R)$ policy does not specify the action for state $s_{ap,v_{2}}^{v_{1}}$ when $v_{1}\leq W$. Recall that we consider that the initial state of the system is $s_{np}^{1}$, i.e., $(1,\infty)$, thus a state $s_{ap,v_{2}}^{v_{1}}$, where $v_{1}\leq W$ is not reachable under the policy $(W,R)$. Thus, for such states, we can assign any admissible action from the set $\mathcal{A}$. A pictorial illustration of a $(W,R)$ policy is provided in Fig.~\ref{fig:(W,R)}.

\begin{figure}[t]
    \centerline{\includegraphics[width =0.5 \columnwidth]{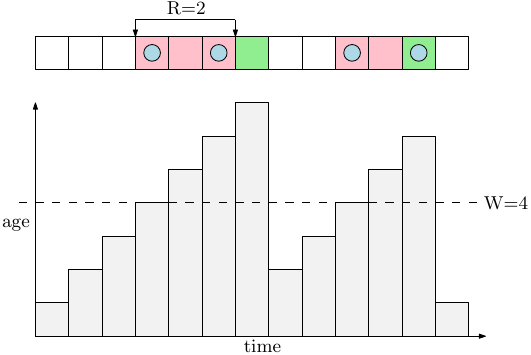}}
    \caption{Illustration of a $(W,R)$ policy with $W=4$ and $R=2$. The cyan circles indicate the sampling of fresh update packets, while the pink and green boxes represent unsuccessful and successful transmissions, respectively.}
    \label{fig:(W,R)}
\end{figure} 

\begin{corollary}\label{corr:2}
    For a fixed $\lambda\geq 0$ and $\mu\geq 0$, there exists a policy in $\bar{\Pi}$ that is optimal for the MDP in (\ref{eq:20}). 
\end{corollary}

The proof of this corollary is provided in Appendix~\ref{appen:corr:2}.

\begin{remark}\label{rem:5}
    In \cite{gong2022sleep}, the authors consider sensing and transmission costs and establish an optimal policy with a two-threshold structure closely related to ours. However, our proof techniques differ from theirs. Our proof of Theorem~\ref{th:4} uses a contraction argument to establish the affine representation of $V_\alpha(s_{ap,v_{2}}^{v_1})$ directly for $v_1\geq\bar v_1$. In \cite{gong2022sleep}, the corresponding property is first established for sufficiently large monitor ages by induction on the value iteration index and is then extended to smaller monitor ages by backward induction. Moreover, our arguments retain the original state space, whereas their analysis introduces auxiliary states where the packet age exceeds the monitor age. Our presentation also differs: we first establish the structural consequences of the no-packet threshold property and then prove that property. Beyond these proof differences, our problem imposes separate average sampling- and transmission-rate constraints, whereas the constrained formulation discussed in \cite{gong2022sleep} imposes a single constraint on the combined average sensing and transmission energy consumption. Our subsequent optimization results and performance guarantees therefore concern a different constrained problem and differ from their results.
\end{remark}

For a policy $\pi=(W,R)\in\bar{\Pi}$, we next derive the average sampling rate, average transmission rate, and average age at the monitor, denoted by $\nu_{s}(W,R)$, $\nu_{r}(W,R)$, and $\Delta(W,R)$, respectively. For $R=\infty$, the corresponding quantities are obtained by taking the limit as $R\rightarrow\infty$ in the expressions derived below for finite $R$. We partition the entire time horizon into a sequence of frames. Each frame ends when a packet is successfully delivered to the monitor, and the subsequent frame begins immediately thereafter. Let $G_n$ denote the number of transmissions in frame $n$, and let $V_{R,n}$ denote the age of the monitor at its beginning. For finite $R$, $n\geq 2$, we have $V_{R,n}=1+((G_{n-1}-1)\bmod R)$. The sequence $\{G_n\}$ is i.i.d.~geometric with success probability $q$, $\{V_{R,n}\}_{n\geq 2}$ is identically distributed, and $V_{R,n}$ is independent of $G_n$. Let $L_{n}(W,R)$ and $H_{n}(W,R)$ denote the frame length and idling duration for frame $n$. Thus,
\begin{align}
    H_{n}(W,R) &= (W-V_{R,n})^{+},\label{eq:n147} \\ 
    L_{n}(W,R) &= H_{n}(W,R) + G_{n}.\label{eq:144} 
\end{align}
Let $V_R$ and $G$ denote independent random variables distributed as $V_{R,n}$ and $G_n$, respectively, and define $H(W,R)$ and $L(W,R)$ analogously to $H_n(W,R)$ and $L_n(W,R)$ using $V_R$ and $G$. Now, for $i\in\{1,\ldots,R\}$,
\begin{align}\label{eq:132}
    \mathbb{P}(V_{R}= i) = \sum_{j=0}^{\infty}\mathbb{P}(G=j R +i) = \frac{q (1-q)^{i-1}}{1-(1-q)^{R}}.
\end{align}
Let $N_{R,n}=\left\lceil \frac{G_n}{R}\right\rceil$ denote the number of fresh samples generated in frame $n$, and let $N_R$ denote a random variable having the same distribution as $N_{R,n}$. Then, 
\begin{align}
    \mathbb{E}[{N}_{R}] = \frac{1}{1-(1-q)^{R}}.
\end{align}
We can show that from the second frame onward, the frame lengths and the corresponding frame rewards are jointly stationary and ergodic. Hence, by applying the ergodic theorem, taking the ratio of the corresponding sample averages, and using the bounded convergence theorem, we have
\begin{align}
    \nu_{r}(W,R) &= \frac{\mathbb{E}[G]}{\mathbb{E}[L(W,R)]} = \frac{1}{q \mathbb{E}[L(W,R)]}, \label{eq:149}\\ 
    \nu_{s}(W,R) &= \frac{\mathbb{E}[N_{R}]}{\mathbb{E}[L(W,R)]} = \frac{\frac{1}{1-(1-q)^{R}}}{ \mathbb{E}[L(W,R)]}. \label{eq:150}
\end{align}
Thus,
\begin{align}
    \nu_{s}(W,R) = \rho_{R} \nu_{r}(W,R),
\end{align}
where we define $\rho_{R}$ to be $\frac{q}{1-(1-q)^{R}}$. Note that $\rho_{R}$ only depends on $R$ and is independent of $W$. Let $Y_n(W,R)$ denote the cumulative age of the monitor in frame $n$, and let $Y(W,R)$ denote a random variable having the same distribution as $Y_n(W,R)$ for $n\geq2$. Thus,
\begin{align}\label{eq:n157}
    Y(W,R) = V_{R} H({W,R}) + V_{R} G + \frac{H(W,R)(H(W,R)-1)}{2}  + H(W,R)G + \frac{G(G-1)}{2}.
\end{align}
We can show that the time averages of the age are uniformly integrable. Thus, by applying the ergodic theorem, taking the ratio of the corresponding sample averages, and using Vitali's convergence theorem, we have
\begin{align}\label{eq:n158}
    \Delta(W,R) = \frac{\mathbb{E}[Y(W,R)]}{\mathbb{E}[L(W,R)]}.
\end{align}

In \cite{banerjee2025equidistant}, the authors showed that with an average sampling constraint $f_{s}$ and unconstrained transmission, an optimal policy $\pi_{s}^{*}$ lies in the set of equidistant-sampling policies. An equidistant-sampling policy with period $R$ generates consecutive samples exactly $R$ slots apart. In particular, if $\frac{1}{f_{s}}$ is not an integer, $\pi_{s}^{*}$ is obtained by randomizing between the equidistant-sampling policies, $\pi_{s,1}^{*}$ and $\pi_{s,2}^{*}$, with periods $R_{1}=\left\lfloor\frac{1}{f_s}\right\rfloor$ and $R_{2}=\left\lceil\frac{1}{f_s}\right\rceil$, respectively. More precisely, at time slot $1$, choose $\pi_{s,1}^{*}$ with probability
\begin{align}\label{eq:theta_s}
    \theta_{s} = \frac{f_s-\frac{1}{R_2}}{\frac{1}{R_1}-\frac{1}{R_2}},
\end{align}
and choose $\pi_{s,2}^{*}$ with probability $1-\theta_{s}$, and stick to the chosen policy over the entire time horizon. When $\frac{1}{f_s}\in\mathbb{N}$, we get $\pi_{s}^{*}$ by choosing $R_1=R_2=\frac{1}{f_s}$ and $\theta_s=1$. Thus, the average sampling rate of this policy is $f_{s}$. Note that, in the notation of this paper, the equidistant-sampling policy with period $R$ is exactly the $(R,R)$ policy.

Let us assume that $\frac{1}{f_{s}}$ is not an integer, then the average transmission rate for the policy $\pi_{s}^{*}$ is
\begin{align}
    \theta_{s} \nu_{r}(R_{1},R_{1}) + (1-\theta_{s}) \nu_{r}(R_{2},R_{2}) \leq \frac{\theta_{s} \nu_{s}(R_{1},R_{1}) + (1-\theta_{s}) \nu_{s}(R_{2},R_{2})}{q}=\frac{f_{s}}{q}.    
\end{align}
Thus, if the constraints satisfy $\frac{f_{s}}{q}\leq f_{r}$, then $\pi_{s}^{*}$ is feasible for the optimization problem in (\ref{eq:opt_prob}). Moreover, $\pi_{s}^{*}$ is optimal for the problem with sampling-rate constraint $f_s$ and no transmission-rate constraint. Therefore, $\pi_s^{*}$ is also optimal for (\ref{eq:opt_prob}) whenever $f_s\leq qf_{r}$. Similarly, we can show that when $\frac{1}{f_{s}}$ is an integer, $\pi^{*}_{s}$ is optimal for (\ref{eq:opt_prob}). 

Next, we consider the case $f_s\geq f_{r}$. Now, consider a system with only an average transmission constraint $f_{r}$ and no sampling constraint. We can show that it is optimal to always sample a fresh packet when transmitting. Thus, after every successful transmission, the age of the monitor $V_{R}$ becomes $1$. Following the same steps as in~\cite{ceran2019average}, we can show that there exists an optimal policy for the Lagrangian-relaxed MDP corresponding to the system with only the transmission constraint, which follows a threshold structure. That is, there exists a threshold $W$, such that, if the age of the monitor is greater than or equal to $W$, then it is always optimal to transmit a fresh update packet; otherwise, staying idle is optimal. Note that this policy corresponds to the $(W,1)$ policy in $\bar{\Pi}$. Now, from (\ref{eq:149}),
\begin{align}\label{eq:141}
  \nu_{r}(W,1)  = \frac{1}{1+q(W-1)}. 
\end{align}
If $1+\frac{1-f_{r}}{qf_{r}}$ is not an integer, then we obtain an optimal policy for the system with only the transmission constraint $f_{r}$, denoted by $\pi_t^{*}$, by randomizing between two threshold policies, $\pi_{t,1}^{*}$ and $\pi_{t,2}^{*}$, with waiting thresholds $W_1=\left\lfloor1+\frac{1-f_{r}}{qf_{r}}\right\rfloor$ and $W_2=\left\lceil1+\frac{1-f_{r}}{qf_{r}}\right\rceil$, respectively. More precisely, at time slot $1$, we choose $\pi_{t,1}^{*}$ with probability
\begin{align}\label{eq:142}
    \theta_r = \frac{f_{r}-\nu_{r}(W_{2},1)}{\nu_{r}(W_{1},1)-\nu_{r}(W_{2},1)},
\end{align}
and choose $\pi_{t,2}^{*}$ with probability $(1-\theta_r)$, and then stick to the chosen policy over the entire time horizon. When $1+\frac{1-f_{r}}{qf_{r}}\in\mathbb{N}$, then we get $\pi_{t}^{*}$ by choosing $W_{1}=W_{2}= 1+\frac{1-f_{r}}{qf_{r}}$ and $\theta_{r}$=1. Note that, since we always sample whenever we transmit, the average sampling rate is the same as the average transmission rate, which can also be seen formally from~(\ref{eq:149}) and~(\ref{eq:150}), i.e., $\nu_{r}(W,1)=\nu_{s}(W,1)$. Now, since we consider $f_s\geq f_{r}$, the policy $\pi_t^{*}$ is feasible for~(\ref{eq:opt_prob}), and since it is optimal for the system without any sampling constraint, it is also optimal for~(\ref{eq:opt_prob}). 

Next, we derive two sharper lower bounds on the optimal average age of the monitor $\Delta^{*}$ by separately considering the systems with only the sampling-rate constraint and only the transmission-rate constraint.

\begin{theorem}\label{th:7}
    The optimal average age $\Delta^{*}$ for the problem in (\ref{eq:opt_prob}) satisfies
    \begin{align}
        \Delta^{*} &\geq \frac{1}{q} +\frac{\theta_sR_1+(1-\theta_s)R_2-1}{2}, \label{eq:546}\\
        \Delta^{*} &\geq \frac{1}{2} +\frac{1}{2qf_r}+\frac{(1-q)f_r}{2q}.\label{eq:547}
    \end{align}
\end{theorem}

The proof of this theorem is provided in Appendix~\ref{appen:th:7}. 

Now, by Jensen's inequality,
\begin{align}
    \theta_sR_1+(1-\theta_s)R_2 \geq \frac{1}{\frac{\theta_s}{R_1}+\frac{1-\theta_s}{R_2}}=\frac{1}{f_s}\label{eq:76},
\end{align}
where (\ref{eq:76}) follows from (\ref{eq:theta_s}). Therefore, the right-hand side of (\ref{eq:546}) satisfies
\begin{align}
    \frac{1}{q} +\frac{\theta_sR_1+(1-\theta_s)R_2-1}{2} \geq \frac{1}{2}\left(1+\frac{1}{f_s}\right).
\end{align}
Moreover, the right-hand side of (\ref{eq:547}) satisfies
\begin{align}
    \frac{1}{2} +\frac{1}{2qf_r} +\frac{(1-q)f_r}{2q} \geq \frac{1}{2}\left(1+\frac{1}{qf_r}\right).
\end{align}
Thus, from (\ref{eq:nn7}) and Theorem~\ref{th:LB}, we have
\begin{align}
    \max\left\{\frac{1}{2}\left(1+\frac{1}{f_s}\right),\frac{1}{2}\left(1+\frac{1}{qf_r}\right)\right\}=\frac{1}{2}\left(1+\frac{1}{\bar{\eta}_{\min}}\right).
\end{align}
Therefore, the two lower bounds in Theorem~\ref{th:7} jointly strengthen the universal lower bound in Theorem~\ref{th:LB}. 

The preceding results characterize the regimes $f_s\leq qf_r$ and $f_s\geq f_r$. Thus, for the remainder of this section, we focus on the non-trivial region $qf_{r}<f_{s}<f_{r}$.  

From Corollary~\ref {corr:2}, we know that, for every pair of Lagrange multipliers, the corresponding Lagrangian-relaxed MDP admits an optimal policy in $\bar{\Pi}$. This motivates optimizing over randomized mixtures of policies in $\bar{\Pi}$. Let us consider a probability distribution $\bm{x}$, on the policy set $\bar{\Pi}$, i.e., $\bm{x} = \{x(W,R)\}$. As the policy set $\bar{\Pi}$ is countably infinite, $\bm{x}$ is an infinite-dimensional probability vector. Let us consider the following linear program (LP),
\begin{align}\label{eq:158}
    \Delta_{\textrm{LP}} = \inf_{\bm{x}} & \quad \sum_{(W,R)\in\bar{\Pi}} x(W,R) \Delta(W,R) \nonumber\\ 
    \textrm{s.t.} & \quad \sum_{(W,R)\in\bar{\Pi}} x(W,R) \nu_{s}(W,R) \leq f_{s},\nonumber\\ 
    & \quad \sum_{(W,R)\in\bar{\Pi}} x(W,R) \nu_{r}(W,R) \leq f_{r}, \nonumber\\
    & \quad \sum_{(W,R)\in\bar{\Pi}}x(W,R) =1, \nonumber \\
    & \quad x(W,R) \geq 0, \ \forall (W,R)\in\bar{\Pi}.
\end{align}
In the next theorem, we connect the solution of this LP to an optimal policy for (\ref{eq:opt_prob}).

\begin{theorem}\label{th:6}
   The LP in (\ref{eq:158}) admits an optimal solution $\bm{x}^{*}$ that assigns positive probability to at most three $(W,R)$ policies. A policy $\pi^{*}$ that selects a $(W,R)$ policy according to $\bm{x}^{*}$ at time $t=1$ and follows the selected policy for the entire time horizon is optimal for the problem in (\ref{eq:opt_prob}). Moreover, we have
   \begin{align}
       \Delta^{*}=\Delta_{\textrm{LP}}.
   \end{align}
\end{theorem}

The proof of this theorem is provided in Appendix~\ref{appen:th:6}.

We next restrict the countably infinite policy set $\bar{\Pi}$ to a finite policy set. We choose $W_0$ sufficiently large such that $(W_0,\infty)$ is strictly feasible, and define
\begin{align}
    \eta_{\min}=\min\left\{f_s-\nu_s(W_0,\infty),f_r-\nu_r(W_0,\infty)\right\}>0.
\end{align} 
For any $\epsilon>0$, we define
\begin{align}\label{eq:xi_epsilon}
    \xi_{\epsilon}=\frac{\epsilon\eta_{\min}}{4\Delta(W_{0},\infty)}.
\end{align}
We choose
\begin{align}\label{eq:W_epsilon}
    W_{\epsilon}=\max\left\{W_{0}, \left\lceil2\Delta(W_{0},\infty)\right\rceil, \left\lceil \frac{2\Delta(W_{0},\infty)}{\xi_{\epsilon}}\right\rceil\right\}.
\end{align}
For every fixed $W$, from (\ref{eq:149}), (\ref{eq:150}), and (\ref{eq:n158}), we have
\begin{align}
    \lim_{R\rightarrow\infty}\nu_{s}(W,R) &=\nu_{s}(W,\infty),\\
    \lim_{R\rightarrow\infty}\nu_{r}(W,R) &=\nu_{r}(W,\infty),\\
    \lim_{R\rightarrow\infty}\Delta(W,R)  &=\Delta(W,\infty).
\end{align}
Since the set $\{1,2,\ldots,W_{\epsilon}\}$ is finite, there exists a finite $R_{\epsilon}$ such that, for all $1\leq W\leq W_{\epsilon}$ and $R>R_{\epsilon}$,
\begin{align}
    |\nu_{s}(W,R)-\nu_{s}(W,\infty)|&\leq\xi_{\epsilon},\label{eq:R_epsilon_conditions1}\\
    |\nu_{r}(W,R)-\nu_{r}(W,\infty)|&\leq\xi_{\epsilon},\label{eq:R_epsilon_conditions2}\\
    |\Delta(W,R)-\Delta(W,\infty)|&\leq\frac{\epsilon}{2}\label{eq:R_epsilon_conditions3}.
\end{align}
We define the following finite policy set, 
\begin{align}\label{eq:Pi_epsilon}
    \bar{\Pi}_{\epsilon}=\{(W,R):1\leq W\leq W_{\epsilon},1\leq R\leq R_{\epsilon}\}\cup\{(W,\infty):1\leq W\leq W_{\epsilon}\}.
\end{align}
In the next theorem, we study an $\epsilon$-optimal policy for (\ref{eq:opt_prob}).

\begin{theorem}\label{th:epsilon_optimal}
    Let $\Delta_{\epsilon}$ denote the optimal value of the LP obtained from (\ref{eq:158}) by restricting the policy set $\bar{\Pi}$ to $\bar{\Pi}_{\epsilon}$. For every $\epsilon>0$, we have
    \begin{align}\label{eq:269}
        \Delta^{*} \leq\Delta_{\epsilon}\leq\Delta^{*}+\epsilon.
    \end{align}
    Moreover, the restricted LP admits an optimal solution $\bm{x}_{\epsilon}^{*}$ that assigns positive probability to at most three $(W,R)$ policies in $\bar{\Pi}_{\epsilon}$. A policy $\pi_{\epsilon}^{*}$ that selects one of these $(W,R)$ policies according to $\bm{x}_{\epsilon}^{*}$ at time $t=1$ and follows the selected policy for the entire time horizon is feasible for (\ref{eq:opt_prob}) and achieves an average age
    \begin{align}
        \Delta(\pi_{\epsilon}^{*})=\Delta_{\epsilon}.
    \end{align}
\end{theorem}

The proof of this theorem is provided in Appendix~\ref{append:th:epsilon_optimal}.

The finite-dimensional LP in Theorem~\ref{th:epsilon_optimal} provides an $\epsilon$-optimal policy, but its construction requires determining the finite set $\bar{\Pi}_{\epsilon}$ and solving the corresponding LP. We now construct a simpler and fully explicit policy for the jointly constrained regime $qf_r<f_s<f_r$ by randomizing among four policies in $\bar{\Pi}$. Recall that $\rho_{R}$ denotes the ratio of the average sampling rate to the average transmission rate for a policy $(W,R)$. Note that, $\rho_{R}$ is strictly decreasing with $R$, with $\rho_{1}=1$ and $\lim_{R\rightarrow\infty}\rho_{R} = q$. Thus, for $q f_{r}< f_{s}<f_{r}$, there exists a finite $R_a\in\mathbb{N}$, such that
\begin{align}\label{eq:Ra_Rb}
    \rho_{R_{a}}\geq\frac{f_{s}}{f_{r}}>\rho_{R_{a}+1}.
\end{align}
We define $R_{b}=R_{a}+1$. Next, we find two waiting thresholds corresponding to each $R\in\{R_a,R_b\}$. We define
\begin{align}\label{eq:208}
    \ell^{*}=\frac{1}{qf_r}.
\end{align}
For a fixed $R$, from (\ref{eq:144}), we have
\begin{align}
    \mathbb{E}[L(1,R)]=\frac{1}{q}.
\end{align} 
Furthermore, $\mathbb{E}[L(W,R)]$ is strictly increasing with $W$, and it diverges to $\infty$. Since $f_r\leq 1$, from (\ref{eq:208}) we have $\ell^{*}\geq\frac{1}{q}$. Thus, for every $R\in\{R_a,R_b\}$, there exists a finite positive integer $W_R$, such that
\begin{align}\label{eq:210}
    \mathbb{E}[L(W_R,R)]\leq \ell^{*} <\mathbb{E}[L(W_R+1,R)].
\end{align}
Equivalently, from (\ref{eq:149}),
\begin{align}\label{eq:WR_rates}
    \nu_r(W_R,R)\geq f_r>\nu_r(W_R+1,R).
\end{align}
For each $R\in\{R_a,R_b\}$, we define
\begin{align}\label{eq:kappa_R}
      \kappa_{R,0} = \frac{f_r-\nu_r(W_R+1,R)}{\nu_r(W_R,R) - \nu_r(W_R+1,R)}, \qquad \kappa_{R,1} = 1 - \kappa_{R,0}.
\end{align}
From (\ref{eq:WR_rates}), we have $0\leq\kappa_{R,0}\leq1$ and $0\leq\kappa_{R,1}\leq1$. Moreover,
\begin{align}\label{eq:R_block_transmission}
    \kappa_{R,0}\nu_r(W_R,R)+\kappa_{R,1}\nu_r(W_R+1,R)=f_r.
\end{align}
Thus, if we randomize between the policies $(W_R,R)$ and $(W_R+1,R)$ according to the probabilities $\kappa_{R,0}$ and $\kappa_{R,1}$, respectively, we get the average transmission rate $f_r$. Since $\nu_s(W,R)=\rho_R\nu_r(W,R)$, the corresponding average sampling rate is
\begin{align}\label{eq:R_block_sampling}
    \kappa_{R,0}\nu_s(W_R,R)+\kappa_{R,1}\nu_s(W_R+1,R)= \rho_R f_r.
\end{align}
Now, we define
\begin{align}\label{eq:gamma_four}
    \gamma=\frac{\frac{f_s}{f_r}-\rho_{R_b}} {\rho_{R_a}-\rho_{R_b}}.
\end{align}
From (\ref{eq:Ra_Rb}), we have $0<\gamma\leq1$, and
\begin{align}\label{eq:gamma_rate}
    \gamma\rho_{R_a}+(1-\gamma)\rho_{R_b} =\frac{f_s}{f_r}.
\end{align}
Now, we define a policy $\pi_{4}$, under which at time $t=1$ the sampler-transmitter chooses policy $(W_{R_{a}},R_{a})$ with probability $ \gamma\kappa_{R_a,0}$, policy $(W_{R_a}+1,R_a)$ with probability $\gamma\kappa_{R_a,1}$, policy $(W_{R_b},R_b)$ with probability $(1-\gamma)\kappa_{R_b,0}$, and policy $(W_{R_b}+1,R_b)$ with probability $(1-\gamma)\kappa_{R_b,1}$. Note that the average sampling rate $\nu_s(\pi_4)$ and the average transmission rate $\nu_r(\pi_4)$ satisfy the sampling-rate constraint $f_s$ and the transmission-rate constraint $f_r$, respectively, with equality. In the next theorem, we provide a performance guarantee for the policy $\pi_{4}$.

\begin{theorem}\label{th:four_approx}
    For $q<\frac{f_s}{f_r}<1$, policy $\pi_{4}$ satisfies both the sampling and transmission constraints in (\ref{eq:opt_prob}). Moreover,
    \begin{align}
        \Delta(\pi_{4})\leq 2 \Delta^{*} - \frac{1}{q \ell^{*}} <2 \Delta^{*}.
    \end{align}
\end{theorem}

The proof of this theorem is provided in Appendix~\ref{appen:th:four_approx}.

\section{Numerical Results}\label{sec:num}
For numerical comparisons across different parameter regimes, we define a proposed policy $\pi_{\mathrm{prop}}$. It uses $\pi_t^{*}$ when $f_s\geq f_r$, $\pi_s^{*}$ when $f_s\leq qf_r$, and $\pi_4$ when $qf_r<f_s<f_r$. Thus, at every parameter point, the selected policy satisfies both resource constraints. This policy is optimal in the two outer regimes and has a $2$-approximation guarantee in the intermediate regime. We use this selection rule throughout the numerical comparisons, allowing $q$, $f_s$, or $f_r$ to vary across regime boundaries while maintaining a feasible proposed policy. In all figures comparing the policies for the average age of the monitor, we plot the lower bound as the maximum of $\Delta_{\mathrm{LB}}=\frac{1}{2}\left(1+\frac{1} {\bar{\eta}_{\min}}\right)$ and the right-hand sides of (\ref{eq:546}) and (\ref{eq:547}).

For all numerical comparisons, we also solve a CMDP on a truncated state space by truncating both the age of the packet and the age of the monitor at $100$, using an occupation-measure LP~\cite{altman2021constrained}. We denote the resulting optimal policy by $\pi_{\mathrm{tr}}^{*}$, which provides a useful numerical benchmark for the considered model. We do not establish a truncation error bound for this benchmark. At the parameter values considered in the figures, $\pi_{\mathrm{tr}}^{*}$ achieves the lowest average age of the monitor among the compared policies. However, these numerical solutions do not establish any analytical structural properties of optimal policies, either for the original CMDP with the countably infinite state space or the truncated CMDP. In contrast, our analysis proves that an optimal policy for the original CMDP can be implemented by randomizing among at most three $(W,R)$ policies, provides a finite-policy LP with a prescribed $\epsilon$-optimality guarantee, and constructs an explicit feasible policy with a uniform $2$-approximation guarantee. These results identify the roles of waiting and packet replacement and yield compact policy implementations. The resulting structural insights can also provide a principled starting point for policy design in extensions involving feedback delays or multiple monitors, where the curse of dimensionality makes direct dynamic programming or occupation-measure optimization substantially more demanding.

\begin{figure}[t]
    \centering
    \subfloat[Average age of the monitor\label{fig:fr_age}]{%
        \includegraphics[width=0.5\columnwidth]{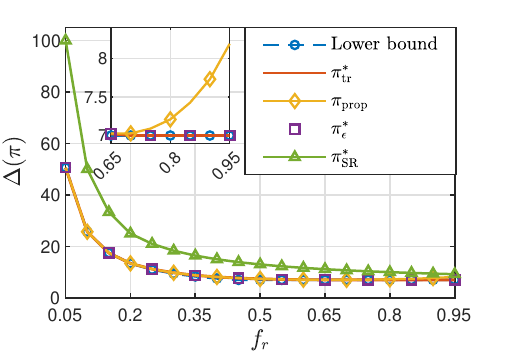}}%
    \hfill
    \subfloat[Average transmission rate\label{fig:fr_usage}]{%
        \includegraphics[width=0.5\columnwidth]{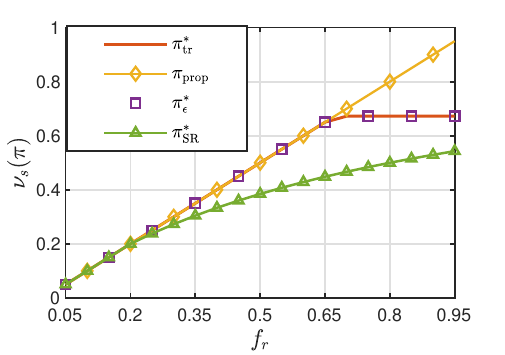}}
    \caption{Average age and transmission rate versus $f_r$ for $q=f_s=0.2$ and $\epsilon=0.1$. The smaller plot in (a) shows an enlarged view of the region $0.65\leq f_r\leq0.95$.}
    \label{fig:fr_comparison}
\end{figure}

In Figs.~\ref{fig:fr_age} and~\ref{fig:fr_usage}, we fix $q=f_s=0.2$ and vary $f_r$ from $0.05$ to $0.95$, with $\epsilon=0.1$. For the considered points with $f_r\leq0.2$, the proposed policy $\pi_{\mathrm{prop}}$ corresponds to $\pi_t^{*}$, whereas for $f_r>0.2$, it corresponds to $\pi_4$. In Fig.~\ref{fig:fr_age}, $\pi_{\epsilon}^{*}$ and $\pi_{\mathrm{tr}}^{*}$ achieve nearly identical average ages. The proposed policy achieves almost the same average age as $\pi_{\epsilon}^{*}$ and $\pi_{\mathrm{tr}}^{*}$ for $f_r\leq0.65$; however, both its average age and its gap relative to $\pi_{\mathrm{tr}}^{*}$ increase as the transmission budget is increased further. For the considered points with $f_r\geq0.7$, $\pi_{\epsilon}^{*}$ and $\pi_{\mathrm{tr}}^{*}$ achieve the lower bound of $7$. Note that the equidistant-sampling policy $(5,5)$ achieves average age $7$, sampling rate $0.2$, and transmission rate $0.67232$. Since it attains the lower bound in (\ref{eq:546}), obtained by considering the system with only the sampling-rate constraint, this policy is optimal whenever $f_r\geq0.67232$. Thus, additional transmission resources beyond this value are not necessary for attaining the optimal age of the monitor. Fig.~\ref{fig:fr_usage} shows that $\pi_{\epsilon}^{*}$ and $\pi_{\mathrm{tr}}^{*}$ use approximately $0.67232$ transmissions per slot at the considered points with $f_r\geq0.7$, whereas $\pi_{\mathrm{prop}}=\pi_4$ continues to use the entire transmission budget. To satisfy both rate constraints with equality, $\pi_4$ uses increasingly large packet-replacement thresholds as $f_r$ increases: it uses replacement thresholds $R_a=5$ and $R_b=6$ at $f_r=0.7$, $R_a=7$ and $R_b=8$ at $f_r=0.8$, and $R_a=13$ and $R_b=14$ at $f_r=0.95$. These larger packet-replacement thresholds result in older delivered updates. In contrast, the optimal equidistant-sampling policy $(5,5)$ retains replacement threshold $5$ and does not use part of the transmission budget. Finally, $\pi_{\mathrm{SR}}^{*}$ has the highest average age throughout the considered range, while using fewer transmissions than the other policies when $f_r>0.2$.

\begin{figure}[t]
    \centering
    \subfloat[Average age of the monitor\label{fig:q_age}]{%
        \includegraphics[width=0.5\columnwidth]{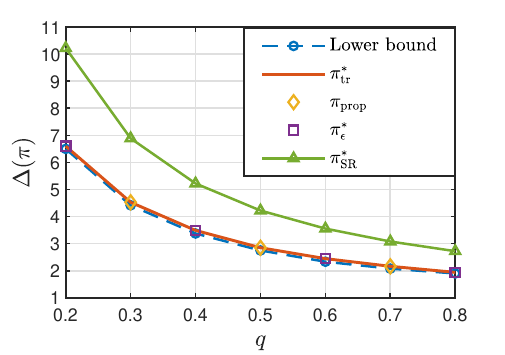}}%
    \hfill
    \subfloat[Average transmission rate\label{fig:q_usage}]{%
        \includegraphics[width=0.5\columnwidth]{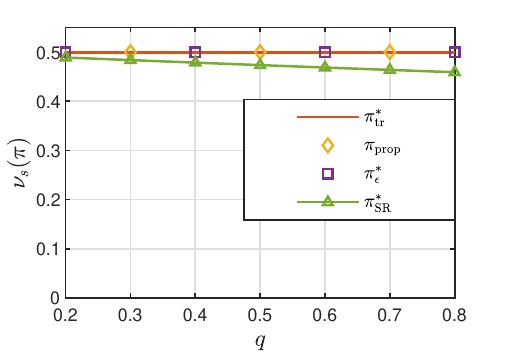}}
    \caption{Average age and transmission rate versus $q$ for
    $f_s=0.45$, $f_r=0.5$, and $\epsilon=0.1$.}
    \label{fig:comparison}
\end{figure}

To study the effect of channel reliability under fixed resource constraints, in Figs.~\ref{fig:q_age} and~\ref{fig:q_usage}, we set $f_s=0.45$ and $f_r=0.5$ and vary $q$ from $0.2$ to $0.8$, with $\epsilon=0.1$. All considered points satisfy $qf_r<f_s<f_r$, thus $\pi_{\mathrm{prop}}$ corresponds to $\pi_4$ throughout this comparison. In Fig.~\ref{fig:q_age}, the average age decreases with $q$ for every compared policy. The policies $\pi_{\mathrm{prop}}$, $\pi_{\epsilon}^{*}$, and $\pi_{\mathrm{tr}}^{*}$ achieve the same average age at all considered points. Fig.~\ref{fig:q_usage} shows that $\pi_{\mathrm{prop}}$, $\pi_{\epsilon}^{*}$ and $\pi_{\mathrm{tr}}^{*}$ use the full transmission budget of $0.5$ at every considered point. In contrast, the actual transmission rate of $\pi_{\mathrm{SR}}^{*}$ decreases from approximately $0.4891$ to $0.4592$ as $q$ increases from $0.2$ to $0.8$. Under $\pi_{\mathrm{SR}}^{*}$, from Theorem~\ref{th:sr_four_approx}, we see that the action-selection probabilities remain fixed because the resource constraints are unchanged. A higher channel success probability causes stored packets to be successfully delivered and removed more quickly from the transmitter, making a packet less likely to be available when retransmission is selected. Such selections result in idling, explaining the decrease in actual transmission usage even as the average age improves. Nevertheless, $\pi_{\mathrm{SR}}^{*}$ has a higher average age than the other compared policies throughout this range of $q$. All compared policies fully utilize the sampling-rate budget $f_s=0.45$ throughout the considered range of $q$.

\section{Conclusion}
We studied age of information minimization in a generate-at-will status update system subject to separate sampling- and transmission-rate constraints. We first derived a simple universal lower bound on the optimal average age of the monitor. We then analyzed a class of stationary randomized policies and constructed a feasible policy with a $4$-approximation guarantee. We identified two regimes in which one constraint becomes redundant and developed a joint sampling and transmission design for the remaining regime. By separately considering the systems with only the sampling-rate constraint and only the transmission-rate constraint, we derived two additional lower bounds on the average age of the monitor that jointly strengthen the universal lower bound. We established the optimality of the proposed $(W,R)$ policy class for the Lagrangian-relaxed problem and formulated an LP whose optimal solution randomizes among at most three $(W,R)$ policies and induces an optimal policy for our problem. We also constructed a finite policy set, and showed that the corresponding LP on that policy set yields an $\epsilon$-optimal policy. Moreover, we developed a policy by randomizing among four $(W,R)$ policies, which provides a $2$-approximation guarantee. Future work will consider the problem under deterministic and random delayed feedback scenarios.

\begin{appendices}

\section{Proof of Theorem~\ref{th:LB}}\label{appen:th1}
\begin{Proof}
Consider an arbitrary policy $\pi\in\Pi$, which satisfies the constraints in (\ref{eq:opt_prob}). Let $D^{\pi}(T)$ denote the total number of successful transmissions till time horizon $T$. As we assume that initially the transmitter has no packet to transmit, 
\begin{align}
    D^{\pi}(T) & \leq \sum_{t=1}^{T} a_{s}^{\pi}(t), 
\end{align}
and therefore,
\begin{align}
    \limsup_{T\rightarrow \infty} \frac{1}{T}\mathbb{E}_{\pi}\big[D^{\pi}(T)\big]  \leq \limsup_{T\rightarrow\infty} \frac{1}{T} \mathbb{E}_{\pi}\left[\sum_{t=1}^{T}a_{s}^{\pi}(t)\right] \leq f_{s}.\label{eq:4}
\end{align}
Recall that the channel success at time $t$ is denoted by the indicator variable $Z(t)$. Thus,
\begin{align}
    D^{\pi}(T) = \sum_{t=1}^{T} a_{r}^{\pi} (t)Z (t).
\end{align}
The transmission action $a_r^\pi(t)$ is taken before realizing $Z(t)$, i.e., without knowing $Z(t)$. Since $Z(t)$ is independent of $\{Z(t_1)\}_{t_1=1}^{t-1}$, $a_r^\pi(t)$ is independent of $Z(t)$. Thus,
\begin{align}
    \limsup_{T\rightarrow\infty} \frac{1}{T} \mathbb{E}_{\pi}\big[D^{\pi}(T)\big]  = \limsup_{T\rightarrow\infty} \frac{q}{T} \sum_{t=1}^{T} \mathbb{E}_{\pi}\big[a_{r}^{\pi}(t)\big] \leq q f_{r}. \label{eq:7}
\end{align}
Therefore, from (\ref{eq:4}) and (\ref{eq:7}), we have,
\begin{align}
    \limsup_{T\rightarrow\infty} \frac{1}{T} \mathbb{E}_{\pi}\big[D^{\pi}(T)\big] \leq \bar{\eta}_{\min}.
\end{align}

Now, let us assume that $D^{\pi}(T) = N$, i.e., there are $N$ successful transmissions till time $T$. We denote the $i$th successful transmission epoch by $\tau_{i}$, $i\in\{1,2,\cdots,N\}$. We assume that $\tau_{0}=0$ and $\tau_{N+1}=T$. We define,
\begin{align}
    L_{i} = \tau_{i} -\tau_{i-1}, \quad i\in\{1,2,\cdots,N+1\}.
\end{align}
Thus, $L_{i}$ denotes the inter-delivery time for $i\leq N$, and $L_{N+1}$ is the residual interval after the delivery of the last packet. After each packet delivery, the age of the monitor is lower bounded by $1$. Hence,
\begin{align}
    \sum_{t=1}^{T} v_{1}^{\pi}(t) & \geq \sum_{i=1}^{N+1} \frac{L_{i}(L_{i}+1)}{2} \\
    & \geq \frac{1}{2(N+1)} \left(\sum_{i=1}^{N+1} L_{i}\right)^{2} + \frac{\sum_{i=1}^{N+1}L_{i}}{2} \\ 
    & \geq \frac{T^{2}}{2(N+1)} + \frac{T}{2}. 
\end{align}
Therefore,
\begin{align}
    \mathbb{E}_{\pi}\left[\frac{\sum_{t=1}^{T}v_{1}^{\pi}(t)}{T}\right] 
    & \geq \frac{1}{2} +\frac{T}{2} \mathbb{E}_{\pi}\left[\frac{1}{D^{\pi}(T)+1}\right] \\ 
    & \geq \frac{1}{2} + \frac{1}{2} \frac{1}{\mathbb{E}_{\pi}\left[\frac{D^{\pi}(T)}{T}\right]+\frac{1}{T}},
\end{align}
and thus,
\begin{align}
    \limsup_{T\rightarrow\infty}\mathbb{E}_{\pi} \left[\frac{\sum_{t=1}^{T}v_{1}^{\pi}(t)}{T}\right] 
    & \geq \frac{1}{2} + \frac{1}{2} \frac{1}{\liminf_{T\rightarrow\infty}\mathbb{E}_{\pi}\left[\frac{D^{\pi}(T)}{T}\right]} \\  
    & \geq \frac{1}{2}\left( 1+ \frac{1}{\bar{\eta}_{\min}}\right).
\end{align}
As $\pi$ is an arbitrary causal policy that satisfies the constraints, the theorem follows.
\end{Proof}

\section{Proofs of Lemma~\ref{lemma:n1} and Theorem~\ref{th:sr_four_approx}} \label{appen:th:sr_four_approx}
\subsection{Proof of Lemma~\ref{lemma:n1}}
\begin{Proof}
A fresh packet is sampled in each slot with probability $\omega_{2}$. Thus, for every $T\geq1$,
\begin{align}
    \nu_s(\pi_{\textrm{SR}};T)=\frac{1}{T}\sum_{t=1}^{T}\mathbb{E}_{\pi_{\textrm{SR}}}\left[a_{s}^{\pi_{\mathrm{SR}}}(t)\right]=\frac{1}{T}\sum_{t=1}^{T}\omega_{2}=\omega_{2},
\end{align}
which proves (\ref{eq:n11}).  Similarly,
\begin{align}
    \nu_{r}(\pi_{\textrm{SR}};T) \leq  \frac{1}{T}\sum_{t=1}^{T} \mathbb{E}_{\pi_{\textrm{SR}}}\left[a_{r}^{\pi_{\mathrm{SR}}}(t)\right] \leq \frac{1}{T}\sum_{t=1}^{T} (\omega_{1}+\omega_{2}) = (\omega_{1}+\omega_{2}),
\end{align}
which proves (\ref{eq:n12}). Now, we prove (\ref{eq:n13}). Let $v_s(t)$ denote the age, at the beginning of slot $t$, of the most recently sampled packet, regardless of whether it is still available at the transmitter or not. Since the initial state is $s(1)=(1,\infty)$, we assume that the update initially available at the monitor is a packet that was sampled at time $t=0$ and successfully delivered by the end of slot $t=0$. Thus,
\begin{align}
    v_s(1)=v_1^{\pi_{\textrm{SR}}}(1)=1.
\end{align}
This virtual sampling and delivery occur before the considered time horizon. Thus, it affects neither the age process from slot $t=1$ onward nor the sampling and transmission rates.

For this proof we consider the following assumption. Whenever the policy chooses to retransmit but no packet is available at the transmitter, we assume that the transmitter retransmits the most recently delivered packet. This is a virtual transmission and is not counted as an actual transmission. If no packet is available at the beginning of slot $t$, the most recently sampled packet must already have been successfully delivered, and no packet has been sampled since that delivery. Consequently, this packet is also the freshest packet available at the monitor, and hence,
\begin{align}\label{eq:n16}
    v_1^{\pi_{\textrm{SR}}}(t)=v_s(t).
\end{align}
Under the original policy, when the retransmission selection is made and no packet is available, the transmitter remains idle. Therefore,
\begin{align}\label{eq:n17}
    v_1^{\pi_{\textrm{SR}}}(t+1) = v_1^{\pi_{\textrm{SR}}}(t)+1.
\end{align}
Under the virtual-transmission assumption, the age evolves as
\begin{align}
    v_1^{\pi_{\textrm{SR}}}(t+1) =
    \begin{cases}
        v_1^{\pi_{\textrm{SR}}}(t)+1, & \text{with probability }1-q,\\
        v_s(t)+1, & \text{with probability }q.
    \end{cases}
\end{align}
By (\ref{eq:n16}), both cases give $v_1^{\pi_{\textrm{SR}}}(t+1)=v_1^{\pi_{\textrm{SR}}}(t)+1$, exactly as in (\ref{eq:n17}). Therefore, the virtual-transmission assumption does not change the sample-path dynamics of the age of the monitor. Now,
\begin{align}
    \mathbb{E}_{\pi_{\textrm{SR}}}[v_{s}(t)] = & \sum_{i=0}^{t-1} \mathbb{P}(v_{s}(t)>i) \\ 
    = & \sum_{i=0}^{t-1} (1-\omega_{2})^{i} \\ 
    = & \frac{1-(1-\omega_{2})^{t}}{\omega_{2}}.
\end{align}
We now obtain a recursion for the expected age of the monitor as,
\begin{align}
    \mathbb{E}_{\pi_{\textrm{SR}}}[v_{1}^{\pi_{\textrm{SR}}}(t+1)] = & \omega_{0} \left(\mathbb{E}_{\pi_{\textrm{SR}}} \left[v_1^{\pi_{\textrm{SR}}}(t)\right]+1\right) + \omega_{2} \left( (1-q)\mathbb{E}_{\pi_{\textrm{SR}}}\left[v_1^{\pi_{\textrm{SR}}}(t)\right]+1 \right) \nonumber\\ 
    & + \omega_{1} \left((1-q) \mathbb{E}_{\pi_{\textrm{SR}}} \left[v_1^{\pi_{\textrm{SR}}}(t)\right]+1 + q   \mathbb{E}_{\pi_{\textrm{SR}}}[v_{s}(t)] \right) \\ 
    =& \left(1-q(\omega_{1}+\omega_{2})\right) \mathbb{E}_{\pi_{\textrm{SR}}} \left[v_1^{\pi_{\textrm{SR}}}(t)\right] +1 + \frac{q\omega_{1}}{\omega_{2}} \left(1-(1-\omega_{2})^{t}\right).\label{eq:n24}
\end{align}
Recursively applying (\ref{eq:n24}) gives 
\begin{align}\label{eq:n25}
    \mathbb{E}_{\pi_{\textrm{SR}}} \left[v_1^{\pi_{\textrm{SR}}}(t)\right]=& \left(1-q(\omega_{1}+\omega_{2})\right)^{t-1} \mathbb{E}_{\pi_{\textrm{SR}}}
    \left[v_1^{\pi_{\textrm{SR}}}(1)\right]+\left(1+\frac{q\omega_{1}}{\omega_{2}}\right)
    \sum_{k=1}^{t-1} \left(1-q(\omega_{1}+\omega_{2})\right)^{t-1-k}\nonumber\\
    & - \frac{q\omega_{1}}{\omega_{2}} \sum_{k=1}^{t-1} \left(1-q(\omega_{1}+\omega_{2})\right)^{t-1-k}(1-\omega_{2})^{k}.
\end{align}
Since $q,\omega_{2}>0$, we have,
\begin{align}\label{eq:lab26}
    0\leq1-q(\omega_{1}+\omega_{2})<1, \qquad 0\leq1-\omega_{2}<1.
\end{align}
Therefore, the first term on the right side of (\ref{eq:n25}) converges to zero. Moreover,
\begin{align}\label{eq:n27}
    \lim_{t\rightarrow\infty} \sum_{k=1}^{t-1} \left(1-q(\omega_{1}+\omega_{2})\right)^{t-1-k}= \frac{1}{q(\omega_{1}+\omega_{2})}.
\end{align}
If $1-q(\omega_{1}+\omega_{2})\neq1-\omega_{2}$, the last sum in (\ref{eq:n25}) is
\begin{align}
    \sum_{k=1}^{t-1} \left(1-q(\omega_{1}+\omega_{2})\right)^{t-1-k} (1-\omega_{2})^{k} = \frac{(1-\omega_{2}) \left[\left(1-q(\omega_{1}+\omega_{2})\right)^{t-1}
    -(1-\omega_{2})^{t-1} \right]} {\left(1-q(\omega_{1}+\omega_{2})\right)-(1-\omega_{2})},
\end{align}
which converges to zero by (\ref{eq:lab26}). If $1-q(\omega_{1}+\omega_{2})=1-\omega_{2}$, the same sum is
\begin{align}
    (t-1)(1-\omega_{2})^{t-1},
\end{align}
which also converges to zero. Thus, from (\ref{eq:n25}) and (\ref{eq:n27}), we have
\begin{align}\label{eq:n30}
    \lim_{t\rightarrow\infty} \mathbb{E}_{\pi_{\textrm{SR}}} \left[v_1^{\pi_{\textrm{SR}}}(t)\right] =& \frac{1+\frac{q\omega_{1}}{\omega_{2}}}
    {q(\omega_{1}+\omega_{2})}= \frac{1}{\omega_{2}}+\frac{1-q}{q(\omega_{1}+\omega_{2})}.
\end{align}
As the sequence $\left\{\mathbb{E}_{\pi_{\textrm{SR}}} [v_1^{\pi_{\textrm{SR}}}(t)]\right\}_{t\geq1}$ converges, the Cesaro mean theorem gives
\begin{align}
    \lim_{T\rightarrow\infty} \frac{1}{T}\sum_{t=1}^{T} \mathbb{E}_{\pi_{\textrm{SR}}} \left[v_1^{\pi_{\textrm{SR}}}(t)\right]=& \lim_{t\rightarrow\infty} \mathbb{E}_{\pi_{\textrm{SR}}} \left[v_1^{\pi_{\textrm{SR}}}(t)\right]=
    \frac{1}{\omega_{2}}+\frac{1-q}{q(\omega_{1}+\omega_{2})}.
\end{align}
Thus, from the definition of $\Delta(\pi_{\textrm{SR}})$, (\ref{eq:n13}) follows.
\end{Proof}

\subsection{Proof of Theorem~\ref{th:sr_four_approx} }
\begin{Proof}
If $(\omega_0, \omega_1, \omega_2)$ is a feasible solution of (\ref{eq:n35}), then we have
\begin{align}
    \omega_{2}\leq \min\{f_{s},f_{r}\}, \quad \omega_{1}\leq f_{r}.
\end{align}
Thus, if $(\omega_{0},\omega_{1},\omega_{2})$ is a feasible solution of (\ref{eq:n35}), then the following holds:
\begin{align}\label{eq:n42}
    \frac{1}{\omega_{2}}+\frac{1-q}{q(\omega_{1}+\omega_{2})} \geq \frac{1}{\min\{f_{s},f_{r}\}} + \frac{1-q}{q f_{r}}.
\end{align}
Note that, the probabilities in (\ref{eq:n36}), (\ref{eq:n37}), and (\ref{eq:n38}) attain the lower bound of (\ref{eq:n42}). Moreover, as $0< f_{s},f_{r}\leq 1$, these probabilities are non-negative and satisfy (\ref{eq:n35}). Therefore, they form an optimal solution of (\ref{eq:n35}). Thus, $\pi_{\textrm{SR}}^{*}$ satisfies both constraints in (\ref{eq:opt_prob}). Finally, (\ref{eq:n39}) follows by substituting (\ref{eq:n37}) and (\ref{eq:n38}) into (\ref{eq:n13}). 

Recall that $\bar{\eta}_{\min}=\min\{f_s,qf_r\}$. Thus, we have the following two inequalities
\begin{align}
     \frac{1}{\min\{f_s,f_r\}} \leq \frac{1}{\bar{\eta}_{\min}}, \qquad \frac{1-q}{qf_r}\leq \frac{1}{\bar{\eta}_{\min}}.
\end{align}
Therefore, inserting these to (\ref{eq:n39}), we get
\begin{align}\label{eq:n45}
    \Delta(\pi_{\textrm{SR}}^{*}) \leq \frac{2}{\bar{\eta}_{\min}}.
\end{align}
Combining Theorem~\ref{th:LB} and (\ref{eq:n45}), we obtain
\begin{align}
    \frac{\Delta(\pi_{\textrm{SR}}^{*})}{\Delta^{*}} &\leq \frac{\frac{2}{\bar{\eta}_{\min}}} {\frac{1}{2} \left( 1+\frac{1}{\bar{\eta}_{\min}} \right)},\\
    &=\frac{4}{1+\bar{\eta}_{\min}} \leq4,
\end{align}
which proves (\ref{eq:n40}).
\end{Proof}

\section{Proofs of Lemma~\ref{lemma:2}, Lemma~\ref{lemma:1} and Lemma~\ref{lemma:4}} \label{appen:discounted_properties}
\subsection{Proof of Lemma~\ref{lemma:2}}
\begin{Proof}
Let $\pi$ be an arbitrary policy, and let $\tau^{\pi}$ be the first successful delivery time under policy $\pi$. Note that, during the first $\tau^{\pi}$ slots, the age of the monitor is at least $v_{1}$, if we begin from state $s_{ap,v_{2}}^{v_{1}}$. Thus,
\begin{align}
    V_{\alpha}^{\pi} (s_{ap,v_{2}}^{v_{1}}) 
    \geq& v_{1} \mathbb{E}_{\pi}\left[\sum_{t=0}^{\tau^{\pi}-1} \alpha^{t}\right] \\ 
    =& v_{1} \mathbb{E}_{\pi} \left[\sum_{t=0}^{\infty}\alpha^{t} \bm{1}_{\{\tau^{\pi}>t\}}\right]\\
    \geq & v_{1} \sum_{t=0}^{\infty}\alpha^{t} \mathbb{P}(\tau^{\pi}>t) \\ 
    \geq & v_{1} \sum_{t=0}^{\infty} \alpha^{t} (1-q)^{t} \\ 
    =&  \frac{v_{1}}{1-\alpha(1-q)}.\label{eq:44}
\end{align}
Now, taking the infimum over $\pi$ on both sides of (\ref{eq:44}), we have the first part of this lemma.

Now, let us consider a policy $\pi$, under which the sampler always samples a fresh packet, and the transmitter transmits it till there is a successful transmission and after that the policy $\pi$ follows the optimal policy for the discounted cost problem. We denote the duration for the first successful transmission by $\tau^{\pi}$. Thus,
\begin{align}\label{eq:45}
      V^{\pi}_{\alpha}(s_{ap,v_{2}}^{v_{1}})  =  \mathbb{E}_{\pi}  \left[ \sum_{t=0}^{\tau^{\pi}-1}   \alpha^{t} (v_{1} + t + \lambda + \mu)  +  \alpha^{\tau^{\pi}}  V_{\alpha}(s_{np}^{1})  \right] .  
\end{align}
First, consider a policy $\pi_{1}$ under which the sampler always samples a fresh update packet and the transmitter always transmits the freshly sampled packet. Thus,
\begin{align}
    V_{\alpha}^{\pi_{1}} (s_{np}^{1}) & \leq  \sum_{t=0}^{\infty} \alpha^{t} (1+t+\mu+\lambda),\\
    V_{\alpha} (s_{np}^{1}) & < \infty.\label{eq:47}
\end{align}
First, note that in (\ref{eq:45}),
\begin{align}\label{eq:48}
    \mathbb{E}_{\pi}[\alpha^{\tau^{\pi}}] = \frac{\alpha q}{1-\alpha(1-q)}.
\end{align}
Now,
\begin{align}
    \mathbb{E}_{\pi}\left[\sum_{t=0}^{\tau^{\pi}-1}\alpha^{t}(v_{1}+t+\lambda+\mu)\right] 
    &= \sum_{t=0}^{\infty} \alpha^{t} (v_{1}+t+\lambda+\mu) \mathbb{P}(\tau^{\pi}>t)\\
    & = \frac{v_{1}+\lambda+\mu}{1-\alpha(1-q)} + \frac{\alpha(1-q)}{(1-\alpha(1-q))^{2}}.\label{eq:50}
\end{align}
Thus, from (\ref{eq:45}), (\ref{eq:48}), and (\ref{eq:50}), we have,
\begin{align}
    &V_{\alpha}^{\pi}(s_{ap,v_{2}}^{v_{1}}) = \frac{v_{1}+\lambda+\mu+\alpha q V_{\alpha}(s_{np}^{1})}{1-\alpha(1-q)} +  \frac{\alpha(1-q)}{(1-\alpha(1-q))^{2}}.   
\end{align}
Now, we define
\begin{align}
    K =  \frac{\lambda + \mu + \alpha q V_{\alpha}(s_{np}^{1})}{1 - \alpha(1 - q)} +  \frac{\alpha(1 - q)}{(1 - \alpha(1 - q))^{2}}.
\end{align}
From (\ref{eq:47}) we have $K<\infty$, which completes the proof of this lemma.
\end{Proof}

\subsection{Proof of Lemma~\ref{lemma:1}}
\begin{Proof}
For $n\geq 0$, by induction, we show that
\begin{align}\label{eq:c146}
    V_{\alpha,n}(s_{ap,v_{1}}^{v_{1}})=V_{\alpha,n}(s_{np}^{v_{1}}).
\end{align}
Then, the proof of the lemma follows directly from (\ref{eq:26}). For $n=0$, (\ref{eq:c146}) follows immediately as $V_{\alpha,0}(s)=0$, for all $s\in\mathcal{S}$. Let us assume that (\ref{eq:c146}) holds for the induction step $n$. Now, from the $n$th induction step we have,
\begin{align}\label{eq:n43}
    V_{\alpha,n}(s_{ap,v_{1}+1}^{v_{1}+1}) = V_{\alpha,n}(s_{np}^{v_{1}+1}).
\end{align}
Replacing $v_{2}=v_{1}$ in (\ref{eq:31}) and (\ref{eq:34}), and from (\ref{eq:n43}), we have,
\begin{align}\label{eq:37}
    V_{\alpha,n + 1}(s_{ap,v_{1}}^{v_{1}};0) = V_{\alpha,n + 1}(s_{np}^{v_{1}};0).
\end{align}
From (\ref{eq:33}) and (\ref{eq:35}), we have,
\begin{align}\label{eq:38}
    V_{\alpha,n+1}(s_{ap,v_{1}}^{v_{1}};2) = V_{\alpha,n+1}(s_{np}^{v_{1}};2).
\end{align}
As $\lambda\geq 0$, from (\ref{eq:n43}), (\ref{eq:31}), and (\ref{eq:32}), we have,
\begin{align}\label{eq:399}
    V_{\alpha,n+1}(s_{ap,v_{1}}^{v_{1}};1) \geq V_{\alpha,n+1}(s_{ap,v_{1}}^{v_{1}};0).
\end{align}
Thus, from (\ref{eq:n26}), (\ref{eq:37}), (\ref{eq:38}), and (\ref{eq:399}), we have,
\begin{align}
    V_{\alpha,n+1}(s_{ap,v_{1}}^{v_{1}}) = V_{\alpha,n+1}(s_{np}^{v_{1}}),
\end{align}  
which completes the proof of this lemma.
\end{Proof}

\subsection{Proof of Lemma~\ref{lemma:4}}
\begin{Proof}
We show by induction that for all $n\in\mathbb{N}$, $V_{\alpha,n}(s_{np}^{v_{1}})$ and $V_{\alpha,n}(s_{ap,v_{2}}^{v_{1}})$ are increasing functions of $v_{1}$ and $V_{\alpha,n}(s_{ap,v_{2}}^{v_{1}})$ is an increasing function of $v_{2}$, then the first part of this lemma follows from (\ref{eq:26}). For $n=0$, $V_{\alpha,0}(s_{np}^{v_{1}})=V_{\alpha,0}(s_{ap,v_{2}}^{v_{1}})=0$, thus, the statement of this lemma immediately holds for $n=0$. Now, we assume that for the induction step $n$, the statement of this lemma holds. Now, from (\ref{eq:31}), (\ref{eq:32}), (\ref{eq:33}), (\ref{eq:34}), (\ref{eq:35}), the $n$th induction step, and (\ref{eq:n26}), we have $V_{\alpha,n+1}(s_{np}^{v_{1}})$ and $V_{\alpha,n+1}(s_{ap,v_{2}}^{v_{1}})$ as increasing functions of $v_{1}$. From (\ref{eq:33}), we see that $V_{\alpha,n+1}(s_{ap,v_{2}}^{v_{1}};2)$ is independent of $v_{2}$, and from (\ref{eq:31}) and (\ref{eq:32}), and the $n$th induction step we see that, $V_{\alpha,n+1}(s_{ap,v_{2}}^{v_{1}};0)$ and $V_{\alpha,n+1}(s_{ap,v_{2}}^{v_{1}};1)$ are increasing functions of $v_{2}$. Thus, from (\ref{eq:n26}), we establish that $V_{\alpha,n+1}(s_{ap,v_{2}}^{v_{1}})$ is an increasing function of $v_{2}$.  

Now, again by induction, we first show that for $x\geq 0$,
\begin{align}
    V_{\alpha,n}(s_{ap,v_{2}}^{v_{1}+x}) &\geq V_{\alpha,n}(s_{ap,v_{2}}^{v_{1}}) + x,\label{eq:383}\\ V_{\alpha,n}(s_{np}^{v_{1}+x}) &\geq V_{\alpha,n}(s_{np}^{v_{1}}) + x.\label{eq:384}
\end{align}
Then, from (\ref{eq:26}), we get (\ref{eq:nn25}) and (\ref{eq:nn26}). Recall that we have initialized $V_{\alpha,0}(s)=0$, for all $s\in\mathcal{S}$. Thus, for $n=1$, (\ref{eq:383}) and (\ref{eq:384}) follow immediately from (\ref{eq:n26}). Let us assume that (\ref{eq:383}) and (\ref{eq:384}) hold for the induction step $n$. From (\ref{eq:31}), we have,
\begin{align}\label{eq:385}
    V_{\alpha,n+1}(s_{ap,v_{2}}^{v_{1}+x};0)-  V_{\alpha,n+1}(s_{ap,v_{2}}^{v_{1}};0)=
    x+\alpha\left(V_{\alpha,n}(s_{ap,v_{2}+1}^{v_{1}+x+1})-V_{\alpha,n}(s_{ap,v_{2}+1}^{v_{1}+1})\right)\geq x.
\end{align}
Similarly, from (\ref{eq:32}),
\begin{align}\label{eq:386}
    V_{\alpha,n+1}(s_{ap,v_{2}}^{v_{1}+x};1)-V_{\alpha,n+1}(s_{ap,v_{2}}^{v_{1}};1)=
    x+\alpha(1-q)\left(V_{\alpha,n}(s_{ap,v_{2}+1}^{v_{1}+x+1})-V_{\alpha,n}(s_{ap,v_{2}+1}^{v_{1}+1})\right)\geq x,
\end{align}
and from (\ref{eq:33}),
\begin{align}\label{eq:387}
    V_{\alpha,n+1}(s_{ap,v_{2}}^{v_{1}+x};2)-V_{\alpha,n+1}(s_{ap,v_{2}}^{v_{1}};2)=
    x+\alpha(1-q)\left(V_{\alpha,n}(s_{ap,1}^{v_{1}+x+1})-V_{\alpha,n}(s_{ap,1}^{v_{1}+1})\right)\geq x.
\end{align}
Note that, (\ref{eq:385}), (\ref{eq:386}), and (\ref{eq:387}) follow as $V_{\alpha,n}(s_{ap,v_{2}}^{v_{1}})$ monotonically increases with $v_{1}$. Now, from (\ref{eq:385}), (\ref{eq:386}), (\ref{eq:387}), and (\ref{eq:n26}), we have (\ref{eq:383}) for the $(n+1)$th induction step. Following the same argument and using (\ref{eq:34}) and
(\ref{eq:35}), we get (\ref{eq:384}) for the $(n+1)$th induction step, which completes the proof of this lemma.
\end{Proof}

\section{Proof of Theorem~\ref{th:2}}\label{appen:th:2}
\begin{Proof}
First, we prove this theorem for the discounted cost formulation. Now, for state $s_{ap,v_{2}}^{v_{1}}$,
\begin{align}\label{eq:53}
    V_{\alpha}(s_{ap,v_{2}}^{v_{1}};0) - V_{\alpha}(s_{ap,v_{2}}^{v_{1}};2)= \alpha V_{\alpha} (s_{ap,v_{2}+1}^{v_{1}+1}) -\lambda -\mu -\alpha q V_{\alpha}(s_{np}^{1}) - \alpha (1-q) V_{\alpha}(s_{ap,1}^{v_{1}+1}). 
\end{align}
From Lemma~\ref{lemma:4}, we have,
\begin{align}
    V_{\alpha}(s_{ap,v_{2}+1}^{v_{1}+1})\geq V_{\alpha}(s_{ap,1}^{v_{1}+1}).
\end{align}
Thus,
\begin{align}
    V_{\alpha}(s_{ap,v_{2}}^{v_{1}};0)-V_{\alpha}(s_{ap,v_{2}}^{v_{1}};2)\geq \alpha q\left(V_{\alpha}(s_{ap,1}^{v_{1}+1})-V_{\alpha}(s_{np}^{1})\right)-\lambda-\mu.
\end{align}
From Lemma~\ref{lemma:1}, we have,
\begin{align}
    V_{\alpha}(s_{ap,1}^{1})=V_{\alpha}(s_{np}^{1}).
\end{align}
Thus, from Lemma~\ref{lemma:4}, we have,
\begin{align}
    V_{\alpha}(s_{ap,1}^{v_{1}+1})-V_{\alpha}(s_{np}^{1})\geq v_{1}.
\end{align}
Therefore,
\begin{align}\label{eq:th2_discount_packet}
    V_{\alpha}(s_{ap,v_{2}}^{v_{1}};0)-V_{\alpha}(s_{ap,v_{2}}^{v_{1}};2)\geq \alpha qv_{1}-\lambda-\mu.
\end{align}

Following a similar argument, we get,
\begin{align}\label{eq:th2_discount_np}
    V_{\alpha}(s_{np}^{v_{1}};0)-V_{\alpha}(s_{np}^{v_{1}};2)\geq \alpha qv_{1}-\lambda-\mu.
\end{align}
We define
\begin{align}\label{eq:th2_vbar}
    \bar{v}_{1}=\max\left\{1,\left\lceil\frac{2(\lambda+\mu)}{q}\right\rceil\right\}.
\end{align}
For every $\alpha\geq\frac{1}{2}$ and $v_{1}\geq\bar{v}_{1}$, from (\ref{eq:th2_discount_packet}) and (\ref{eq:th2_discount_np}), we have
\begin{align}
    V_{\alpha}(s_{ap,v_{2}}^{v_{1}};2) &\leq V_{\alpha}(s_{ap,v_{2}}^{v_{1}};0),\\
    V_{\alpha}(s_{np}^{v_{1}};2) &\leq V_{\alpha}(s_{np}^{v_{1}};0).
\end{align}

Now, consider a sequence $\{\alpha_n\}_{n=1}^{\infty}$ such that $\alpha_{n}\geq \frac{1}{2}$ for all $n$ and $\alpha_n\rightarrow 1$. Let $\pi_n$ be a deterministic stationary policy that is optimal for the discounted-cost problem with discount factor $\alpha_n$. We choose $\pi_n$ such that, for every $v_1\geq\bar{v}_1$,
\begin{align}
    \pi_n(s_{np}^{v_1})&=2,\\
    \pi_n(s_{ap,v_2}^{v_1})&\in\{1,2\}, \quad 1\leq v_2\leq v_1.
\end{align}
Such a choice is possible since, for these states, action $a=2$ is no worse than action $a=0$. Since the state space $\mathcal{S}$ is countable, we enumerate its elements as
\begin{align}
    \mathcal{S}=\{s_1,s_2,\cdots\},
\end{align}
and represent each policy $\pi_n$ by its infinite action vector
\begin{align}
    \bm{\pi}_{n} = \big(\pi_{n}(s_{1}),\pi_{n}(s_{2}),\cdots\big),
\end{align}
where the $k$th coordinate specifies the action selected by policy $\pi_{n}$ at the corresponding state $s_{k}$. Throughout the paper, we refer to the infinite action vector associated with a stationary policy as the policy itself. Thus, we use $\bm{\pi}$ and $\pi$ interchangeably to denote the same policy. From Lemma~\ref{lemma:2}, we have $V_{\alpha}(s_{ap,v_{2}}^{v_{1}})<\infty$, for all $1\leq v_{2}\leq v_{1}$. From Lemma~\ref{lemma:1}, we have $V_{\alpha}(s_{np}^{v_{1}}) = V_{\alpha}(s_{ap,v_{1}}^{v_{1}})$. Thus, for all $s\in\mathcal{S}$, we have $V_{\alpha}(s)<\infty$. 

Now, for $0<\alpha<1$ and for all $s\in\mathcal{S}$, we define
\begin{align}
    h_{\alpha}(s) = V_{\alpha}(s) - V_{\alpha}(s_{np}^{1}).
\end{align}
Combining Lemma~\ref{lemma:1} and Lemma~\ref{lemma:4}, we get $V_{\alpha}(s)\geq V_{\alpha}(s_{np}^{1})$ for all $s\in\mathcal{S}$. Thus, we get $h_{\alpha}(s)\geq0$, for all $s\in\mathcal{S}$.

Note that the $\alpha$-dependent coefficients in the upper bound of Lemma~\ref{lemma:2} can be upper bounded by taking $\alpha\rightarrow1$. Thus,
\begin{align}\label{eq:h_sbound}
    V_{\alpha}(s_{ap,v_{2}}^{v_{1}}) \leq \frac{v_{1} + \lambda + \mu}{q} + \frac{1-q}{q^{2}} + V_{\alpha}(s_{np}^{1}).
\end{align}
From Lemma~\ref{lemma:1}, the bound in (\ref{eq:h_sbound}) also holds for $s_{np}^{v_{1}}$. Thus, for all $s\in\mathcal{S}$, we get
\begin{align}
    h_{\alpha}(s) \leq \frac{v_{1} + \lambda + \mu}{q} + \frac{1-q}{q^{2}}.
\end{align}

Thus, the conditions required by the vanishing-discount result in \cite{sennott1989average} are satisfied. Therefore, there exists a subsequence $\{\bm{\pi}_{n_i}\}_{i=1}^{\infty}$ that converges pointwise to a deterministic stationary policy $\bm{\pi}^{*}_{L}$ that is optimal for the Lagrangian-relaxed MDP in (\ref{eq:20}), i.e., for every fixed state $s_k$, we have,
\begin{align}
    \pi_{n_i}(s_k)\rightarrow\pi_{L}^{*}(s_k).
\end{align}
Since the action space is finite, this implies that $\pi_{n_i}(s_k)=\pi^{*}_{L}(s_k)$ for all sufficiently large $i$. Therefore, for $v_1\geq\bar{v}_1$,
\begin{align}
    \pi^{*}_{L}(s_{np}^{v_1})&=2,\\
    \pi^{*}_{L}(s_{ap,v_2}^{v_1})&\in\{1,2\},\quad 1\leq v_2\leq v_1,
\end{align}
which concludes the proof of this theorem.
\end{Proof}

\section{Proof of Theorem~\ref{th:3}}\label{appen:th:3}
\begin{Proof}
Let us first consider a discount factor $\alpha$, where $\bar{\alpha}(\mu,\lambda)<\alpha<1$. By the no-packet threshold property, there exists an optimal policy under which the set of no-packet states in which action $a=2$ is optimal has a threshold structure. Moreover, from the discounted-cost result in the proof of Theorem~\ref{th:2}, action $a=2$ is optimal for all sufficiently large values of $v_{1}$. Thus, there exists a finite no-packet threshold $\bar{v}_{1}$, such that
\begin{align}
    V_{\alpha}(s_{np}^{v_{1}};0) &\leq V_{\alpha}(s_{np}^{v_{1}};2), \qquad v_{1}<\bar{v}_{1},\nonumber \\
    V_{\alpha}(s_{np}^{v_{1}};2) &\leq V_{\alpha}(s_{np}^{v_{1}};0), \qquad v_{1}\geq\bar{v}_{1}. \label{eq:58}
\end{align}
First, we show that if 
\begin{align}\label{eq:59}
    \lambda \leq \frac{\alpha q}{1-\alpha},
\end{align}
then, for every $v_{1} \geq \bar{v}_{1}$, the following holds,
\begin{align}\label{eq:60}
    \min\{V_{\alpha}(s_{ap,v_{2}}^{v_{1}};1), V_{\alpha}(s_{ap,v_{2}}^{v_{1}};2)\} \leq V_{\alpha}(s_{ap,v_{2}}^{v_{1}};0).
\end{align}
From Theorem~\ref{th:2}, we know that there exists $\tilde{v}_{1}$ such that for all $v_{1}\geq \tilde{v}_{1}$, we have,
\begin{align}
    V_{\alpha}(s_{ap,v_{2}}^{v_{1}};2) \leq V_{\alpha}(s_{ap,v_{2}}^{v_{1}};0).
\end{align}
If $\bar{v}_{1}\geq \tilde{v}_{1}$, then (\ref{eq:60}) follows immediately. Thus, now we assume that $\tilde{v}_{1} > \bar{v}_{1}$. Now, we show that (\ref{eq:60}) holds for $v_{1}\in\{\tilde{v}_{1}-1,\tilde{v}_{1}-2\cdots,\bar{v}_{1}\}$, by backward induction. Let us consider a fixed $v_{1}$, where $\bar{v}_{1}\leq v_{1}<\tilde{v}_{1} $, and assume the induction hypothesis that (\ref{eq:60}) holds for monitor's age $v_{1}+1$ and for all $1\leq v_{2}\leq v_{1}+1$. Now, we show that (\ref{eq:60}) also holds for monitor's age $v_{1}$ and for all $1 \leq v_{2}\leq v_{1}$. For a fixed $v_{2}$, from the induction hypothesis, we know that either action $a=1$ or $a=2$ is optimal for state $s_{ap,v_{2}+1}^{v_{1}+1}$.

\textbf{Case 1: Action $a=2$ is optimal for state $s_{ap,v_{2}+1}^{v_{1}+1}$:} As, $v_{1}+1\geq \bar{v}_{1}$, from (\ref{eq:58}), we have,
\begin{align}\label{eq:63}
    V_{\alpha}(s_{np}^{v_{1}+1};2) \leq V_{\alpha}(s_{np}^{v_{1}+1};0).
\end{align}
Taking $n\rightarrow\infty$ in (\ref{eq:33}) and (\ref{eq:35}) and using the assumption of Case $1$, (\ref{eq:24}), and (\ref{eq:63}), we obtain,
\begin{align}\label{eq:64}
    V_{\alpha}(s_{ap,v_{2}+1}^{v_{1}+1}) = V_{\alpha}(s_{np}^{v_{1}+1}).
\end{align}
Taking $n\rightarrow\infty$ in (\ref{eq:31}) and (\ref{eq:34}), and from (\ref{eq:64}) we have,
\begin{align}\label{eq:65}
    V_{\alpha}(s_{ap,v_{2}}^{v_{1}};0) = V_{\alpha}(s_{np}^{v_{1}};0).
\end{align}
Since $v_{1}\geq\bar{v}_{1}$, taking $n\rightarrow\infty$ in (\ref{eq:33}) and (\ref{eq:35}) and using (\ref{eq:58}) and (\ref{eq:65}), we have,
\begin{align}
    V_{\alpha}(s_{ap,v_{2}}^{v_{1}};2) \leq V_{\alpha}(s_{ap,v_{2}}^{v_{1}};0).
\end{align}
Thus, (\ref{eq:60}) holds for Case $1$.

\textbf{Case 2: Action $a=1$ is optimal for state $s_{ap,v_{2}+1}^{v_{1}+1}$:} If $v_{2}=v_{1}$, then from Lemma~\ref{lemma:1}, (\ref{eq:nnn26}) and (\ref{eq:58}), we have (\ref{eq:60}). Now, we prove (\ref{eq:60}) for $v_{2}\leq v_{1}-1$. From the assumption of case $2$, we have,
\begin{align}\label{eq:67}
    V_{\alpha}(s_{ap,v_{2}+1}^{v_{1}+1};1) \leq V_{\alpha}(s_{ap,v_{2}+1}^{v_{1}+1};0).
\end{align}
Similar to (\ref{eq:31}) and (\ref{eq:32}), expanding both sides of (\ref{eq:67})  and rearranging it, we get,
\begin{align}\label{eq:68}
    V_{\alpha}(s_{ap,v_{2}+2}^{v_{1}+2}) - V_{\alpha}(s_{np}^{v_{2}+2}) \geq \frac{\lambda}{\alpha q}.
\end{align}
Now,
\begin{align}
    V_{\alpha}(s_{ap,v_{2} + 1}^{v_{1} + 1})  -  V_{\alpha}(s_{np}^{v_{2} + 1})  \geq  &V_{\alpha}(s_{ap,v_{2} + 1}^{v_{1}+1}; 1)  -  V_{\alpha}(s_{np}^{v_{2} + 1}; 0) \\ 
    =& v_{1} -v_{2} + \lambda +\alpha(1-q) (V_{\alpha}(s_{ap,v_{2}+2}^{v_{1}+2}) -  V_{\alpha}(s_{np}^{v_{2}+2}))\\ 
    \geq & v_{1}-v_{2}  + \frac{\lambda}{q} \label{eq:71}\\ 
    \geq & 1+\frac{\lambda}{q}, \label{eq:72}
\end{align}
where (\ref{eq:71}) follows from (\ref{eq:68}), and (\ref{eq:72}) follows from the fact that $v_{2}\leq v_{1}-1$. Now, from the condition in (\ref{eq:59}) and (\ref{eq:72}), we say that,
\begin{align}\label{eq:73}
    V_{\alpha}(s_{ap,v_{2}+1}^{v_{1}+1}) - V_{\alpha}(s_{np}^{v_{2}+1}) \geq \frac{\lambda}{\alpha q}.
\end{align}
Now, from (\ref{eq:73}), we have,
\begin{align}
    &V_{\alpha}(s_{ap,v_{2}}^{v_{1}};1) - V_{\alpha}(s_{ap,v_{2}}^{v_{1}};0) = \lambda -\alpha q(V_{\alpha}(s_{ap,v_{2}+1}^{v_{1}+1}) - V_{\alpha}(s_{np}^{v_{2}+1})) \leq 0.
\end{align}
Thus, under the condition (\ref{eq:59}), (\ref{eq:60}) holds.     

We now show that the same structural property holds for an optimal policy for the Lagrangian-relaxed MDP in (\ref{eq:20}). Consider a sequence $\{\alpha_{n}\}_{n=1}^{\infty}$ such that $\alpha_{n}\rightarrow1$ and
\begin{align}
    \alpha_{n}>\bar{\alpha}(\mu,\lambda), \quad \text{and} \quad \lambda\leq\frac{\alpha_{n}q}{1-\alpha_{n}}, \quad \forall n\in\mathbb{N}.
\end{align}
For each $\alpha_{n}$, we choose an optimal stationary policy $\pi_{n}$ for the corresponding discounted-cost problem such that its actions at the no-packet states follow the no-packet threshold structure, and denote its no-packet threshold by $\bar{v}_{1}(\alpha_{n})$. Therefore,
\begin{align}
    \pi_{n}(s_{np}^{v_{1}})=2 \quad\Longleftrightarrow\quad v_{1}\geq\bar{v}_{1}(\alpha_{n}).
\end{align}
From the discounted-cost result proved above, (\ref{eq:60}) holds for every $v_{1}\geq\bar{v}_{1}(\alpha_{n})$. Thus, we can choose $\pi_{n}$ such that, for every $v_{1}$,
\begin{align}\label{eq:th3_discount_structure}
    \pi_{n}(s_{np}^{v_{1}}) = 2   \quad  \Longrightarrow  \quad   \pi_{n}(s_{ap,v_{2}}^{v_{1}})\in\{1,2\}, \quad 1\leq v_{2}\leq v_{1}.
\end{align}

We have already verified the conditions required by the vanishing-discount result in \cite{sennott1989average} in the proof of Theorem~\ref{th:2}. Thus, as in the proof of Theorem~\ref{th:2}, there exists a subsequence $\{\bm{\pi}_{n_{i}}\}_{i=1}^{\infty}$ that converges pointwise to a deterministic stationary policy $\bm{\pi}_{L}^{*}$ that is optimal for the Lagrangian-relaxed MDP in (\ref{eq:20}). Now, consider a state $s_{np}^{v_{1}}$, for which
\begin{align}
    \pi_{L}^{*}(s_{np}^{v_{1}})=2.
\end{align}
By pointwise convergence, there exists $i_{0}<\infty$, such that
\begin{align}
    \pi_{n_{i}}(s_{np}^{v_{1}})=2, \quad i\geq i_{0}.
\end{align}
Therefore, from (\ref{eq:th3_discount_structure}), for every $i\geq i_{0}$,
\begin{align}
    \pi_{n_{i}}(s_{ap,v_{2}}^{v_{1}}) \in\{1,2\}, \quad 1\leq v_{2}\leq v_{1}.
\end{align}
For any fixed $1\leq v_{2}\leq v_{1}$, pointwise convergence at the state $s_{ap,v_{2}}^{v_{1}}$ then gives,
\begin{align}
    \pi_{L}^{*}(s_{ap,v_{2}}^{v_{1}})\in\{1,2\},
\end{align}
which concludes the proof of this theorem.
\end{Proof}

\section{Proof of Theorem~\ref{th:4}}\label{appen:th_4}
\begin{Proof}
Let us consider the following set of states,
\begin{align}
    \bar{\mathcal{S}}= \{s_{ap,v_{2}}^{v_{1}}| v_{1}\geq \bar{v}_{1}, \ 1\leq v_{2}\leq v_{1}\}.
\end{align}
From Remark~\ref{rem:2}, for a state $s_{ap,v_{2}}^{v_{1}}\in\bar{\mathcal{S}}$, we have,
\begin{align}\label{eq:77}
    V_{\alpha}(s_{ap,v_{2}}^{v_{1}}) = \min\{V_{\alpha}(s_{ap,v_{2}}^{v_{1}};1), V_{\alpha}(s_{ap,v_{2}}^{v_{1}};2)\}.
\end{align}
We define,
\begin{align}\label{eq:78}
    U_{\alpha}(s_{ap,v_{2}}^{v_{1}}) = V_{\alpha}(s_{ap,v_{2}}^{v_{1}}) - \frac{v_{1}}{1-\alpha(1-q)}.
\end{align}
Taking $n\rightarrow\infty$ in (\ref{eq:32}) and (\ref{eq:33}), and using (\ref{eq:78}), we have,
\begin{align}
    V_{\alpha}(s_{ap,v_{2}}^{v_{1}};1) & = \frac{v_{1}}{1-\alpha(1 - q)}  +  \lambda  +  \frac{\alpha (1 - q)}{1 - \alpha (1 - q)}   +  \alpha q V_{\alpha}(s_{np}^{v_{2}+1}) +\alpha(1-q) U_{\alpha}(s_{ap,v_{2}+1}^{v_{1}+1}),\label{eq:79}\\ 
    V_{\alpha}(s_{ap,v_{2}}^{v_{1}};2) & = \frac{v_{1}}{1 - \alpha(1 - q)}  +   \lambda  +   \mu  +  \frac{\alpha (1 - q)}{1 - \alpha(1 - q)}  + \alpha q V_{\alpha}(s_{np}^{1})  +\alpha (1-q) U_{\alpha}(s_{ap,1}^{v_{1}+1}). \label{eq:80}
\end{align}
We define,
\begin{align}
    A_{v_{2}} & = \lambda + \alpha q V_{\alpha}(s_{np}^{v_{2}+1}) + \frac{\alpha(1-q)}{1-\alpha(1-q)}, \label{eq:81}\\ 
    B & = \lambda + \mu + \alpha q V_{\alpha}(s_{np}^{1}) + \frac{\alpha (1-q)}{1-\alpha(1-q)}.\label{eq:82}
\end{align}
Thus, from (\ref{eq:77}), (\ref{eq:78}), (\ref{eq:79}) and (\ref{eq:80}), we have
\begin{align}\label{eq:83}
    U_{\alpha}(s_{ap,v_{2}}^{v_{1}}) = \min\left\{A_{v_{2}} + \alpha (1-q)U_{\alpha}(s_{ap,v_{2}+1}^{v_{1}+1}), B+\alpha(1-q)U_{\alpha}(s_{ap,1}^{v_{1}+1})\right\}.
\end{align}
Now, consider a constant $\beta$, such that,
\begin{align}\label{eq:84}
    \beta>1, \quad \alpha (1-q)\beta<1.
\end{align}
Then, define a function space $\mathcal{U}_{\beta}$ as,
\begin{align}\label{eq:85}
    \mathcal{U}_{\beta} = \{U:\bar{\mathcal{S}}\rightarrow\mathbb{R} \ | \ ||U||_{\beta}<\infty\},
\end{align}
where
\begin{align}\label{eq:86}
    ||U||_{\beta} = \sup_{(v_{1},v_{2})\in \bar{\mathcal{S}}} \frac{|U(s_{ap,v_{2}}^{v_{1}})|}{\beta^{v_{1}}}.
\end{align}
Thus, a function in $\mathcal{U}_{\beta}$ does not grow faster than $\beta^{v_{1}}$. We can show that $||\cdot||_{\beta}$ is a valid norm defined on the set $\mathcal{U}_{\beta}$.

First, we show that the set (\ref{eq:85}) is complete with the norm defined in (\ref{eq:86}). Consider a Cauchy sequence $\{\bar{U}_{n}\}_{n=1}^{\infty}$ in $\mathcal{U}_{\beta}$ with the norm in (\ref{eq:86}). Let us define,
\begin{align}\label{eq:nn86}
    \tilde{U}_{n} (s_{ap,v_{2}}^{v_{1}}) = \frac{\bar{U}_{n}(s_{ap,v_{2}}^{v_{1}})}{\beta^{v_{1}}}.
\end{align}
Now,
\begin{align}
    ||\tilde{U}_{n} - \tilde{U}_{m}||_{\infty} &= \sup_{(v_{1},v_{2}) \in\bar{\mathcal{S}}} \frac{|\bar{U}_{n}(s_{ap,v_{2}}^{v_{1}}) - \bar{U}_{m}(s_{ap,v_{2}}^{v_{1}})|}{\beta^{v_{1}}} \label{eq:n90}\\ 
    &= ||\bar{U}_{n} -\bar{U}_{m}||_{\beta}.\label{eq:n91}
\end{align}
Note that $||\cdot||_{\infty}$ is the \textit{sup norm} as defined in (\ref{eq:n90}). Let us define the following set of bounded functions,
\begin{align}
    \mathcal{U}_{\infty} = \{U: \bar{\mathcal{S}}\rightarrow\mathbb{R} \ | \ ||U||_{\infty}<\infty \}.
\end{align}
As $\bar{U}_{n} \in \mathcal{U}_{\beta}$, we have $\tilde{U}_{n}\in\mathcal{U}_{\infty}$. Moreover, from (\ref{eq:n91}) we see that, $\{\tilde{U}_{n}\}_{n=1}^{\infty}$ is a Cauchy sequence in $\mathcal{U}_{\infty}$ with the sup norm. We know that the set of bounded functions $\mathcal{U}_{\infty}$ is complete under the sup norm. Therefore, there exists a bounded function $\tilde{U}\in\mathcal{U}_{\infty}$, such that,
\begin{align}\label{eq:n93}
    ||\tilde{U} - \tilde{U}_{n}||_{\infty} \rightarrow 0.
\end{align}
Now, we define,
\begin{align}\label{eq:nn91}
    \bar{U}(s_{ap,v_{2}}^{v_{1}}) = \beta^{v_{1}} \tilde{U}(s_{ap,v_{2}}^{v_{1}}).
\end{align}
Then, $||\bar{U}||_{\beta}<\infty$, thus, $\bar{U}\in\mathcal{U}_{\beta}$. Also, from (\ref{eq:nn86}), (\ref{eq:n93}) and (\ref{eq:nn91}), we have,
\begin{align}
    ||\bar{U}_{n} - \bar{U}||_{\beta}\rightarrow 0. 
\end{align}
Thus, the set of functions $\mathcal{U}_{\beta}$ is complete with the norm defined in (\ref{eq:86}). 

Now, we show that the function $U_{\alpha}$ in (\ref{eq:83}) belongs to the set $\mathcal{U}_{\beta}$. From Lemma~\ref{lemma:2}, we have,
\begin{align}
    \sup_{(v_{1},v_{2})\in\bar{\mathcal{S}}} \frac{|V_{\alpha}(s_{ap,v_{2}}^{v_{1}}) - \frac{v_{1}}{1-\alpha(1-q)}|} {\beta^{v_{1}}} \leq  \sup_{(v_{1},v_{2})\in\bar{\mathcal{S}}} \frac{|K|}{\beta^{v_{1}}} <\infty.
\end{align}
Thus, from (\ref{eq:78}), we have $||U_{\alpha}||_{\beta} <\infty$, implying that $U_{\alpha}\in \mathcal{U}_{\beta}$.

For any function ${U}\in\mathcal{U}_{\beta}$, we define an operator $\mathcal{T}$ as,
\begin{align}\label{eq:92}
    (\mathcal{T} {U}) (s_{ap,v_{2}}^{v_{1}}) = \min\left\{A_{v_{2}} + \alpha (1-q) {U}(s_{ap,v_{2}+1}^{v_{1}+1}), B+ \alpha(1-q){U}(s_{ap,1}^{v_{1}+1}) \right\}.
\end{align}
As $U_{\alpha}\in\mathcal{U}_{\beta}$, from (\ref{eq:83}), we see that, 
\begin{align}
    U_{\alpha}(s_{ap,v_{2}}^{v_{1}}) = (\mathcal{T}U_{\alpha}) (s_{ap,v_{2}}^{v_{1}}).
\end{align}
We show that the operator $\mathcal{T}$ maps a function ${U} \in \mathcal{U}_{\beta}$ back to the set $\mathcal{U}_{\beta}$. From (\ref{eq:92}), we have,
\begin{align}\label{eq:n103}
    |(\mathcal{T} {U}) (s_{ap,v_{2}}^{v_{1}})| \leq \max\left\{|A_{v_{2}} + \alpha (1-q) {U}(s_{ap,v_{2}+1}^{v_{1}+1})|, |B+\alpha(1-q) {U}(s_{ap,1}^{v_{1}+1})|\right\}.
\end{align}
From the definition of norm in (\ref{eq:86}), we have,
\begin{align}
    \frac{|{U}(s_{ap,v_{2}+1}^{v_{1}+1})|} {\beta^{v_{1}}} \leq \beta ||{U}||_{\beta}, \quad  \frac{|{U}(s_{ap,1}^{v_{1}+1})|} {\beta^{v_{1}}}\leq \beta ||{U}||_{\beta}.
\end{align}
Now, as $||{U}||_{\beta}<\infty$, we have,
\begin{align}\label{eq:n106}
    \max\left\{\sup_{(v_{1},v_{2})\in\bar{\mathcal{S}}} \frac{|{U}(s_{ap,v_{2}+1}^{v_{1}+1})|} {\beta^{v_{1}}}, \sup_{(v_{1},v_{2})\in\bar{\mathcal{S}}} \frac{|{U}(s_{ap,1}^{v_{1}+1})|} {\beta^{v_{1}}} \right\} <\infty.
\end{align}
Now, from Lemma~\ref{lemma:4} and (\ref{eq:81}), we have $A_{v_{2}}\leq A_{v_{1}}$. From (\ref{eq:81}), Lemma~\ref{lemma:2} and Lemma~\ref{lemma:1}, we see that an upper bound of $A_{v_{1}}$ grows linearly with $v_{1}$ and is independent of $v_{2}$. Thus,
\begin{align}\label{eq:n107}
    \sup_{(v_{1},v_{2})\in\bar{\mathcal{S}}} \frac{|A_{v_{1}}|}{\beta^{v_{1}}} = \sup_{v_{1}\geq \bar{v}_{1}} \frac{|A_{v_{1}}|}{\beta^{v_{1}}} <\infty.
\end{align}
Note that $B$ is a finite constant, and from (\ref{eq:n103}), (\ref{eq:n106}) and (\ref{eq:n107}), we have $||(\mathcal{T} U)||_{\beta}<\infty$, implying that $(\mathcal{T} U) \in \mathcal{U}_{\beta}$. 

We now show that the operator $\mathcal{T}:\mathcal{U}_{\beta}\rightarrow\mathcal{U}_{\beta}$ is a contraction. Let us consider two functions $\hat{U}, \check{U} \in \mathcal{U}_{\beta}$. Let
\begin{align}
    \hat{x}_{1} =& A_{v_{2}} + \alpha(1-q) \hat{U}(s_{ap,v_{2}+1}^{v_{1}+1}), \quad  \hat{x}_{2} = B+ \alpha(1-q) \hat{U}(s_{ap,1}^{v_{1}+1}),\nonumber \\ 
    \check{x}_{1} =& A_{v_{2}} + \alpha(1-q) \check{U}(s_{ap,v_{2}+1}^{v_{1}+1}),\quad \check{x}_{2} = B + \alpha(1-q) \check{U}(s_{ap,1}^{v_{1}+1}).
\end{align}
Now, from (\ref{eq:92}), we have,
\begin{align}
  (\mathcal{T}\hat{U})(s_{ap,v_{2}}^{v_{1}}) =\min\{\hat{x}_{1},\hat{x}_{2}\}, \quad(\mathcal{T}\check{U})(s_{ap,v_{2}}^{v_{1}}) = \min\{\check{x}_{1},\check{x}_{2}\}.
\end{align}
With the following identity,
\begin{align}\label{eq:n112}
    |\min\{\hat{x}_{1},\hat{x}_{2}\} - \min\{\check{x}_{1},\check{x}_{2}\}| \leq \max\{|\hat{x}_{1}-\check{x}_{1}|, |\hat{x}_{2}-\check{x}_{2}|\},
\end{align}
we have,
\begin{align}\label{eq:102}
    |( \mathcal{T}\hat{U} )(s_{ap,v_{2}}^{v_{1}})  -  ( \mathcal{T}\check{U} )(s_{ap,v_{2}}^{v_{1}})|  \leq  \alpha(1 - q) \max\left\{ |\hat{U}(s_{ap,v_{2}+1}^{v_{1}+1}) - \check{U}(s_{ap,v_{2}+1}^{v_{1}+1})| , |\hat{U}(s_{ap,1}^{v_{1}+1}) - \check{U}(s_{ap,1}^{v_{1}+1})|\right\}.
\end{align}
From the definition of the norm in (\ref{eq:86}) and (\ref{eq:102}), we have,
\begin{align}\label{eq:103}
    \frac{|(\mathcal{T}\hat{U})(s_{ap,v_{2}}^{v_{1}}) - (\mathcal{T}\check{U})(s_{ap,v_{2}}^{v_{1}})|}{\beta^{v_{1}}}  \leq  \alpha(1-q) \beta ||\hat{U}-\check{U}||_{\beta},
\end{align}
thus,
\begin{align}
    \sup_{(v_{1},v_{2})\in\bar{\mathcal{S}}}\frac{|(\mathcal{T}\hat{U})(s_{ap,v_{2}}^{v_{1}})  -  (\mathcal{T}\check{U})(s_{ap,v_{2}}^{v_{1}})|}{\beta^{v_{1}}} \leq  \alpha(1-q) \beta ||\hat{U}-\check{U}||_{\beta}, 
\end{align}
therefore,
\begin{align}
    ||(\mathcal{T}\hat{U})  - (\mathcal{T}\check{U})||_{\beta}  \leq  \alpha(1-q)\beta ||\hat{U}-\check{U}||_{\beta}.\label{eq:104}
\end{align}
From (\ref{eq:84}), we have $\alpha(1-q)\beta<1$. Hence, (\ref{eq:104}) shows that $\mathcal{T}$ is a contraction.

Now, we define another function space $\mathcal{M}_{\beta}$ as,
\begin{align}\label{eq:87}
    \mathcal{M}_{\beta} = \left\{M:\mathbb{N}\rightarrow\mathbb{R}\ | \ ||M||_{\beta,v_{2}}<\infty\right\},
\end{align}
where 
\begin{align}\label{eq:88}
    ||M||_{\beta,v_{2}} = \sup_{v_{2}\geq 1 } \frac{|M(v_{2})|}{\beta^{v_{2}}}.
\end{align}
We can show that $||\cdot||_{\beta,v_{2}}$ is a valid norm. Using a similar technique to the one we used to show that the set $\mathcal{U}_{\beta}$ is complete, we can show that the set of functions $\mathcal{M}_{\beta}$ is also complete under the norm in \eqref{eq:88}. For every function ${M}\in\mathcal{M}_{\beta}$, we define an operator $\mathcal{T}_{0}$ as,
\begin{align}\label{eq:n118}
    (\mathcal{T}_{0}{M})(v_{2}) = \min\left\{A_{v_{2}}+\alpha(1-q){M}(v_{2}+1), \ B+\alpha(1-q) {M}(1)\right\}.
\end{align}
Similar to the operator $\mathcal{T}:\mathcal{U}_{\beta}\rightarrow\mathcal{U}_{\beta}$, we can show that the operator $\mathcal{T}_{0}$ maps any function $\bar{M}\in\mathcal{M}_{\beta}$ back into $\mathcal{M}_{\beta}$. Also, following a similar argument as before, we now show that the operator $\mathcal{T}_{0}$ is a contraction. Let us consider two functions $\bar{M},\tilde{M}\in\mathcal{M}_{\beta}$. Now, using the inequality of (\ref{eq:n112}) and (\ref{eq:n118}), we have, 
\begin{align}\label{eq:119}
    |(\mathcal{T}_{0}\bar{M})(v_{2}) - (\mathcal{T}_{0}\tilde{M})(v_{2})| \leq \alpha (1-q) \max\left\{|\bar{M}(v_{2}+1)-\tilde{M}(v_{2}+1)|, |\bar{M}(1)-\tilde{M}(1)|\right\}.
\end{align}
Now, from the definition of the norm in (\ref{eq:88}), we have,
\begin{align}
    |\bar{M}(v_{2}+1) -\tilde{M}(v_{2}+1)| & \leq ||\bar{M}-\tilde{M}||_{\beta,v_{2}} \beta^{v_{2}+1}, \\
    |\bar{M}(1)-\tilde{M}(1)| & \leq ||\bar{M}-\tilde{M}||_{\beta,v_{2}} \beta. 
\end{align}
As $\beta>1$, from (\ref{eq:119}), we get,
\begin{align}\label{eq:122}
    \frac{|(\mathcal{T}_{0}\bar{M})(v_{2})  -  (\mathcal{T}_{0}\tilde{M})(v_{2})|}{\beta^{v_{2}}} \leq  \alpha(1 - q) \beta ||\bar{M} - \tilde{M}||_{\beta,v_{2}}.
\end{align}
Taking supremum of both sides on (\ref{eq:122}) for $v_{2}\geq 1$, we get,
\begin{align}
    ||\mathcal{T}_{0}\bar{M} - \mathcal{T}_{0}\tilde{M}||_{\beta,v_{2}} \leq \alpha (1-q) \beta ||\bar{M}-\tilde{M}||_{\beta,v_{2}}.
\end{align}
Now, as from (\ref{eq:84}) we have $\alpha(1-q)\beta<1$, we see that the operator $\mathcal{T}_{0}:\mathcal{M}_{\beta}\rightarrow\mathcal{M}_{\beta}$ is a contraction. Moreover, as $(\mathcal{M}_{\beta}, ||M||_{\beta,v_{2}})$ is a complete metric space, from the Banach fixed-point theorem, we say that, the operator $\mathcal{T}_{0}$ admits a unique fixed point $\hat{M}\in\mathcal{M}_{\beta}$, i.e., 
\begin{align}\label{eq:124}
    \hat{M}(v_{2}) = \min\left\{A_{v_{2}}+\alpha(1-q)\hat{M}(v_{2}+1),   B+\alpha(1-q) \hat{M}(1)\right\}.
\end{align}
Now, we define a function $\hat{U}:\bar{\mathcal{S}}\rightarrow\mathbb{R}$ as,
\begin{align}\label{eq:125}
    \hat{U}(s_{ap,v_{2}}^{v_{1}}) = \hat{M}(v_{2}).
\end{align}
Now, we first show that $\hat{U}\in\mathcal{U}_{\beta}$. Note,
\begin{align}\label{eq:126}
    |\hat{U}(s_{ap,v_{2}}^{v_{1}})|  =  |\hat{M}(v_{2})| \leq  ||\hat{M}||_{\beta,v_{2}} \beta^{v_{2}}  \leq  ||\hat{M}||_{\beta,v_{2}} \beta^{v_{1}},
\end{align}
where the last inequality follows from $v_{2}\leq v_{1}$. Thus, from (\ref{eq:126}) we have,
\begin{align}
    \sup_{(v_{1},v_{2})\in\bar{\mathcal{S}}} \frac{|\hat{U}(s_{ap,v_{2}}^{v_{1}})|}{\beta^{v_{1}}} \leq ||\hat{M}||_{\beta,v_{2}}<\infty.
\end{align}
Thus, $\hat{U}(s_{ap,v_{2}}^{v_{1}})\in\mathcal{U}_{\beta}$. Now, from (\ref{eq:92}), we have,
\begin{align}
    (\mathcal{T} \hat{U}) (s_{ap,v_{2}}^{v_{1}}) & = \min\{A_{v_{2}} + \alpha (1-q) \hat{U}(s_{ap,v_{2}+1}^{v_{1}+1}),  \ \ B+ \alpha(1-q)\hat{U}(s_{ap,1}^{v_{1}+1}) \} \\ 
    & = \min\{A_{v_{2}} + \alpha (1-q) \hat{M}(v_{2}+1),  \ \ B+ \alpha(1-q)\hat{M}(1) \} \label{eq:129}\\ 
    & = \hat{M}(v_{2}) \label{eq:130}\\
    & = \hat{U}(s_{ap,v_{2}}^{v_{1}}),\label{eq:131}
\end{align}
where (\ref{eq:129}) follows from (\ref{eq:125}), (\ref{eq:130}) follows from (\ref{eq:124}), and (\ref{eq:131}) follows from (\ref{eq:125}). Thus, $\hat{U}$ is a fixed point of the operator $\mathcal{T}$. As $\mathcal{T}:\mathcal{U}_{\beta}\rightarrow\mathcal{U}_{\beta}$ is a contraction and $(\mathcal{U}_{\beta}, ||U||_{\beta})$ is a complete metric space, from the Banach fixed point theorem, we see that $\hat{U}$ is the unique fixed point of $\mathcal{T}$. Now, we have already shown that $U_{\alpha}\in\mathcal{U}_{\beta}$. Moreover from (\ref{eq:83}) and (\ref{eq:92}), we see that $U_{\alpha}(s_{ap,v_{2}}^{v_{1}})$ is also a fixed point of the operator $\mathcal{T}$. Thus, from the uniqueness of the fixed point of $\mathcal{T}$, we have, 
\begin{align}
    U_{\alpha}(s_{ap,v_{2}}^{v_{1}}) = \hat{U}(s_{ap,v_{2}}^{v_{1}}) = \hat{M}(v_{2}).
\end{align}
Thus, for a state $s_{ap,v_{2}}^{v_{1}}\in\bar{\mathcal{S}}$, from (\ref{eq:78}), we have,
\begin{align}
    V_{\alpha}(s_{ap,v_{2}}^{v_{1}}) = \frac{v_{1}}{1-\alpha(1-q)} + \hat{M}(v_{2}),
\end{align}
which completes the proof of this theorem.
\end{Proof}

\section{Proof of Theorem~\ref{th:5}}\label{appen:th:5}
\begin{Proof}
Let us consider the following set of states,
\begin{align}
    \bar{\mathcal{S}}= \left\{s_{ap,v_{2}}^{v_{1}}| v_{1}\geq \bar{v}_{1}, \ 1\leq v_{2}\leq v_{1}\right\}.
\end{align}
From Remark~\ref{rem:2}, we know that for a state $s_{ap,v_{2}}^{v_{1}}\in\bar{\mathcal{S}}$, if  
\begin{align}\label{eq:135}
    V_{\alpha}(s_{ap,v_{2}}^{v_{1}};1) - V_{\alpha}(s_{ap,v_{2}}^{v_{1}};2) \geq 0,
\end{align}
then action $a=2$ is optimal, otherwise action $a=1$ is optimal. Now, from Theorem~\ref{th:4}, we have,
\begin{align}\label{eq:136}
    V_{\alpha}(s_{ap,v_{2}}^{v_{1}};1) - V_{\alpha}(s_{ap,v_{2}}^{v_{1}};2) =  \alpha q(V_{\alpha}(s_{np}^{v_{2}+1}) - V_{\alpha}(s_{np}^{1}))  + \alpha(1-q) (\hat{M}(v_{2}+1) - \hat{M}(1)) -\mu.
\end{align}
Thus, the optimal action that we get from the comparison of (\ref{eq:135}) is independent of the age of the monitor $v_{1}$. 

For a state $s_{ap,v_{2}}^{v_{1}}\in\bar{\mathcal{S}}$, from Theorem~\ref{th:4}, we have
\begin{align}
    \hat{M}(v_{2}) =  V_{\alpha}(s_{ap,v_{2}}^{v_{1}}) - \frac{v_{1}}{1-\alpha(1-q)}. 
\end{align}
Thus, from Lemma~\ref{lemma:4}, we see that $\hat{M}(v_{2})$ is an increasing function of $v_{2}$, similarly from Lemma~\ref{lemma:4} we have $V_{\alpha}(s_{np}^{v_{2}})$ as an increasing function of $v_{2}$. Thus, from (\ref{eq:136}), we see that if for a state $s_{ap,v_{2}'}^{v_{1}'}\in\bar{\mathcal{S}}$, action $a=2$ is optimal, then for state $s_{ap,v_{2}}^{v_{1}}\in\bar{\mathcal{S}}$, $a=2$ is still optimal, where $v_{2}\geq v_{2}'$. 
    
Now, from Lemma~\ref{lemma:4}, we have,
\begin{align}\label{eq:139}
    V_{\alpha}(s_{np}^{v_{2}+1}) \geq v_{2}+1.
\end{align}
Thus, there exists a finite $\bar{v}_{2}$, such that,
\begin{align}
    \alpha (q(V_{\alpha}(s_{np}^{\bar{v}_{2}+1}) - V_{\alpha}(s_{np}^{1})) + (1 - q) (\hat{M}(\bar{v}_{2} + 1) - \hat{M}(1))) \geq \mu,
\end{align}
That is, for state $s_{ap,\bar{v}_{2}}^{v_{1}}\in\bar{\mathcal{S}}$, action $a=2$ is optimal, which concludes the proof of this theorem.
\end{Proof}

\section{Proof of Theorem~\ref{th:no_packet_threshold}}\label{appen:no_packet_threshold}
For simplicity of presentation, we suppress the dependence of $\hat{v}_{1}(\alpha)$ on $\alpha$ and denote it by $\hat{v}_{1}$. For a fixed discount factor $0<\alpha<1$, we define the following quantities,
\begin{align}
    \bar{\phi}_\alpha(v_{1}) &= V_\alpha(s_{np}^{v_{1}+1})-V_\alpha(s_{np}^{v_{1}}), \ v_{1}\geq1, \label{eq:npt_D_definition}\\
    \tilde{\phi}_\alpha(v_1,v_2) &= V_\alpha(s_{ap,v_2+1}^{v_1})-V_\alpha(s_{ap,v_2}^{v_1}),\ 1\leq v_2<v_1, \label{eq:npt_L_definition}\\
    \hat{\phi}_\alpha(v_{1}) &= V_\alpha(s_{np}^{v_{1}};0)-V_\alpha(s_{np}^{v_{1}};2), \ v_{1}\geq1. \label{eq:420}
\end{align}
We first state and prove Lemma~\ref{lemma:npt_discrete_concavity}, Lemma~\ref{lemma:7}, Lemma~\ref{lemma:8}, Lemma~\ref{lemma:9}, Lemma~\ref{lemma:10}, Lemma~\ref{lemma:npt_boundary_increment}, Lemma~\ref{lemma:12} and Lemma~\ref{lemma:13} below. These will be used to prove Theorem~\ref{th:no_packet_threshold}.

\begin{lemma}\label{lemma:npt_discrete_concavity}
    For every $0<\alpha<1$, the following relations hold:
    \begin{align}
        \bar{\phi}_\alpha(v_{1}) & \geq 1, \label{eq:npt_D_positive} \\
        \tilde{\phi}_\alpha(v_1,v_2) & \geq 0, \label{eq:npt_L_positive} \\
        \bar{\phi}_\alpha(v_{1}+1) & \leq \bar{\phi}_\alpha(v_{1}). \label{eq:npt_D_decreasing}
    \end{align}
\end{lemma}
\begin{Proof}
The first two inequalities follow directly from Lemma~\ref{lemma:4}. It remains to prove~(\ref{eq:npt_D_decreasing}). We prove by induction that $V_{\alpha,n}(s_{np}^{v_{1}})$ and $V_{\alpha,n}(s_{ap,v_{2}}^{v_{1}})$ are discretely concave functions of $v_{1}$ for $n\geq 0$. Then, from (\ref{eq:26}), we see that $V_{\alpha}(s_{np}^{v_{1}})$ is a discretely concave function of $v_{1}$ and (\ref{eq:npt_D_decreasing}) follows immediately. It is immediate that $V_{\alpha,0}(s_{np}^{v_{1}})$ and $V_{\alpha,0}(s_{ap,v_{2}}^{v_{1}})$ are discretely concave functions as $V_{\alpha,0}(s_{np}^{v_{1}})=V_{\alpha,0}(s_{ap,v_{2}}^{v_{1}})= 0$. Let us assume that $V_{\alpha,n}(s_{np}^{v_{1}})$ and $V_{\alpha,n}(s_{ap,v_{2}}^{v_{1}})$ are discretely concave functions. Now, from (\ref{eq:34}) and (\ref{eq:35}), we see that $V_{\alpha,n+1}(s_{np}^{v_{1}};0)$ and $V_{\alpha,n+1}(s_{np}^{v_{1}};2)$ are discretely concave functions. Since the pointwise minimum of one-dimensional discretely concave functions is discretely concave, from (\ref{eq:n26}), we see that $V_{\alpha,n+1}(s_{np}^{v_{1}})$ is a discretely concave function of $v_{1}$. Similarly, from (\ref{eq:31}), (\ref{eq:32}) and (\ref{eq:33}), we see that $V_{\alpha,n+1}(s_{ap,v_{2}}^{v_{1}};0)$, $V_{\alpha,n+1}(s_{ap,v_{2}}^{v_{1}};1)$ and $V_{\alpha,n+1}(s_{ap,v_{2}}^{v_{1}};2)$ are discretely concave functions. Thus, $V_{\alpha,n+1}(s_{ap,v_{2}}^{v_{1}})$ is also a discretely concave function. 
\end{Proof}

\begin{lemma}\label{lemma:7}
    For $\frac{\lambda}{\lambda+q}<\alpha<1$, $v_{1}\geq \hat{v}_{1}$, and $1 \leq v_{2}< v_{1}$, the following relation holds,
    \begin{align}\label{eq:424}
         \tilde{\phi}_{\alpha}(v_1,v_2) = \tilde{\phi}_{\alpha}(v_1+1,v_2)< \bar{\phi}_\alpha(v_2).
    \end{align}
\end{lemma}
\begin{Proof}
The first equality follows immediately from (\ref{eq:affine_above_hat_v1}) in Remark~\ref{rem:4}. We prove the last inequality by backward induction on $v_{2}$. First, let us assume that $v_{1}> \bar{v}_{2}$. From the definition of $\bar{v}_{2}$ in Remark~\ref{rem:4}, we know that if $v_{2}\geq \bar{v}_{2}$, action $a=2$ is optimal at both states $s_{ap,v_{2}+1}^{v_{1}}$ and $s_{ap,v_{2}}^{v_{1}}$. Thus, from (\ref{eq:33}) and (\ref{eq:npt_D_positive}), for $\bar{v}_{2}\leq v_{2}< v_{1}$, we have
\begin{align}\label{eq:425}
    \tilde{\phi}_\alpha(v_1,v_2)=0<\bar{\phi}_\alpha(v_2).
\end{align}
If $\bar{v}_{2}\geq 2$, we extend the inequality in (\ref{eq:425}) to $v_{2}\in\{1,2,\cdots,\bar{v}_{2}-1\}$ as follows, 
\begin{align}
    \tilde{\phi}_{\alpha}(v_{1},\bar{v}_{2}-1) &= V_{\alpha}(s_{ap,\bar{v}_{2}}^{v_{1}};2) -V_{\alpha}(s_{ap,\bar{v}_{2}-1}^{v_{1}};1) \\ 
    &\leq  V_{\alpha}(s_{ap,\bar{v}_{2}}^{v_{1}};1) -V_{\alpha}(s_{ap,\bar{v}_{2}-1}^{v_{1}};1) \\  
    & = \alpha(q\bar{\phi}_{\alpha}(\bar{v}_{2}) + (1 - q)\tilde{\phi}_{\alpha}(v_{1}+1,\bar{v}_{2}))\label{eq:429}\\
    & \leq \alpha \bar{\phi}_{\alpha}(\bar{v}_{2})\label{eq:n429} \\ 
    & < \bar{\phi}_{\alpha}(\bar{v}_{2})\label{eq:nn430} \\
    & \leq \bar{\phi}_{\alpha}(\bar{v}_{2}-1),\label{eq:n430}
\end{align}
where (\ref{eq:n429}) follows from (\ref{eq:425}), and (\ref{eq:n430}) follows from Lemma~\ref{lemma:npt_discrete_concavity}. If $\bar{v}_{2}\geq 3$, then, for $v_{2}=\bar{v}_{2}-2$, we have
\begin{align}
    \tilde{\phi}_{\alpha}(v_{1},\bar{v}_{2}-2) & = V_{\alpha}(s_{ap,\bar{v}_{2}-1}^{v_{1}};1) -V_{\alpha}(s_{ap,\bar{v}_{2}-2}^{v_{1}};1) \\
    & < \bar{\phi}_{\alpha}(\bar{v}_{2}-1) \label{eq:n432}\\
    & \leq \bar{\phi}_{\alpha}(\bar{v}_{2}-2),\label{eq:n433}
\end{align}
where (\ref{eq:n432}) follows by repeating the steps used to obtain (\ref{eq:nn430}), and (\ref{eq:n433}) follows from Lemma~\ref{lemma:npt_discrete_concavity}. Continuing this backward recursion in $v_{2}$, we get
\begin{align}
    \tilde{\phi}_{\alpha}(v_{1},v_{2}) < \bar{\phi}_{\alpha}(v_{2}), \quad 1\leq v_{2}\leq \bar{v}_{2}-1.
\end{align}

Next, consider the case $v_{1}\leq \bar{v}_{2}$. Then, we show that for the boundary case, the second inequality in (\ref{eq:424}) holds, i.e., we consider $v_{2}=v_{1}-1$. As $\hat{v}_{1}\leq v_{1}\leq \bar{v}_{2}$, from Remark~\ref{rem:4}, we know that action $a=1$ is optimal at the state $s_{ap,v_{1}-1}^{v_{1}}$ and either action $a=1$ or $a=2$ is optimal at the state $s_{ap,v_{1}}^{v_{1}}$. Thus,
\begin{align}\label{eq:435}
    \tilde{\phi}_{\alpha}(v_{1},v_{1}-1)= V_{\alpha}(s_{ap,v_{1}}^{v_{1}})-V_{\alpha}(s_{ap,v_{1}-1}^{v_{1}};1). 
\end{align}
From (\ref{eq:32}), we have
\begin{align}
    V_{\alpha}(s_{ap,v_{1}-1}^{v_{1}};1)  =  v_{1}  +  \lambda  +  \alpha qV_{\alpha}(s_{np}^{v_{1}}) + \alpha (1-q)V_{\alpha}(s_{ap,v_{1}}^{v_{1}+1}).
\end{align}
Now, from (\ref{eq:34}), we get
\begin{align}
    V_{\alpha}(s_{np}^{v_{1}-1};0) = v_{1} -1+ \alpha V_{\alpha}(s_{np}^{v_{1}}).  
\end{align}
Thus, from Lemma~\ref{lemma:1} and Lemma~\ref{lemma:4}, we get 
\begin{align}\label{eq:438}
    V_{\alpha}(s_{ap,v_{1}-1}^{v_{1}};1) \geq V_{\alpha}(s_{np}^{v_{1}-1};0)+\lambda+1. 
\end{align}
Now, 
\begin{align}
    \bar{\phi}_{\alpha}(v_{1}-1) \geq  V_{\alpha}(s_{np}^{v_{1}}) - V_{\alpha}(s_{np}^{v_{1}-1};0). 
\end{align}
Since $\lambda\geq 0$, Lemma~\ref{lemma:1} and (\ref{eq:438}) give
\begin{align}
    \bar{\phi}_{\alpha}(v_{1}-1) \geq V_{\alpha}(s_{ap,v_{1}}^{v_{1}}) - V_{\alpha}(s_{ap,v_{1}-1}^{v_{1}}) +1.
\end{align}
Thus, from (\ref{eq:435}), we have
\begin{align}
    \tilde{\phi}_{\alpha}(v_{1},v_{1}-1) < \bar{\phi}_{\alpha}(v_{1}-1).
\end{align}
Having established the inequality in (\ref{eq:424}) at the boundary
when $v_{1}\leq\bar{v}_{2}$, we apply the same backward recursion as
in the case $v_{1}>\bar{v}_{2}$. This establishes (\ref{eq:424}) for
$\hat{v}_{1}\leq v_{1}\leq\bar{v}_{2}$ and
$1\leq v_{2}<v_{1}$.                
\end{Proof}

\begin{lemma}\label{lemma:8}
    If $\hat{\phi}_\alpha(v_{1})<0$, then action $2$ is strictly suboptimal at every packet-available state $s_{ap,v_2}^{v_{1}}$, $1\leq v_2\leq v_{1}$.
\end{lemma}

\begin{Proof}
From (\ref{eq:31}), (\ref{eq:33}), and Lemma~\ref{lemma:4}, we have
\begin{align}
    V_\alpha(s_{ap,v_2}^{v_{1}};0)-V_\alpha(s_{ap,v_2}^{v_{1}};2)\leq V_\alpha(s_{ap,v_{1}}^{v_{1}};0) - V_\alpha(s_{ap,v_{1}}^{v_{1}};2)= \hat{\phi}_\alpha(v_{1})<0, \label{eq:npt_action2_excluded}
\end{align}
where the equality follows from Lemma~\ref{lemma:1}.  Hence, action $a=0$ is strictly better than action $a=2$ for state $s_{ap,v_{2}}^{v_{1}}$.
\end{Proof}

\begin{lemma}\label{lemma:9}
    If $\hat{\phi}_\alpha(v_{1})<0$ and
    \begin{align}\label{eq:430}
        \tilde{\phi}_\alpha(v_{1}+1,v_{2})<\bar{\phi}_\alpha(v_{2}), \quad 1\leq v_{2}<v_{1}+1.
    \end{align}
    Then, there exists $1\leq \tilde{v}_{2} \leq v_{1}$ such that the following holds,
    \begin{align}
        \begin{cases}\label{eq:431}
        V_{\alpha}(s_{ap,v_{2}}^{v_{1}};1) \leq V_{\alpha}(s_{ap,v_{2}}^{v_{1}};0),  &\textrm{ if } 1 \leq v_{2} <\tilde{v}_{2}, \\ V_{\alpha}(s_{ap,v_{2}}^{v_{1}};0) \leq V_{\alpha}(s_{ap,v_{2}}^{v_{1}};1), &\textrm{ if } \tilde{v}_{2}\leq v_{2}\leq v_{1} . \end{cases}
    \end{align}
\end{lemma}

\begin{Proof}
Since $\lambda\geq 0$, (\ref{eq:31}) and (\ref{eq:32}) imply that
\begin{align}
    V_{\alpha}(s_{ap,v_{1}}^{v_{1}};0) \leq V_{\alpha}(s_{ap,v_{1}}^{v_{1}};1).
\end{align}
Thus, the threshold $\tilde{v}_{2}$ defined in (\ref{eq:431}) exists. We prove this lemma by contradiction. Suppose that the threshold structure in (\ref{eq:431}) does not hold. Thus, there exists a state $s_{ap,v_{2}}^{v_{1}}$, $1\leq v_{2}<v_{1}$, such that
\begin{align}
    V_{\alpha}(s_{ap,v_{2}}^{v_{1}};0) \leq V_{\alpha}(s_{ap,v_{2}}^{v_{1}};1), \quad V_{\alpha}(s_{ap,v_{2}+1}^{v_{1}};1) < V_{\alpha}(s_{ap,v_{2}+1}^{v_{1}};0),\label{eq:432}
\end{align}
both hold. By Lemma~\ref{lemma:8}, an optimal action at each of the states $s_{ap,v_{2}}^{v_{1}}$ and $s_{ap,v_{2}+1}^{v_{1}}$ belongs to $\{0,1\}$. Thus, from (\ref{eq:432}) we see that action $a=0$ is optimal for state $s_{ap,v_{2}}^{v_{1}}$ and action $a=1$ is optimal for state $s_{ap,v_{2}+1}^{v_{1}}$. Thus,
\begin{align}
    \tilde{\phi}_{\alpha}(v_{1},v_{2}) & \leq V_{\alpha}(s_{ap,v_{2}+1}^{v_{1}};0) - V_{\alpha}(s_{ap,v_{2}}^{v_{1}};0)\\
    &=\alpha \tilde{\phi}_{\alpha}(v_{1}+1,v_{2}+1).\label{eq:n436}
\end{align}
Also, from (\ref{eq:31}), (\ref{eq:32}) and (\ref{eq:432}), we get,
\begin{align}\label{eq:436}
    \lambda \geq \alpha q \left[ V_\alpha(s_{ap,v_2+1}^{v_{1}+1}) - V_\alpha(s_{np}^{v_2+1}) \right].
\end{align}
Now,
\begin{align}
    \tilde{\phi}_{\alpha}(v_{1},v_{2}) &= V_{\alpha}(s_{ap,v_{2}+1}^{v_{1}};1) - V_{\alpha}(s_{ap,v_{2}}^{v_{1}};0), \\
    &= \lambda+\alpha qV_\alpha(s_{np}^{v_2+2}) +\alpha(1-q)V_\alpha(s_{ap,v_2+2}^{v_{1}+1})-\alpha V_\alpha(s_{ap,v_2+1}^{v_{1}+1}). 
\end{align}
Using (\ref{eq:436}), we get 
\begin{align}\label{eq:n454}
\tilde{\phi}_\alpha(v_{1} ,v_2)
 \geq 
\alpha (q\bar{\phi}_\alpha(v_2 + 1)
 + 
(1 - q)\tilde{\phi}_\alpha(v_{1}  + 1,v_2 + 1)).
\end{align}
Now, from (\ref{eq:n436}) and (\ref{eq:n454}) we get
\begin{align}\label{eq:441}
    \tilde{\phi}_{\alpha}(v_{1}+1,v_{2}+1) \geq \bar{\phi}_{\alpha}(v_{2}+1).
\end{align}
Recall that we have assumed $v_{2}<v_{1}$. Therefore, from (\ref{eq:430}),
\begin{align}
    \bar{\phi}_{\alpha}(v_{2}+1) > \tilde{\phi}_\alpha(v_{1}+1,v_{2}+1),
\end{align}
which contradicts (\ref{eq:441}). Thus, the threshold structure on the age of the packet in (\ref{eq:431}) holds. 
\end{Proof}

\begin{lemma}\label{lemma:10}
    For $\frac{\lambda}{\lambda+q}<\alpha<1$, the following properties hold for the discounted-cost problem.
    \begin{enumerate}
    \item If action $a=2$ is optimal at $s_{ap,v_2}^{\hat v_1}$, then action $a=
    0$ is strictly optimal at $s_{ap,v_2}^{\hat v_1-1}$, for $1\leq v_{2}<\hat{v}_{1}$.  
    
    \item Let $v_{1}+1<\hat v_1$.  Suppose action $a=2$ is not optimal for state $s_{ap,v_{2}}^{v_{1}}$. If action $a=0$ is optimal at $s_{ap,v_2}^{v_{1}+1}$, then action $a=0$ is also optimal at $s_{ap,v_2}^{v_{1}}$, for $1\leq v_{2}\leq v_{1}$.  
    \end{enumerate}
\end{lemma}

\begin{Proof}
Since action $a=2$ is optimal at the state $s_{ap,v_{2}}^{\hat{v}_{1}}$, Remark~\ref{rem:4} shows that it is also optimal at the state $s_{ap,v_{2}+1}^{\hat{v}_{1}}$. From the definition of $\hat{v}_{1}$, action $a=2$ is also optimal for state $s_{np}^{\hat{v}_{1}}$. Thus, from (\ref{eq:33}) and (\ref{eq:35}), we get
\begin{align}
    V_{\alpha}(s_{ap,v_{2}+1}^{\hat{v}_{1}}) = V_{\alpha}(s_{np}^{\hat{v}_{1}}).
\end{align}
Thus, we have
\begin{align}
    V_{\alpha}(s_{ap,v_{2}}^{\hat{v}_{1}-1};0) = V_{\alpha}(s_{np}^{\hat{v}_{1}-1};0), \quad V_{\alpha}(s_{ap,v_{2}}^{\hat{v}_{1}-1};2) =  V_{\alpha}(s_{np}^{\hat{v}_{1}-1};2).\label{eq:445}
\end{align}
From the definition of $\hat{v}_{1}$ in Remark~\ref{rem:3}, we see that $\hat{\phi}_{\alpha}(\hat{v}_{1}-1)<0$. Thus, from (\ref{eq:445}), we have
\begin{align}\label{eq:446}
    V_{\alpha}(s_{ap,v_{2}}^{\hat{v}_{1}-1};0) < V_{\alpha}(s_{ap,v_{2}}^{\hat{v}_{1}-1};2).
\end{align}
From (\ref{eq:affine_above_hat_v1}) in Remark~\ref{rem:4}, we have
\begin{align}\label{eq:447}
    V_{\alpha}(s_{ap,v_{2}}^{\hat{v}_{1}-1};1)-V_{\alpha}(s_{ap,v_{2}}^{\hat{v}_{1}-1};2) = V_{\alpha}(s_{ap,v_{2}}^{\hat{v}_{1}};1)- V_{\alpha}(s_{ap,v_{2}}^{\hat{v}_{1}};2).
\end{align}
Since action $a=2$ is optimal at the state
$s_{ap,v_{2}}^{\hat{v}_{1}}$, (\ref{eq:446}) and (\ref{eq:447}) give
\begin{align}
    V_{\alpha}(s_{ap,v_{2}}^{\hat{v}_{1}-1};0)< V_{\alpha}(s_{ap,v_{2}}^{\hat{v}_{1}-1};2) \leq V_{\alpha}(s_{ap,v_{2}}^{\hat{v}_{1}-1};1),
\end{align}
i.e., action $a=0$ is strictly optimal for state $s_{ap,v_{2}}^{\hat{v}_{1}-1}$.

Now, let $v_{1}+1<\hat{v}_{1}$. If action $a=0$ is optimal at the state $s_{ap,v_{2}}^{v_{1}+1}$, then it is no worse than action $a=1$ at that state. It follows from (\ref{eq:31}), (\ref{eq:32}) and Lemma~\ref{lemma:4} that action $a=0$ is also no
worse than action $a=1$ at the state $s_{ap,v_{2}}^{v_{1}}$. Now, from the statement of this lemma, action $a=2$ is not optimal at $s_{ap,v_{2}}^{v_{1}}$. Therefore, action $a=0$ is optimal at $s_{ap,v_{2}}^{v_{1}}$.
\end{Proof}

\begin{lemma}\label{lemma:npt_boundary_increment}
    For $\frac{\lambda}{\lambda+q}<\alpha<1$, the following relation holds,
    \begin{align}\label{eq:449}
        \tilde{\phi}_\alpha(\hat v_1 - 1,v_2) \leq \tilde{\phi}_\alpha(\hat v_1,v_2) < \bar{\phi}_\alpha(v_2), \quad 1 \leq  v_2 < \hat v_1 - 1.
    \end{align}
\end{lemma}

\begin{Proof}
For any fixed monitor age $v_{1}$, we represent a choice of optimal actions at $s_{ap,v_{2}}^{v_{1}}$ and $s_{ap,v_{2}+1}^{v_{1}}$ by the ordered pair $(a_{1},a_{2})$, where $a_{1}$ is an optimal action at $s_{ap,v_{2}}^{v_{1}}$ and $a_{2}$ is an optimal action at $s_{ap,v_{2}+1}^{v_{1}}$. For the age of the monitor $\hat{v}_{1}$, the collection
\begin{align}\label{eq:450}
    (1,1),\quad (1,2),\quad (2,2)
\end{align}
is a complete collection of optimal action pairs. We call a collection of optimal action pairs complete if, regardless of any ties between the actions at the two states, the collection contains at least one pair whose first action is optimal at the first state and whose second action is optimal at the second state. Thus, at least one pair from (\ref{eq:450}) can always be selected as an optimal action pair for the two states $s_{ap,v_{2}}^{\hat{v}_{1}}$ and $s_{ap,v_{2}+1}^{\hat{v}_{1}}$. Note that, a complete collection may not contain every possible optimal action pair. For example, if actions $a=1$ and $a=2$ are both optimal at both states, then $(2,1)$ is also an optimal action pair. Nevertheless, $(1,1)$ is an optimal action pair and is already included in the collection.  From Lemma~\ref{lemma:7}, we have $\tilde{\phi}_{\alpha}(\hat{v}_{1},v_{2}) < \bar{\phi}_{\alpha}(v_{2})$. From the definition of $\hat{v}_{1}$, we have $\hat{\phi}_{\alpha}(\hat{v}_{1}-1)<0$. Thus, Lemmas~\ref{lemma:8} and~\ref{lemma:9} show that, for the states $s_{ap,v_{2}}^{\hat{v}_{1}-1}$ and $s_{ap,v_{2}+1}^{\hat{v}_{1}-1}$, the collection 
\begin{align}\label{eq:451}
    (1,0),\quad (1,1),\quad (0,0)
\end{align}
is a complete collection of optimal action pairs, in the same sense defined above.

We now consider the combinations obtained by selecting one optimal action pair from the collection in (\ref{eq:450}) and one optimal action pair from the collection in (\ref{eq:451}). In each of the following combinations, the first pair corresponds to the age of the monitor $\hat{v}_{1}$, while the second pair corresponds to the age of the monitor $\hat{v}_{1}-1$. Part~1 of Lemma~\ref{lemma:10} shows that it is sufficient to consider the following six combinations:
\begin{align}
    &\text{(i)}\ (2,2)\ \text{and}\ (0,0),\qquad
    \text{(ii)}\ (1,2)\ \text{and}\ (0,0),\qquad
    \text{(iii)}\ (1,2)\ \text{and}\ (1,0),\nonumber\\
    &\text{(iv)}\ (1,1)\ \text{and}\ (0,0),\qquad
    \text{(v)}\ (1,1)\ \text{and}\ (1,0),\qquad
    \text{(vi)}\ (1,1)\ \text{and}\ (1,1).
\end{align}
Although ties between actions may permit additional combinations, at
least one of these six combinations can always be obtained by
selecting an optimal action at each of the aforementioned four states.
Thus, it is sufficient to verify the inequality in (\ref{eq:449}) for
these six combinations.

\emph{Case (i): $(2,2)$ and $(0,0)$}. From Remark~\ref{rem:4}, action $a=2$ is also optimal for state $s_{ap,v_{2}+2}^{\hat{v}_{1}}$. Thus, from (\ref{eq:33}) and the definition of $\tilde{\phi}_{\alpha}(v_{1},v_{2})$ in (\ref{eq:npt_L_definition}), we have
\begin{align}
    \tilde{\phi}_{\alpha}(\hat{v}_{1},v_{2}) = \tilde{\phi}_{\alpha}(\hat{v}_{1},v_{2}+1) = 0. 
\end{align}
Similarly, from (\ref{eq:31}) we have
\begin{align}\label{eq:454}
    \tilde{\phi}_{\alpha}(\hat{v}_{1}-1,v_{2}) = \alpha \tilde{\phi}_{\alpha}(\hat{v}_{1},v_{2}+1). 
\end{align}
Thus, we have
\begin{align}
    \tilde{\phi}_{\alpha}(\hat{v}_{1}-1,v_{2}) =  \tilde{\phi}_{\alpha}(\hat{v}_{1},v_{2}).
\end{align}

\emph{Case (ii): $(1,2)$ and $(0,0)$}. Again, Remark~\ref{rem:4} shows that action $a=2$ is optimal at the state $s_{ap,v_{2}+2}^{\hat{v}_{1}}$. From (\ref{eq:31}) and (\ref{eq:33}), we have
\begin{align}
   \tilde{\phi}_\alpha(\hat v_1-1,v_2) = \alpha \tilde{\phi}_\alpha(\hat v_1,v_2+1) =0 \leq \tilde{\phi}_\alpha(\hat v_1,v_2),
\end{align}
where the last inequality follows from Lemma~\ref{lemma:npt_discrete_concavity}. 

\emph{Case (iii): $(1,2)$ and $(1,0)$}.
\begin{align}
    \tilde{\phi}_{\alpha}(\hat{v}_{1}-1,v_{2}) & = V_{\alpha}(s_{ap,v_{2}+1}^{\hat{v}_{1}-1};0) - V_{\alpha}(s_{ap,v_{2}}^{\hat{v}_{1}-1};1) \\ 
    & \leq  V_{\alpha}(s_{ap,v_{2}+1}^{\hat{v}_{1}-1};2) - V_{\alpha}(s_{ap,v_{2}}^{\hat{v}_{1}-1};1).\label{eq:458}
\end{align}
From (\ref{eq:32}), (\ref{eq:33}), and (\ref{eq:affine_above_hat_v1}) in Remark~\ref{rem:4}, we have 
\begin{align}
    V_\alpha(s_{ap,v_2+1}^{\hat v_1};2) - V_\alpha(s_{ap,v_2+1}^{\hat v_1-1};2) = \frac{1}{1-\alpha(1-q)},\quad V_\alpha(s_{ap,v_2}^{\hat v_1};1) - V_\alpha(s_{ap,v_2}^{\hat v_1-1};1) = \frac{1}{1-\alpha(1-q)}. \label{eq:460}
\end{align}
Thus, we have
\begin{align}
    V_{\alpha}(s_{ap,v_{2}+1}^{\hat{v}_{1}};2)  -  V_{\alpha}(s_{ap,v_{2}}^{\hat{v}_{1}};1) = V_{\alpha}(s_{ap,v_{2} + 1}^{\hat{v}_{1} - 1};2)  -  V_{\alpha}(s_{ap,v_{2}}^{\hat{v}_{1} - 1};1).
\end{align}
Thus, from (\ref{eq:458}), we have
\begin{align}
    \tilde{\phi}_{\alpha}(\hat{v}_{1}-1,v_{2}) \leq \tilde{\phi}_{\alpha}(\hat{v}_{1},v_{2}).
\end{align}

\emph{Case (iv): $(1,1)$ and $(0,0)$}. From (\ref{eq:454}) and Lemma~\ref{lemma:7} we have
\begin{align}\label{eq:463}
    \tilde{\phi}_{\alpha}(\hat{v}_{1},v_{2}+1) = \tilde{\phi}_{\alpha}(\hat{v}_{1}+1,v_{2}+1). 
\end{align}
Now,
\begin{align}\label{eq:464}
    \tilde{\phi}_{\alpha}(\hat{v}_{1},v_{2}) = \alpha (q \bar{\phi}_{\alpha}(v_{2}+1)+ (1-q) \tilde{\phi}_{\alpha} (\hat{v}_{1}+1,v_{2}+1)).
\end{align}
Now, from Lemma~\ref{lemma:7}, we get
\begin{align}
    \tilde{\phi}_{\alpha}(\hat{v}_{1},v_{2}) \geq \alpha \tilde{\phi}_{\alpha}(\hat{v}_{1}+1,v_{2}+1) = \tilde{\phi}_{\alpha}(\hat{v}_{1}-1,v_{2}),
\end{align}
where the last equality follows from (\ref{eq:463}) and (\ref{eq:454}).

\emph{Case (v): $(1,1)$ and $(1,0)$}.
\begin{align}
    \tilde{\phi}_{\alpha}(\hat{v}_{1}-1,v_{2}) & = V_{\alpha}(s_{ap,v_{2}+1}^{\hat{v}_{1}-1};0) - V_{\alpha}(s_{ap,v_{2}}^{\hat{v}_{1}-1};1) \\ 
    & \leq V_{\alpha}(s_{ap,v_{2}+1}^{\hat{v}_{1}-1};1) - V_{\alpha}(s_{ap,v_{2}}^{\hat{v}_{1}-1};1) \\ 
    & = \alpha(q\bar{\phi}_{\alpha}(v_{2}+1) + (1-q)\tilde{\phi}_{\alpha}(\hat{v}_{1},v_{2}+1)).\label{eq:468}
\end{align}
Now, from Lemma~\ref{lemma:7} and (\ref{eq:464}), we get
\begin{align}
    \tilde{\phi}_{\alpha}(\hat{v}_{1}-1,v_{2}) \leq \tilde{\phi}_{\alpha}(\hat{v}_{1},v_{2}).
\end{align}

\emph{Case (vi): $(1,1)$ and $(1,1)$}.
\begin{align}\label{eq:n484}
    \tilde{\phi}_{\alpha}(\hat{v}_{1}-1,v_{2}) =\alpha (q\bar{\phi}_{\alpha}(v_{2}+1)+(1-q) \tilde{\phi}_{\alpha}(\hat{v}_{1},v_{2}+1)).
\end{align}
From Lemma~\ref{lemma:7}, (\ref{eq:464}), and  (\ref{eq:n484}) we have
\begin{align}
    \tilde{\phi}_{\alpha}(\hat{v}_{1}-1,v_{2}) \leq \tilde{\phi}_{\alpha}(\hat{v}_{1},v_{2}).
\end{align}
Thus, for all six cases we have verified the first inequality of (\ref{eq:449}), and the final inequality follows directly from Lemma~\ref{lemma:7}.
\end{Proof}

\begin{lemma}\label{lemma:12}
    For $v_{1}\geq 2$, if
    \begin{align}\label{eq:471}
        \tilde{\phi}_{\alpha}(v_{1},v_{2}) \leq \tilde{\phi}_{\alpha}(v_{1}+1,v_{2}), \quad 1\leq v_{2}<v_{1},  
    \end{align}
    then
    \begin{align}\label{eq:472}
        \hat{\phi}_\alpha(v_1-1) < \hat{\phi}_\alpha(v_1).
    \end{align}
    Consequently, $\hat\phi_\alpha(v_1)<0$ implies $\hat\phi_\alpha(v_1-1)<0$.
\end{lemma}

\begin{Proof}
Lemma~\ref{lemma:1} and the definition of $\tilde{\phi}_\alpha$ give
\begin{align}
    \bar{\phi}_\alpha(v_{1}) &= V_\alpha(s_{np}^{v_1+1}) - V_\alpha(s_{np}^{v_1})\\
    &= V_\alpha(s_{ap,v_1+1}^{v_1+1}) - V_\alpha(s_{ap,v_1}^{v_1})\\
    &=( V_\alpha(s_{ap,1}^{v_1+1}) - V_\alpha(s_{ap,1}^{v_1}))+\tilde{\phi}_\alpha(v_1+1,v_1) +\sum_{v_2=1}^{v_1-1}( \tilde{\phi}_\alpha(v_1+1,v_2)-\tilde{\phi}_\alpha(v_1,v_2)). \label{eq:npt_telescope}
\end{align}
All terms after the first bracket are nonnegative by (\ref{eq:471}) and Lemma~\ref{lemma:4}. Therefore,
\begin{align}\label{eq:npt_D_ge_monitor_increment}
    \bar{\phi}_\alpha(v_1) \geq V_\alpha(s_{ap,1}^{v_1+1}) - V_\alpha(s_{ap,1}^{v_1}).
\end{align}
Expanding $\hat{\phi}_\alpha(v_1-1)$ and $\hat{\phi}_\alpha(v_1)$ using (\ref{eq:34}) and~(\ref{eq:35}) gives
\begin{align}
    \hat{\phi}_\alpha(v_1)-\hat{\phi}_\alpha(v_1-1) &= \alpha\Big( \bar{\phi}_\alpha(v_1) -
    (1-q) \big(V_\alpha(s_{ap,1}^{v_1+1}) - V_\alpha(s_{ap,1}^{v_1}) \big) \Big) \\ 
    & \geq \alpha q (V_{\alpha}(s_{ap,1}^{v_{1}+1})- V_{\alpha}(s_{ap,1}^{v_{1}}))\label{eq:476}\\
    & >0,\label{eq:477}
\end{align}
where (\ref{eq:476}) follows from (\ref{eq:npt_D_ge_monitor_increment}) and (\ref{eq:477}) follows from the fact that $q>0$ and from Lemma~\ref{lemma:4}.
\end{Proof}

\begin{lemma}\label{lemma:13}
    For $\frac{\lambda}{\lambda+q}<\alpha<1$, if $\hat{v}_{1}-2\geq2$, then for $\hat{v}_{1}-2\geq v_{1}\geq 2$ and $1\leq v_{2} <v_1$, the following relation holds,
    \begin{align}\label{eq:478}
        \tilde{\phi}_\alpha(v_{1},v_2) \leq \tilde{\phi}_\alpha(v_{1}+1,v_2) < \bar{\phi}_\alpha(v_2).
    \end{align}
\end{lemma}

\begin{Proof}
First, we consider the boundary case $v_{1}=\hat{v}_{1}-2$. From Lemma~\ref{lemma:npt_boundary_increment}, Lemma~\ref{lemma:12}, and the fact that $\hat{\phi}_{\alpha}(\hat{v}_{1}-1)<0$, we have $\hat{\phi}_{\alpha}(\hat{v}_{1}-2)<0$. Thus, from Lemma~\ref{lemma:8} and Lemma~\ref{lemma:9}, for the age of the monitor $\hat{v}_{1}-2$, a complete collection of optimal action pairs is the same as the complete collection of action pairs in (\ref{eq:451}), corresponding to the age of the monitor $\hat{v}_{1}-1$. Now, part~2 of Lemma~\ref{lemma:10} shows that it is sufficient to verify the first inequality in (\ref{eq:478}) for the following six combinations:
\begin{align}\label{eq:479}
    &\text{(i)}\   (0,0)\ \text{and}\ (0,0), \qquad 
    \text{(ii)}\   (1,0)\ \text{and}\ (0,0), \qquad 
    \text{(iii)}\ (1,0)\ \text{and}\ (1,0), \nonumber\\ 
    &\text{(iv)}\ (1,1)\ \text{and}\ (0,0),\qquad
    \text{(v)}\ (1,1)\ \text{and}\ (1,0),\qquad
    \text{(vi)}\ (1,1)\ \text{and}\ (1,1).
\end{align}
The first and second action pairs in each combination in
(\ref{eq:479}) correspond to the monitor ages $\hat{v}_{1}-1$ and
$\hat{v}_{1}-2$, respectively.

\emph{Case (i): $(0,0)$ and $(0,0)$}.
\begin{align}
    \tilde{\phi}_{\alpha}(\hat{v}_{1}-2,v_{2}) &=\alpha\tilde{\phi}_{\alpha}(\hat{v}_{1}-1,v_{2}+1)\\
    & \leq\alpha\tilde{\phi}_{\alpha}(\hat{v}_{1},v_{2}+1)\\
    & =\tilde{\phi}_{\alpha}(\hat{v}_{1}-1,v_{2}).
\end{align}
The first and last equalities follow from the definition of $\tilde{\phi}_{\alpha}(v_{1},v_{2})$ in (\ref{eq:31}) and (\ref{eq:npt_L_definition}), while the inequality follows from Lemma~\ref{lemma:npt_boundary_increment}.

\emph{Case (ii): $(1,0)$ and $(0,0)$}.
For this case, we have
\begin{align}
    \tilde{\phi}_{\alpha}(\hat{v}_{1}-1,v_{2}) &=  V_{\alpha}(s_{ap,v_{2}+1}^{\hat{v}_{1}-1};0) - V_{\alpha}(s_{ap,v_{2}}^{\hat{v}_{1}-1};1), \\ 
    &\geq V_{\alpha}(s_{ap,v_{2}+1}^{\hat{v}_{1}-1};0) - V_{\alpha}(s_{ap,v_{2}}^{\hat{v}_{1}-1};0).  
\end{align}
Now, from (\ref{eq:31}) and Lemma~\ref{lemma:npt_boundary_increment}, we have
\begin{align}
    \tilde{\phi}_{\alpha}(\hat{v}_{1}-1,v_{2}) \geq \tilde{\phi}_{\alpha}(\hat{v}_{1}-2,v_{2}).
\end{align}

\emph{Case (iii): $(1,0)$ and $(1,0)$}.
\begin{align}\label{eq:484}
    \tilde{\phi}_{\alpha}(\hat{v}_{1}-1,v_{2}) - \tilde{\phi}_{\alpha}(\hat{v}_{1}-2,v_{2}) = V_{\alpha}(s_{ap,v_{2}+1}^{\hat{v}_{1}-1};0) -V_{\alpha}(s_{ap,v_{2}}^{\hat{v}_{1}-1};1) - V_{\alpha}(s_{ap,v_{2}+1}^{\hat{v}_{1}-2};0) + V_{\alpha}(s_{ap,v_{2}}^{\hat{v}_{1}-2};1).
\end{align}
Expanding the right-hand side of (\ref{eq:484}) using (\ref{eq:31}) and (\ref{eq:32}), we get
\begin{align}
     \tilde{\phi}_{\alpha}(\hat{v}_{1} - 1,v_{2})  -  \tilde{\phi}_{\alpha}(\hat{v}_{1} - 2,v_{2})  &=  \alpha(V_{\alpha}(s_{ap,v_{2} + 2}^{\hat{v}_{1}}) - V_{\alpha}(s_{ap,v_{2} + 2}^{\hat{v}_{1} - 1})  - (1-q)( V_{\alpha}(s_{ap,v_{2}+1}^{\hat{v}_{1}}) - V_{\alpha}(s_{ap,v_{2}+1}^{\hat{v}_{1}-1}))), \\ 
     &\geq \alpha(1-q) (\tilde{\phi}_{\alpha}(\hat{v}_{1},v_{2}+1) - \tilde{\phi}_{\alpha}(\hat{v}_{1}-1,v_{2}+1)),\label{eq:486}
\end{align}
where (\ref{eq:486}) follows from Lemma~\ref{lemma:4} and the definition of $\tilde{\phi}_{\alpha}$. Thus, from Lemma~\ref{lemma:npt_boundary_increment} and (\ref{eq:486}) we have
\begin{align}
    \tilde{\phi}_{\alpha}(\hat{v}_{1}-1,v_{2}) \geq \tilde{\phi}_{\alpha}(\hat{v}_{1}-2,v_{2}).
\end{align}

\emph{Case (iv): $(1,1)$ and $(0,0)$}.
\begin{align}
    \tilde{\phi}_{\alpha}(\hat{v}_{1}-1,v_{2}) - \tilde{\phi}_{\alpha}(\hat{v}_{1}-2,v_{2}) = \alpha(q \bar{\phi}_{\alpha}(v_{2}+1)+(1-q) \tilde{\phi}_{\alpha}(\hat{v}_{1},v_{2}+1)  - \tilde{\phi}_{\alpha}(\hat{v}_{1}-1,v_{2}+1)).
\end{align}
Now, from Lemma~\ref{lemma:npt_boundary_increment}, we have 
\begin{align}
    \tilde{\phi}_{\alpha}(\hat{v}_{1}-1,v_{2}) - \tilde{\phi}_{\alpha}(\hat{v}_{1}-2,v_{2})\geq 0.
\end{align}

\emph{Case (v): $(1,1)$ and $(1,0)$}.
\begin{align}
    \tilde{\phi}_{\alpha}(\hat{v}_{1}-2,v_{2}) &\leq V_{\alpha}(s_{ap,v_{2}+1}^{\hat{v}_{1}-2};1) - V_{\alpha}(s_{ap,v_{2}}^{\hat{v}_{1}-2};1) \label{eq:490} \\ 
    &= \alpha(q \bar{\phi}_{\alpha}(v_{2}+1)  +  (1 - q) \tilde{\phi}_{\alpha}(\hat{v}_{1} - 1,v_{2} + 1)) \\ &\leq\alpha(q \bar{\phi}_{\alpha}(v_{2} + 1)  +  (1 - q) \tilde{\phi}_{\alpha}(\hat{v}_{1},v_{2} + 1)) \label{eq:492} \\ 
    &= \tilde{\phi}_{\alpha}(\hat{v}_{1}-1,v_{2}), \label{eq:493}
\end{align}
where (\ref{eq:492}) follows from Lemma~\ref{lemma:npt_boundary_increment}.

\emph{Case (vi): $(1,1)$ and $(1,1)$}. For this case, the first inequality directly follows from (\ref{eq:490})-(\ref{eq:493}).

For all six cases, Lemma~\ref{lemma:npt_boundary_increment} gives the second inequality in (\ref{eq:478}). Thus, for all six possible combinations in (\ref{eq:479}), we have shown that
\begin{align}
    \tilde{\phi}_{\alpha}(\hat{v}_{1}-2,v_{2}) \leq \tilde{\phi}_{\alpha}(\hat{v}_{1}-1,v_{2})<\bar{\phi}_{\alpha}(v_{2}),
\end{align}
for $1 \leq v_{2}<\hat{v}_{1}-2$. Thus, from Lemma~\ref{lemma:12}, we have $\hat{\phi}_{\alpha}(\hat{v}_{1}-3)<0$. Now, if $\hat{v}_{1}-3\geq 2$, then following the same arguments as for the boundary case $v_{1}=\hat{v}_{1}-2$, we get  
\begin{align}
    \tilde{\phi}_{\alpha}(\hat{v}_{1}-3,v_{2}) \leq \tilde{\phi}_{\alpha}(\hat{v}_{1}-2,v_{2}) <\bar{\phi}_{\alpha}(v_{2}),
\end{align}
for $1\leq v_{2} <\hat{v}_{1}-3$. Thus, by recursively applying the same argument, we obtain (\ref{eq:478}) for all $\hat{v}_{1}-2 \geq v_{1} \geq 2$ and $1\leq v_{2}<v_{1}$.
\end{Proof}

\subsection{Proof of Theorem~\ref{th:no_packet_threshold}}
\begin{Proof}
Now, from the definition of $\hat{v}_{1}$, for $v_{1}\geq \hat{v}_{1}$ we have
\begin{align}
    V_{\alpha}(s_{np}^{v_{1}};2) \leq V_{\alpha}(s_{np}^{v_{1}};0).
\end{align}
From Remark~\ref{rem:3}, we have $\hat{v}_{1}<\infty$. When $\hat{v}_{1}\geq 4$, from the definition of $\hat{v}_{1}$, action $a=0$ is optimal for state $s_{np}^{\hat{v}_{1}-1}$. From Lemma~\ref{lemma:npt_boundary_increment} and Lemma~\ref{lemma:12}, we see that $\hat{\phi}_{\alpha}(\hat{v}_{1}-2)<0$, i.e., action $a=0$ is optimal for state $s_{np}^{\hat{v}_{1}-2}$. Now, from Lemma~\ref{lemma:13} and Lemma~\ref{lemma:12}, we get $\hat{\phi}_{\alpha}(\hat{v}_{1}-i)<0$ for all $3\leq i \leq \hat{v}_{1}-1$, i.e., action $a=0$ is optimal for state $s_{np}^{\hat{v}_{1}-i}$. Thus, when $\hat{v}_{1}\geq 4$, we set $\hat{v}_{1}=\bar{v}_{1}$ in the statement of Theorem~\ref{th:no_packet_threshold}, and the statement of the theorem follows. If $\hat{v}_{1}=3$, then, from the definition of $\hat{v}_{1}$, we see that action $a=0$ is optimal for state $s_{np}^{2}$, and from Lemma~\ref{lemma:npt_boundary_increment} and Lemma~\ref{lemma:12}, we see that action $a=0$ is also optimal for state $s_{np}^{1}$. Thus, Theorem~\ref{th:no_packet_threshold} holds in this case with $\bar{v}_{1}=3$. When $\hat{v}_{1}=2$, setting $\bar{v}_{1}=\hat{v}_{1}$ proves Theorem~\ref{th:no_packet_threshold}, since, by the definition of $\hat{v}_{1}$, action $a=0$ is optimal at $s_{np}^{1}$. When $\hat{v}_{1}=1$, it is immediate that, by setting $\bar{v}_{1}=\hat{v}_{1}$, the statement of Theorem~\ref{th:no_packet_threshold} follows.
\end{Proof}

\section{Proof of Corollary~\ref{corr:2}}\label{appen:corr:2}
\begin{Proof}
Note that there exists $0<\tilde{\alpha}<1$ such that $\lambda\leq\frac{\alpha q}{1-\alpha}$ for every $\tilde{\alpha}\leq\alpha<1$. Consider a sequence $\{\alpha_n\}_{n=1}^{\infty}$ satisfying $\max\{\bar{\alpha}(\mu,\lambda),\tilde{\alpha}\} <\alpha_n<1$ for every $n$, and $\lim_{n\rightarrow\infty}\alpha_n=1$. Now, for the discounted-cost problem in (\ref{eq:23}) corresponding to a fixed $\alpha_{n}$, Theorem~\ref{th:no_packet_threshold} guarantees the existence of $\bar{v}_{1,n}<\infty$ such that, if $v_{1}\geq\bar{v}_{1,n}$, action $a=2$ is optimal for state $s_{np}^{v_{1}}$, whereas action $a=0$ is optimal otherwise. Now, from Remark~\ref{rem:2}, we know that if the transmitter has a packet and the age of the monitor is greater than or equal to $\bar{v}_{1,n}$, then irrespective of the age of the packet, either action $a=1$ or $a=2$ is optimal. Thus, we set $W_{n}=\bar{v}_{1,n}$. Now, from Theorem~\ref{th:5}, we know that there exists $\bar{v}_{2,n}$ such that whenever age of the current packet at the transmitter is less than $\bar{v}_{2,n}$, action $a=1$ is optimal and whenever it becomes $\bar{v}_{2,n}$, action $a=2$ is optimal and the sampler samples a new packet and transmitter discards the old packet and transmits this newly sampled packet and if the transmission fails then at the next slot the age of this packet becomes $1$. We choose $R_{n}=\bar{v}_{2,n}$. Thus, the policy $(W_{n},R_{n})=(\bar{v}_{1,n},\bar{v}_{2,n})$ is optimal for the discount criterion when the discount factor is $\alpha_{n}$. Similar to the proof for Theorem~\ref{th:2} and Theorem~\ref{th:3}, we represent policy $\pi_{n}$ by its infinite action vector $\bm{\pi}_{n}$.

We have already verified the conditions required by the vanishing-discount result in \cite{sennott1989average} in the proof of Theorem~\ref{th:2}. Thus, as in the proof of  Theorem~\ref{th:2}, there exists a subsequence  $\{\bm{\pi}_{n_{i}}\}_{i=1}^{\infty}$ that converges pointwise to a  deterministic stationary policy $\bm{\pi}_{L}^{*}$ that is optimal  for the Lagrangian-relaxed MDP in (\ref{eq:20}). Since the action space is finite, pointwise convergence implies that,  for every fixed state $s$, we have $\pi_{n_i}(s)=\pi_{L}^{*}(s)$, for all sufficiently large $i$. We first show that the waiting-threshold  structure is preserved in the limiting policy ${\pi}_{L}^{*}$. Suppose that $\pi_{L}^{*}(s_{np}^{v_{1}})=2$  for some $v_{1}$. Then, for all sufficiently large $i$, we have $\pi_{n_i}(s_{np}^{v_{1}})=2$.  Since $\pi_{n_i}$ has waiting threshold $W_{n_i}$, this implies that  $W_{n_i}\leq v_{1}$ for all sufficiently large $i$. Hence, for every  $x\geq0$, we have $\pi_{n_i}(s_{np}^{v_{1}+x})=2$, for all sufficiently large $i$. By pointwise convergence, we get $\pi_{L}^{*}(s_{np}^{v_{1}+x})=2$, for all $x\geq 0$. Therefore, there exists  $W\in\mathbb{N}\cup\{\infty\}$, such that
\begin{align}
    \pi_{L}^{*}(s_{np}^{v_{1}}) =
    \begin{cases}
        0, & v_{1}<W,\\
        2, & v_{1}\geq W.
    \end{cases}
\end{align}
The case $W=\infty$ cannot occur, since starting from  $s_{np}^{1}$, the system remains idle forever and incurs  infinite average Lagrangian cost, whereas the policy that always employs action $a=2$ incurs a  finite average cost. Hence, $W\in\mathbb{N}$. 

We next show that the packet-age-threshold structure is also preserved for $\pi_{L}^{*}$. Fix $v_{2}\geq1$, and consider any $v_{1},v_{1}'\geq\max\{W,v_{2}\}$. As, $\pi_{L}^{*}(s_{np}^{\min\{v_{1},v_{1}'\}})=2$, pointwise convergence implies that $W_{n_i}\leq \min\{v_{1},v_{1}'\}$ for all sufficiently large $i$. As $\pi_{n_i}$ is a
$(W_{n_i},R_{n_i})$ policy we have, 
\begin{align}
    \pi_{n_i}(s_{ap,v_{2}}^{v_{1}}) = \pi_{n_i}(s_{ap,v_{2}}^{v_{1}'}),
\end{align}
for all sufficiently large $i$. Taking the pointwise limit gives
\begin{align}
    \pi_{L}^{*}(s_{ap,v_{2}}^{v_{1}}) = \pi_{L}^{*}(s_{ap,v_{2}}^{v_{1}'}).
\end{align}
Thus, whenever the age of the monitor is at least $W$, the action selected by $\pi_{L}^{*}$ at a packet-available state is independent of the age of the monitor and depends only on the packet age $v_{2}$. Fix any $v_{2}\geq1$ and $v_{1}\geq\max\{W,v_{2}\}$. Since $\pi_{L}^{*}(s_{np}^{v_{1}})=2$, pointwise convergence implies that, for all sufficiently large $i$,
\begin{align}
    \pi_{n_i}(s_{np}^{v_{1}})=2,\quad W_{n_i}\leq v_{1}.
\end{align}
Since $\pi_{n_i}$ is a $(W_{n_i},R_{n_i})$ policy, we have
\begin{align}
    \pi_{n_i}(s_{ap,v_{2}}^{v_{1}})\in\{1,2\}
\end{align}
for all sufficiently large $i$. Taking the pointwise limit gives
\begin{align}
    \pi_{L}^{*}(s_{ap,v_{2}}^{v_{1}})\in\{1,2\}.
\end{align}

Now, suppose that $\pi_{L}^{*}$ selects action $a=2$ for  some packet age $v_{2}$. Fix  $v_{1}\geq\max\{W,v_{2}\}$. By pointwise convergence,  for all sufficiently large $i$, we have  
\begin{align}
    \pi_{n_i}(s_{ap,v_{2}}^{v_{1}})=2.
\end{align}
As $W_{n_i}\leq v_{1}$ for all sufficiently large $i$, the packet-age-threshold structure of $\pi_{n_i}$ implies
that $R_{n_i}\leq v_{2}$ for all sufficiently large $i$. Consequently, for every integer $x\geq0$ satisfying
$v_2+x\leq v_1$, we have
\begin{align}
    \pi_{n_i}(s_{ap,v_2+x}^{v_1})=2
\end{align}
for all sufficiently large $i$. Taking the pointwise limit gives
\begin{align}
    \pi_L^{*}(s_{ap,v_2+x}^{v_1})=2.
\end{align}

If action $a=2$ is selected for at least one packet age,  let $R$ be the smallest such packet age. Otherwise, let  $R=\infty$. Since $\pi_L^{*}$ selects only actions $a=1$ and $a=2$  at packet-available states with $v_1\geq W$, there exists  $R\in\mathbb{N}\cup\{\infty\}$, such that, for  $v_1\geq W$ and $1\leq v_2\leq v_1$,
\begin{align}
    \pi_{L}^{*}(s_{ap,v_{2}}^{v_{1}})= 
    \begin{cases}
        1, & v_{2}<R,\\
        2, & v_{2}\geq R.
    \end{cases}
\end{align}
The case $R=\infty$ corresponds to retransmitting the same  packet until it is successfully delivered and is permitted  by the definition of $\bar{\Pi}$. Packet-available states  with monitor age below $W$ are unreachable from the initial  state $s_{np}^{1}$, and the definition of $\bar{\Pi}$ permits  arbitrary admissible actions at these states. Therefore,  $\pi_{L}^{*}$ corresponds to a policy $(W,R)\in\bar{\Pi}$  and is optimal for the Lagrangian-relaxed MDP in  (\ref{eq:20}).
\end{Proof}

\section{Proof of Theorem~\ref{th:7}}\label{appen:th:7}
\begin{Proof}
We consider an equidistant-sampling policy $(R,R)$. Since $V_R\in\{1,\cdots,R\}$, from (\ref{eq:n147}) we have
\begin{align}
    H(R,R)=R-V_R. \label{eq:548}
\end{align}
Now, from (\ref{eq:132}), we have
\begin{align}
    \mathbb{E}[V_R]=\frac{1}{q}-\frac{R(1-q)^R}{1-(1-q)^R},\label{eq:549}
\end{align}
and
\begin{align}
    \mathbb{E}[V_R(V_R-1)]&=\sum_{i=1}^{R}i(i-1) \frac{q(1-q)^{i-1}} {1-(1-q)^R}\\
    &=\frac{2(1-q)}{q^2}-\frac{R(1-q)^R}{1-(1-q)^R}\left(R-1+\frac{2}{q}\right). \label{eq:550}
\end{align}
From (\ref{eq:144}) and (\ref{eq:548}), we get
\begin{align}
    \mathbb{E}[L(R,R)]=\frac{R}{1-(1-q)^R}.\label{eq:551}
\end{align}
Now, from (\ref{eq:n157}), (\ref{eq:548}), (\ref{eq:549}), (\ref{eq:550}) and (\ref{eq:551}), we have
\begin{align}
    \mathbb{E}[Y(R,R)]=\frac{R}{1-(1-q)^R}\left(\frac{1}{q}+\frac{R-1}{2}\right). \label{eq:553}
\end{align}
It now follows from (\ref{eq:n158}), (\ref{eq:551}) and (\ref{eq:553}) that
\begin{align}
    \Delta(R,R) = \frac{1}{q}+\frac{R-1}{2}. \label{eq:554}
\end{align}
The policy $\pi_s^{*}$ is optimal for the problem with only the sampling-rate constraint. Therefore, using (\ref{eq:554}), 
\begin{align}
    \Delta(\pi_s^{*})&=\theta_s\Delta(R_1,R_1)+(1-\theta_s)\Delta(R_2,R_2)\\
    &=\frac{1}{q}+\frac{\theta_sR_1+(1-\theta_s)R_2-1}{2}.\label{eq:555}
\end{align}
As $\Delta^{*}\geq \Delta(\pi_{s}^{*})$, we have (\ref{eq:546}). 

We next prove (\ref{eq:547}). Note that $V_{1}=1$. Now, again from (\ref{eq:n147}), (\ref{eq:144}) and (\ref{eq:n157}), we have 
\begin{align}\label{eq:556}
    H(W,1)=W-1, \quad L(W,1)=W-1+G, \quad Y(W,1)=\frac{L(W,1)(L(W,1)+1)}{2}.
\end{align}
Thus, from (\ref{eq:n158}) and (\ref{eq:141}), we have
\begin{align}
    \Delta(W,1) =\frac{1}{2}+\frac{1}{2q\nu_{r}(W,1)}+\frac{(1-q)\nu_{r}(W,1)}{2q}.
\end{align}
We denote $\nu_{r}(W,1)$ with $\bar{r}$ and $\Delta(W,1)$ with $\bar{g}(\bar{r})$. Thus, 
\begin{align}\label{eq:154}
    \bar{g}(\bar{r}) = \frac{1}{2} + \frac{1}{2q\bar{r}} + \frac{(1-q) \bar{r}}{2q},
\end{align}
is a convex function of $\bar{r}$. Now, from the definition of the policy $\pi_{t}^{*}$, we have
\begin{align}
    \Delta(\pi_{t}^{*}) &= \theta_{r} \bar{g}(\bar{r}_{1}) + (1-\theta_{r}) \bar{g}(\bar{r}_{2}) \\ 
    &\geq  \bar{g}(\theta_{r} \bar{r}_{1}+ (1-\theta_{r})\bar{r}_{2}),
\end{align}
where $\bar{r}_{1} = \nu_{r}(W_{1},1)$ and $\bar{r}_{2} = \nu_{r}(W_{2},1)$. Now, from (\ref{eq:142}), we have 
\begin{align}\label{eq:157}
    \theta_{r} \bar{r}_{1} + (1-\theta_{r}) \bar{r}_{2} = f_{r}.
\end{align}
Thus, from (\ref{eq:154}) and (\ref{eq:157}), 
\begin{align}
    \Delta(\pi_{t}^{*}) &\geq \bar{g}(f_{r}) = \frac{1}{2} + \frac{1}{2q f_{r}} + \frac{(1-q) f_{r}}{2q}.
\end{align}
As $\Delta^{*} \geq \Delta(\pi_{t}^{*})$, we have (\ref{eq:547}).
\end{Proof}

\section{Proof of Theorem~\ref{th:6}}\label{appen:th:6}
\begin{Proof}
For Lagrangian multipliers $\mu,\lambda\geq 0$, the Lagrangian of (\ref{eq:opt_prob}) is, \begin{align}\label{eq:164}
    \mathcal{L}_{\mu,\lambda}(\pi) = \limsup_{T\rightarrow\infty}\Delta(\pi;T) + \mu (\limsup_{T\rightarrow\infty}\nu_{s}(\pi;T)-f_{s}) + \lambda (\limsup_{T\rightarrow\infty}\nu_{r}(\pi;T)-f_{r}).
\end{align}
However, for the Lagrangian-relaxed MDP in (\ref{eq:20}), we use,
\begin{align}
    \tilde{\mathcal{L}}_{\mu,\lambda}(\pi) =  \limsup_{T\rightarrow\infty}(\Delta(\pi;T)  + \mu \nu_{s}(\pi;T) + \lambda \nu_{r}(\pi;T))-\mu f_{s} - \lambda f_{r}.
\end{align}
Note that for every policy $\pi\in\Pi$, we have,
\begin{align}\label{eq:166}
        \mathcal{L}_{\mu,\lambda}(\pi)\geq \tilde{\mathcal{L}}_{\mu,\lambda}(\pi).
\end{align}
From Corollary~\ref{corr:2}, we have that for a fixed $\mu,\lambda\geq0$,  under the no-packet threshold property, there exists an optimal policy $(W,R)\in\bar{\Pi}$, which we denote with $\pi_{\mu,\lambda}^{*}$. Thus, for all $\pi\in\Pi$, we have,
\begin{align}\label{eq:167}
    \tilde{\mathcal{L}}_{\mu,\lambda}(\pi)\geq \tilde{\mathcal{L}}_{\mu,\lambda}(\pi_{\mu,\lambda}^{*}).
\end{align}
Now, from (\ref{eq:149}), (\ref{eq:150}), and (\ref{eq:n158}), we observe that the limits of $\nu_r(\pi^{*}_{\mu,\lambda};T)$, $\nu_s(\pi^{*}_{\mu,\lambda};T)$, and $\Delta(\pi^{*}_{\mu,\lambda};T)$ exist and are given by $\nu_r(W,R)$, $\nu_s(W,R)$, and $\Delta(W,R)$, respectively. Thus, 
\begin{align}
    \tilde{\mathcal{L}}_{\mu,\lambda}(\pi^{*}_{\mu,\lambda}) = \Delta(W,R) + \mu (\nu_{s}(W,R) -f_{s})+ \lambda(\nu_{r}(W,R)-f_{r})= \mathcal{L}_{\mu,\lambda}(\pi_{\mu,\lambda}^{*}).\label{eq:169}
\end{align}
From (\ref{eq:166}), (\ref{eq:167}), and (\ref{eq:169}), we have,
\begin{align}\label{eq:170}
    \mathcal{L}_{\mu,\lambda}(\pi) \geq \mathcal{L}_{\mu,\lambda}(\pi^{*}_{\mu,\lambda}).
\end{align}
Note that (\ref{eq:170}) holds for all policies $\pi$. Thus,
\begin{align}
    \inf_{\pi\in{\Pi}} \mathcal{L}_{\mu,\lambda}(\pi) = \mathcal{L}_{\mu,\lambda}(\pi^{*}_{\mu,\lambda})= \min_{(W,R)\in\bar{\Pi}} \Delta(W,R) + \mu (\nu_{s}(W,R) -f_{s}) + \lambda(\nu_{r}(W,R)-f_{r}).\label{eq:172}
\end{align}
Now, consider a policy $(W,\infty)$, from (\ref{eq:149}) and (\ref{eq:150}), we have,
\begin{align}
    \nu_{s}(W,\infty)  =   \frac{1}{\mathbb{E}[L(W,\infty)]}, \qquad \nu_{r}(W,\infty)  =   \frac{1}{q\mathbb{E}[L(W,\infty)]}. 
\end{align}
Now, from (\ref{eq:n147}) and (\ref{eq:144}), we have,
\begin{align}
    L(W,\infty) = (W-V_{\infty})^{+} +G.
\end{align}
As $V_{\infty}$ is geometrically distributed, it is finite almost surely. Thus, as $W\rightarrow\infty$ $L(W,\infty)\rightarrow\infty$ almost surely. Moreover, $(W-V_{\infty})^{+}$ is increasing in $W$, thus from monotone convergence theorem, we have,
\begin{align}\label{eq:n175}
    \lim_{W\rightarrow\infty}\mathbb{E}[L(W,\infty)]=\infty.
\end{align} 
Thus, there exists a sufficiently large $W_{0}$, such that, we have,
\begin{align}
    \nu_{s}(W_{0},\infty) <f_{s}, \ \nu_{r}(W_{0},\infty)<f_{r}.
\end{align}
Thus, the policy $\pi_{0}=(W_{0},\infty)$ is feasible. Let us define,
\begin{align}\label{eq:176}
    \eta_{\min} = \min\{f_{s}-\nu_{s}(W_{0},\infty), f_{r}-\nu_{r}(W_{0},\infty)\}.
\end{align}
Note that the distribution defined by $x(W_{0},\infty)=1$ and $x(W,R)=0$ for all other $(W,R)\in\bar{\Pi}$ is a feasible solution to the LP in (\ref{eq:158}). Thus, we have
\begin{align}\label{eq:177}
    1\leq \Delta_{\textrm{LP}} \leq \Delta(W_{0},\infty)<\infty,
\end{align}
where the leftmost inequality follows from the fact that the age of the monitor is always at least $1$. The LP in (\ref{eq:158}) allows probability distributions $\bm{x}=\{x(W,R)\}$, with countably infinite support. We first show that its value can be approached arbitrarily closely by feasible distributions having finite support. This finite support result follows from the Richter-Rogosinski theorem~\cite[Theorem~7.32]{shapiro2009lectures}. For completeness, we prove it directly, based on truncating the distribution and restoring feasibility through mixing with the strictly feasible policy $(W_0,\infty)$.
   
Because $\bar{\Pi}$ is countable, we choose nested finite sets 
\begin{align}
    \bar{\Pi}_{1} \subseteq \bar{\Pi}_{2} \subseteq \bar{\Pi}_{3}\subseteq \cdots, \qquad
    \bigcup_{n=1}^{\infty} \bar{\Pi}_{n} = \bar{\Pi}. \label{eq:180}
\end{align}
We also consider that $(W_0,\infty)\in\bar{\Pi}_n$ for every $n\geq1$. Let us consider a feasible distribution $\bm{x}$ with a finite objective, i.e.,
\begin{align}
    \sum_{(W,R)\in\bar{\Pi}} x(W,R)\Delta(W,R)<\infty
\end{align}
We denote the tail probability with $\delta_{n}$ as,
\begin{align}
    \delta_{n} = \sum_{(W,R)\notin\bar{\Pi}_{n}} x(W,R).
\end{align}
Since $\bar{\Pi}_{n}$ increases to $\bar{\Pi}$, we have $\delta_{n}\rightarrow 0$. Now, let us define a modified distribution $\hat{\bm{x}}_{n} = \{\hat{x}_{n}(W,R)\}$,
\begin{align}
    \hat{{x}}_{n}(W,R) = 
    \begin{cases}
        x(W,R), & (W,R)\in\bar{\Pi}_{n} \  \textrm{and} \ (W,R) \neq (W_{0},\infty), \\ 
        x(W_{0},\infty) + \delta_{n}, & (W,R) =(W_{0},\infty), \\ 
        0,& (W,R) \notin\bar{\Pi}_{n}.
    \end{cases}
\end{align}
Note that, $\hat{\bm{x}}_{n}$ is a probability distribution supported on $\bar{\Pi}_{n}$. The average age under this distribution is
\begin{align}
    \sum_{(W,R)\in\bar{\Pi}}    \hat{x}_{n}(W,R) \Delta(W,R)  & =  \sum_{(W,R)\in\bar{\Pi}}    x(W,R) \Delta(W,R) +\delta_{n} \Delta(W_{0},\infty)  - \sum_{(W,R)\notin\bar{\Pi}_{n}} x(W,R) \Delta(W,R)\\
    &\leq \sum_{(W,R)\in\bar{\Pi}} x(W,R) \Delta(W,R) +\delta_{n} \Delta(W_{0},\infty).\label{eq:n183}
\end{align}
Note that (\ref{eq:n183}) holds as $\Delta(W,R)\geq 1$, for all $(W,R)\in\bar{\Pi}$. The sampling rate corresponding to the distribution $\hat{\bm{x}}_{n}$ satisfies,
\begin{align}\label{eq:182}
    &\sum_{(W,R)\in\bar{\Pi}}   \hat{x}_{n}(W,R) \nu_{s}(W,R)  =    \sum_{(W,R)\in\bar{\Pi}}    x(W,R) \nu_{s}(W,R) +  \delta_{n}\nu_{s}(W_{0},\infty) \  -        \sum_{(W,R)\notin\bar{\Pi}_{n}}      x(W,R) \nu_{s}(W,R)  \leq  f_{s}  +  \delta_{n},
\end{align}
where we have used the fact that $\nu_{s}(W,R)\geq 0$ and $\nu_{s}(W_{0},\infty)\leq f_{s} \leq 1$. Similarly, 
\begin{align}
    \sum_{(W,R)\in\bar{\Pi}} \hat{x}_{n}(W,R) \nu_{r}(W,R) \leq f_{r} + \delta_{n}.   
\end{align}
Thus, the distribution $\hat{\bm{x}}_{n}$ can violate either constraint by at most $\delta_{n}$. To restore feasibility, we define
\begin{align}
    \eta_{n} = \frac{\delta_{n}}{\eta_{\min}+\delta_{n}}.
\end{align}
Now, we consider the following distribution $\bar{\bm{x}}_{n}$,
\begin{align}
    \bar{x}_{n}(W,R)  =  
    \begin{cases}
        (1 - \eta_{n}) \hat{x}_{n}(W,R), & (W,R)\neq (W_{0},\infty)\\
        (1-\eta_{n}) \hat{x}_{n}(W,R)+ \eta_{n} & (W,R) = (W_{0},\infty).
    \end{cases} 
\end{align}
Now,
\begin{align}
    \sum_{(W,R)\in\bar{\Pi}} \bar{x}_{n}(W,R) \nu_{s}(W,R) -f_{s}  & =  (1 - \eta_{n}) \left(\sum_{(W,R)\in\bar{\Pi}} \hat{x}_{n}(W,R) \nu_{s}(W,R) - f_{s} \right)  - \eta_{n} \left(f_{s} - \nu_{s}(W_{0},\infty)\right)\\
    & \leq (1-\eta_{n}) \delta_{n} - \eta_{n} \eta_{\min} \label{eq:187} \\ 
    &=0,
\end{align}
where (\ref{eq:187}) follows from (\ref{eq:176}) and (\ref{eq:182}). Similarly, we get
\begin{align}
    \sum_{(W,R)\in\bar{\Pi}} \bar{x}_{n}(W,R) \nu_{r}(W,R) -f_{r} \leq 0.
\end{align}
Thus, the distribution $\bar{\bm{x}}_{n}$ is feasible for the LP in (\ref{eq:158}) and has finite support. Now,
\begin{align}
    \sum_{(W,R)\in\bar{\Pi}} \bar{x}_{n}(W,R) \Delta(W,R) & = (1 - \eta_{n}) \sum_{(W,R)\in\bar{\Pi}}      \hat{x}_{n}(W,R) \Delta(W,R) + \eta_{n}\Delta(W_{0},\infty)\\
    & \leq \sum_{(W,R)\in\bar{\Pi}} x(W,R) \Delta(W,R)+ (\delta_{n}+\eta_{n}) \Delta(W_{0},\infty). \label{eq:194}
\end{align}
Recall that $\mathcal{F}$ denotes the set of all feasible distributions $\bm{x}$. We also denote the set of all feasible distributions with finite support by $\mathcal{F}_{\textrm{fin}}$. As $\mathcal{F}_{\textrm{fin}}\subseteq\mathcal{F}$, we have,
\begin{align}
    \Delta_{\textrm{LP}} =&\inf_{\bm{x}\in\mathcal{F}} \sum_{(W,R)\in\bar{\Pi}} x(W,R) \Delta(W,R)\\ 
    \leq & \inf_{\bm{x}\in\mathcal{F}_{\textrm{fin}}} \sum_{(W,R)\in\bar{\Pi}} x(W,R) \Delta(W,R).\label{eq:196}
\end{align}
Now, for a fixed $\epsilon>0$, we can find a distribution $\bm{x}_{\epsilon}\in\mathcal{F}$, such that 
\begin{align}\label{eq:197}
    \sum_{(W,R)\in\bar{\Pi}} x_{\epsilon}(W,R) \Delta(W,R) \leq \Delta_{\textrm{LP}} + \frac{\epsilon}{2}.
\end{align}
From (\ref{eq:177}) and (\ref{eq:197}), we see that $\bm{x}_{\epsilon}$ provides finite objective. Now, as $\delta_{n}+\eta_{n} \rightarrow 0$, we can find a large enough $n_{1}$, such that for all $n\geq n_1$, we have 
\begin{align}
    (\delta_{n}+\eta_{n}) \Delta(W_{0},\infty) \leq \frac{\epsilon}{2}.
\end{align}
Now, from (\ref{eq:194}), there exists a finite support feasible distribution $\bar{x}_{n}$, such that 
\begin{align}\label{eq:199}
    \sum_{(W,R)\in\bar{\Pi}} \bar{x}_{n}(W,R) \Delta(W,R)\leq \sum_{(W,R)\in\bar{\Pi}} x_{\epsilon}(W,R) \Delta(W,R) + \frac{\epsilon}{2}.
\end{align}
Now, from (\ref{eq:197}) and (\ref{eq:199}), we have 
\begin{align}
    \sum_{(W,R)\in\bar{\Pi}}\bar{x}_{n}(W,R) \Delta(W,R)\leq \Delta_{\textrm{LP}} + \epsilon.
\end{align}
Now, as $\bar{\bm{x}}_{n}\in\mathcal{F}_{\textrm{fin}}$, we have 
\begin{align}\label{eq:201}
    \inf_{\bm{x}\in\mathcal{F}_{\textrm{fin}}} \sum_{(W,R)\in\bar{\Pi}} x(W,R) \Delta(W,R) \leq \Delta_{LP} + \epsilon.
\end{align}
Note that (\ref{eq:201}) holds for every $\epsilon>0$, thus taking $\epsilon\rightarrow0$ on both sides of (\ref{eq:201}), we have
\begin{align}\label{eq:202}
    \inf_{\bm{x}\in\mathcal{F}_{\textrm{fin}}} \sum_{(W,R)\in\bar{\Pi}} x(W,R) \Delta(W,R) \leq \Delta_{LP}.
\end{align}
From (\ref{eq:196}) and (\ref{eq:202}), we have
\begin{align}\label{eq:204}
    \Delta_{\textrm{LP}} = \inf_{\bm{x}\in\mathcal{F}_{\textrm{fin}}} \sum_{(W,R)\in\bar{\Pi}} x(W,R) \Delta(W,R).
\end{align}
Let $\Delta_{n}$ be the optimal LP value on the restricted policy set $\bar{\Pi}_{n}$. That is,
\begin{align}\label{eq:n205}
    \Delta_{n} = \min_{\{x(W,R)| (W,R)\in\bar{\Pi}_{n}\}} & \quad \sum_{(W,R)\in\bar{\Pi}_{n}} x(W,R) \Delta(W,R) \nonumber\\ 
    \textrm{s.t.} \qquad & \quad \sum_{(W,R)\in\bar{\Pi}_{n}} x(W,R) \nu_{s}(W,R) \leq f_{s},\nonumber\\ 
    & \quad \sum_{(W,R)\in\bar{\Pi}_{n}} x(W,R) \nu_{r}(W,R) \leq f_{r}, \nonumber\\
    & \quad \sum_{(W,R)\in\bar{\Pi}_{n}}x(W,R) =1, \nonumber \\
    & \quad x(W,R) \geq 0, \quad \forall (W,R)\in {\bar{\Pi}_n}.
\end{align}
As $\bar{\Pi}_{n}$ is finite, the feasible set of the LP in (\ref{eq:n205}) is a closed subset of the finite-dimensional probability simplex, which is compact. The objective is linear and therefore continuous. Moreover, it is non-empty because $(W_{0},\infty)\in\bar{\Pi}_{n}$. Thus, the finite primal LP admits a minimizer. Since this finite LP optimizes over a subset of the feasible distributions of the infinite LP,
\begin{align}\label{eq:206}
    \Delta_{n} \geq \Delta_{\textrm{LP}}.
\end{align}
Fix $\epsilon>0$, from (\ref{eq:204}), we know that there exists a feasible finite-support distribution $\bm{x}_{\epsilon}\in\mathcal{F}_{\textrm{fin}}$ satisfying
\begin{align}
    \sum_{(W,R)\in\bar{\Pi}} x_{\epsilon}(W,R) \Delta(W,R) \leq \Delta_{\textrm{LP}} + \epsilon.
\end{align}
Because of (\ref{eq:180}), there exists $n_{1}$, such that
\begin{align}
    \{(W,R)\in\bar{\Pi}|x_{\epsilon}(W,R)>0\} \subseteq \bar{\Pi}_{n_{1}}.
\end{align}
Thus, for all $n\geq n_{1}$, $x_{\epsilon}$ is feasible for the $n$th finite LP. Thus,
\begin{align}
    \Delta_{n} \leq \sum_{(W,R)\in\bar{\Pi}} x_{\epsilon}(W,R) \Delta(W,R) \leq \Delta_{\textrm{LP}} + \epsilon.
\end{align}
Together with (\ref{eq:206}), we have,
\begin{align}\label{eq:n210}
    |\Delta_{n}-\Delta_{\textrm{LP}}|\leq \epsilon.
\end{align}
The dual problem of the LP in (\ref{eq:n205}) can be equivalently written as,
\begin{align}
    \max_{\mu,\lambda\geq 0} \ g_{n}(\mu,\lambda),
\end{align}
where 
\begin{align}
    g_{n}(\mu,\lambda) = \inf_{(W,R)\in\bar{\Pi}_{n}} \mathcal{L}_{\mu,\lambda}(W,R).
\end{align}
From finite-dimensional LP strong duality, we have
\begin{align}
    \Delta_{n} =  \max_{\mu,\lambda\geq 0} \ g_{n}(\mu,\lambda).
\end{align}
Let $(\mu_{n},\lambda_{n})$ be an optimal dual solution, then
\begin{align}
    \Delta_{n} & = g_{n}(\mu_{n},\lambda_{n}) \\ 
    & \leq \mathcal{L}_{\mu_{n},\lambda_{n}} (W_{0},\infty)\nonumber\\
    & \leq \Delta(W_{0},\infty)-\eta_{\min}(\mu_{n}+\lambda_{n}).
\end{align}
As $\Delta_{n}\geq 1$, 
\begin{align}
    \lambda_{n}+\mu_{n}\leq \frac{\Delta(W_{0},\infty)-1}{\eta_{\min}}.
\end{align}
The sequence $\{\mu_{n},\lambda_{n}\}$ is bounded, thus, from the Bolzano-Weierstrass theorem, there exists a subsequence, still indexed for simplicity by $n$, such that,
\begin{align}\label{eq:217}
    (\mu_{n},\lambda_{n}) \rightarrow (\mu^{*},\lambda^{*}),
\end{align}
where $\mu^{*}$ and $\lambda^{*}$ are finite and non-negative. Now, fix an arbitrary policy $\pi\in\bar{\Pi}$. We can find a large enough $n$, such that, we have $\pi\in\bar{\Pi}_{n}$. Thus,
\begin{align}
    \Delta_{n} = g_{n}(\mu_{n},\lambda_{n})\leq  \Delta(\pi) + \mu_{n}(\nu_{s}(\pi)-f_{s}) + \lambda_{n}(\nu_{r}(\pi)-f_{r}).
\end{align}
Taking $n\rightarrow\infty$ on both sides, from (\ref{eq:n210}) and (\ref{eq:217}), we have,
\begin{align}\label{eq:220}
    \Delta_{\textrm{LP}} \leq \Delta(\pi) + \mu^{*}(\nu_{s}(\pi)-f_{s}) + \lambda^{*}(\nu_{r}(\pi)-f_{r}).
\end{align}
Note that (\ref{eq:220}) holds for all policies $\pi\in\bar{\Pi}$. Thus,
\begin{align}\label{eq:221}
    \Delta_{\textrm{LP}} \leq g(\mu^{*},\lambda^{*}),
\end{align}
where similar to $g_{n}(\mu,\lambda)$, we define
\begin{align}\label{eq:222}
    g(\mu,\lambda) = \inf_{\pi\in\bar{\Pi}} \ \mathcal{L}_{\mu,\lambda} (\pi).
\end{align}
Let $\bm{x}$ be any feasible LP distribution with finite objective. For every $\mu,\lambda\geq 0$,
\begin{align}
    g(\mu,\lambda) \leq & \sum_{(W,R)\in\bar{\Pi}} x(W,R) \mathcal{L}_{\mu,\lambda}(W,R)\\ 
    \leq& \sum_{(W,R)\in\bar{\Pi}} x(W,R)\Delta(W,R).\label{eq:224}
\end{align}
Note that (\ref{eq:224}) holds for all $\mu,\lambda\geq0$ and $\bm{x} \in\mathcal{F}$. Thus,
\begin{align}
    g(\mu^{*},\lambda^{*}) & \leq \inf_{\bm{x}\in\mathcal{F}} \sum_{(W,R)\in\bar{\Pi}} x(W,R)\Delta(W,R) \\ &=\Delta_{\textrm{LP}}.\label{eq:n226}
\end{align}
Thus, from (\ref{eq:221}) and (\ref{eq:n226}), we have
\begin{align}\label{eq:n227}
    \Delta_{\textrm{LP}} =g(\mu^{*},\lambda^{*}) =\max_{\mu,\lambda\geq 0} \ g(\mu,\lambda).
\end{align}
Consider a feasible finite support distribution $\bm{x}\in\mathcal{F}_{\textrm{fin}}$, with the following support,
\begin{align}
    (W_{1},R_{1}), (W_{2},R_{2}), \cdots,(W_{m},R_{m}).
\end{align}
Now, we construct a policy $\pi$ that chooses policy $(W_{i},R_{i})$ with probability $x(W_{i},R_{i})$ at time slot $t=1$, and then follows the selected policy forever. Thus, for a finite horizon $T$,
\begin{align}
    \Delta(\pi;T) = \sum_{i=1}^{m} x(W_{i},R_{i}) \Delta(W_{i},R_{i};T).
\end{align}
As $m$ is finite and $\Delta(W_{i},R_{i};T)$ converges to $\Delta(W_{i},R_{i})$ as $T\rightarrow\infty$, we have,
\begin{align}\label{eq:n230}
    \Delta(\pi) = \sum_{i=1}^{m} x(W_i,R_i) \Delta(W_{i},R_{i}).
\end{align}
Similarly, we have,
\begin{align}
    \nu_{s}(\pi) = \sum_{i=1}^{m} x(W_i,R_i) \nu_{s}(W_{i},R_{i}), \qquad \nu_{r}(\pi) =  \sum_{i=1}^{m} x(W_i,R_i) \nu_{r}(W_i,R_i). 
\end{align}
Thus, $\pi$ is feasible for the problem in (\ref{eq:opt_prob}). Thus, from (\ref{eq:n230}), we have,
\begin{align}
    \Delta^{*} \leq \sum_{i=1}^{m} x(W_i,R_i) \Delta(W_{i},R_{i}).
\end{align}
Now, taking the infimum over all feasible finite-support LP distributions and using (\ref{eq:204}), we get
\begin{align}\label{eq:234}
    \Delta^{*}\leq \Delta_{\textrm{LP}}.
\end{align}
Let us consider an arbitrary feasible policy $\pi_{1}$ for (\ref{eq:opt_prob}). From (\ref{eq:164}), we have
\begin{align}\label{eq:235}
    \mathcal{L}_{\mu,\lambda}(\pi_{1}) \leq \limsup_{T\rightarrow\infty} \Delta(\pi_{1};T).
\end{align}
From (\ref{eq:172}) and (\ref{eq:222}), we have
\begin{align}\label{eq:236}
    g(\mu,\lambda) = \inf_{\pi\in\Pi} \mathcal{L}_{\mu,\lambda}(\pi).
\end{align}
From (\ref{eq:235}), we have
\begin{align}
    \max_{\mu,\lambda\geq 0}  g(\mu,\lambda)\leq \Delta(\pi_{1}).
\end{align}
From (\ref{eq:n227}), we have
\begin{align}
    \Delta_{\textrm{LP}} \leq \Delta(\pi_{1}).
\end{align}
Therefore, taking the infimum over all feasible policies we get,
\begin{align}\label{eq:239}
    \Delta_{\textrm{LP}}\leq \Delta^{*}.
\end{align}
From (\ref{eq:234}) and (\ref{eq:239}), we have,
\begin{align}
    \Delta_{\textrm{LP}} = \Delta^{*}.
\end{align}

It remains to show that the infimum in (\ref{eq:158}) is attained by a distribution that assigns positive probability to at most three policies. For every $n$, let $\bm{x}_{n}^{*}$ be an optimal basic feasible solution of the finite LP in (\ref{eq:n205}). After adding two slack variables, this LP has three equality constraints. Thus, a basic feasible solution has at most three positive variables. Consequently, $\bm{x}_{n}^{*}$ assigns positive probability to at most three policies. We denote these policies by
\begin{align}
    \pi_{n,j}=(W_{n,j},R_{n,j}),\quad j=1,2,3,
\end{align}
and their corresponding probabilities by $\theta_{n,j}$. If fewer than three policies have a positive probability, we add arbitrary policies with zero probability. Thus,
\begin{align}
    \theta_{n,j}\geq0,\quad \sum_{j=1}^{3}\theta_{n,j}=1,
\end{align}
and
\begin{align}\label{eq:n240}
    \Delta_{n}=\sum_{j=1}^{3}\theta_{n,j}\Delta(\pi_{n,j}).
\end{align}
The sequence $\{(\theta_{n,1},\theta_{n,2},\theta_{n,3})\}_{n=1}^{\infty}$ belongs to the compact probability simplex in $\mathbb{R}^{3}$. Therefore, there exists a subsequence, still indexed by $n$ for simplicity, such that,
\begin{align}\label{eq:n241}
    \theta_{n,j}\rightarrow\theta_{j}, \quad j=1,2,3.
\end{align}
Now, consider an index $j$ such that $\theta_{j}>0$. From (\ref{eq:n241}), $\theta_{n,j}$ is bounded away from zero for all sufficiently large $n$. Moreover, from (\ref{eq:n240}),
\begin{align}
    \theta_{n,j}\Delta(\pi_{n,j})\leq\Delta_{n}.
\end{align}
Since $\Delta_{n}\rightarrow\Delta_{\textrm{LP}}<\infty$, the sequence $\{\Delta(\pi_{n,j})\}$ is bounded. Furthermore, from (\ref{eq:144}), (\ref{eq:n147}) and (\ref{eq:n157}), we have
\begin{align}
    \Delta(W,R)\geq\frac{W}{2}.
\end{align}
Thus, $\{W_{n,j}\}$ is bounded. Since $W_{n,j}$ is integer-valued, there exists a further subsequence, still indexed by $n$ for simplicity, such that $W_{n,j}=W_{j}$, for all sufficiently large $n$. Similarly, there exists a further subsequence, still indexed by $n$ for simplicity, such that either $R_{n,j}=R_{j}$ for all sufficiently large $n$, or $R_{n,j}\rightarrow\infty$, in which case we define $R_{j}=\infty$. Since there are at most three such indices, we can choose one common subsequence for which these properties hold simultaneously. Let $\pi_{j}=(W_{j},R_{j})$, for every $j$ such that $\theta_{j}>0$, if $R_j<\infty$, the following limits hold immediately,
\begin{align}\label{eq:n242}
    \nu_{s}(\pi_{n,j})\rightarrow\nu_{s}(\pi_{j}), \qquad \nu_{r}(\pi_{n,j})\rightarrow\nu_{r}(\pi_{j}), \qquad \Delta(\pi_{n,j})\rightarrow\Delta(\pi_{j}).
\end{align}
If for $\theta_j>0$, if $R_{j}=\infty$, then the limits in (\ref{eq:n242}) follow from a dominated convergence theorem argument. If $\theta_j>0$ and $R_j=\infty$, then $W_{n,j}=W_j$ for all sufficiently large $n$ and
$R_{n,j}\rightarrow\infty$. From (\ref{eq:132}), we have 
\begin{align}
    V_{R_{n,j}} = 1+ ((\bar G-1) \mod R_{n,j}),
\end{align}
where $\bar{G}$ is an independent copy of $G$. Thus, $V_{R_{n,j}} \leq \bar G$, and $V_{R_{n,j}}\rightarrow \bar G$. Consequently, $L(W_j,R_{n,j})$ and $Y(W_j,R_{n,j})$ converge almost surely to $L(W_j,\infty)$ and $Y(W_j,\infty)$, respectively. Moreover, from (\ref{eq:144}) and (\ref{eq:n157}), we have 
\begin{align}\label{eq:457}
    0\leq L(W_j,R_{n,j})\leq W_j+G, \qquad 0\leq Y(W_j,R_{n,j})\leq(W_j+\bar{G}+G)^2.
\end{align}
The upper bounds in (\ref{eq:457}) have finite expectations because geometric random variables have finite second moments. Thus, the limits in (\ref{eq:n242}) follow from the dominated convergence theorem. For an index with $\theta_{j}=0$, we assign an arbitrary policy $\pi_{j}$. Since $0\leq\nu_{s}(\pi),\nu_{r}(\pi)\leq1$ for every $\pi\in\bar{\Pi}$, we have
\begin{align}
    \theta_{n,j}\nu_{s}(\pi_{n,j})\rightarrow0=\theta_{j}\nu_{s}(\pi_{j}), \qquad  \theta_{n,j}\nu_{r}(\pi_{n,j}) \rightarrow0=\theta_{j}\nu_{r}(\pi_{j}).
\end{align}
Now, define $\bm{x}^{*}$ by assigning probability $\theta_{j}$ to policy $\pi_{j}$, $j=1,2,3$. Passing to the limit in the resource constraints of $\bm{x}_{n}^{*}$ gives
\begin{align}
    \sum_{j=1}^{3}\theta_{j}\nu_{s}(\pi_{j})=\lim_{n\rightarrow\infty}
    \sum_{j=1}^{3}\theta_{n,j}\nu_{s}(\pi_{n,j})\leq f_{s}, \qquad
    \sum_{j=1}^{3}\theta_{j}\nu_{r}(\pi_{j})=\lim_{n\rightarrow\infty}
    \sum_{j=1}^{3}\theta_{n,j}\nu_{r}(\pi_{n,j})\leq f_{r}.
\end{align}
Thus, $\bm{x}^{*}$ is feasible for the LP in (\ref{eq:158}). Finally,
\begin{align}
    \sum_{j=1}^{3}\theta_{j}\Delta(\pi_{j})&=\sum_{j:\theta_{j}>0}\theta_{j}\Delta(\pi_{j})\\
    &=\lim_{n\rightarrow\infty}\sum_{j:\theta_{j}>0}\theta_{n,j}\Delta(\pi_{n,j})\\
    &\leq\lim_{n\rightarrow\infty} \sum_{j=1}^{3}\theta_{n,j}\Delta(\pi_{n,j})\\
    &=\lim_{n\rightarrow\infty}\Delta_{n}=\Delta_{\textrm{LP}},\label{eq:257}
\end{align}
where (\ref{eq:257}) follows from (\ref{eq:n210}). Since $\bm{x}^{*}$ is feasible for (\ref{eq:158}), its objective value cannot be smaller than $\Delta_{\textrm{LP}}$. Therefore,
\begin{align}
    \sum_{j=1}^{3}\theta_{j}\Delta(\pi_{j})=\Delta_{\textrm{LP}}.
\end{align}
Hence, the LP in (\ref{eq:158}) admits a minimizer $\bm{x}^{*}$ that assigns positive probability to at most three policies.

Finally, we construct a policy $\pi^{*}$ that, at time $t=1$, selects policy $\pi_{j}$ with probability $\theta_{j}$, $j\in\{1,2,3\}$, and follows the selected policy for the entire time horizon. For every finite $T$, we have
\begin{align}
    \Delta(\pi^{*};T)= \sum_{j=1}^{3}\theta_{j}\Delta(\pi_{j};T),\qquad
    \nu_{s}(\pi^{*};T)=\sum_{j=1}^{3}\theta_{j}\nu_{s}(\pi_{j};T),\qquad
    \nu_{r}(\pi^{*};T)=\sum_{j=1}^{3}\theta_{j}\nu_{r}(\pi_{j};T).
\end{align}
Since the above sums contain at most three terms and the corresponding limits exist for every $(W,R)$ policy, taking $T\rightarrow\infty$ gives
\begin{align}
    \Delta(\pi^{*})=\sum_{j=1}^{3}\theta_{j}\Delta(\pi_{j})=\Delta_{\textrm{LP}},\qquad
    \nu_{s}(\pi^{*})=\sum_{j=1}^{3}\theta_{j}\nu_{s}(\pi_{j})\leq f_{s},\qquad
    \nu_{r}(\pi^{*})=\sum_{j=1}^{3}\theta_{j}\nu_{r}(\pi_{j})\leq f_{r}.
\end{align}
Thus, $\pi^{*}$ is feasible for the problem in (\ref{eq:opt_prob}). Moreover, since $\Delta_{\textrm{LP}}=\Delta^{*}$, we have
\begin{align}
    \Delta(\pi^{*})=\Delta^{*}.
\end{align}
Therefore, $\pi^{*}$ is an optimal policy for the problem in (\ref{eq:opt_prob}).
\end{Proof}

\section{Proof of Theorem~\ref{th:epsilon_optimal}}\label{append:th:epsilon_optimal}
\begin{Proof}
Let $\bm{x}^{*}$ be an optimal solution of the LP in (\ref{eq:158}). We define
\begin{align}
    p_{\epsilon}=\sum_{\substack{(W,R)\in\bar{\Pi}:\\W>W_{\epsilon}}} x^{*}(W,R).
\end{align}
Since $\Delta(W,R)\geq \frac{W}{2}$, we have
\begin{align}
    \frac{W_{\epsilon}p_{\epsilon}}{2} \leq \sum_{\substack{(W,R)\in\bar{\Pi}:\\
    W>W_{\epsilon}}} x^{*}(W,R)\Delta(W,R)\leq \Delta_{\textrm{LP}} \leq \Delta(W_{0},\infty).
\end{align}
Thus, from (\ref{eq:W_epsilon}),
\begin{align}\label{eq:p_epsilon}
    p_{\epsilon} \leq \frac{2\Delta(W_{0},\infty)}{W_{\epsilon}} \leq \xi_{\epsilon}.
\end{align}

Starting from $\bm{x}^{*}$, we construct a distribution $\hat{\bm{x}}_{\epsilon}$ as follows. For every policy $(W,R)$ with $W>W_{\epsilon}$, we transfer the probability mass $x^{*}(W,R)$ to the policy $(W_{0},\infty)$. For every remaining policy $(W,R)$ with $W\leq W_{\epsilon}$ and $R>R_{\epsilon}$, we transfer the probability mass $x^{*}(W,R)$ to the policy $(W,\infty)$. The probability masses of all other policies remain unchanged. If multiple policies are mapped to the same policy, their probability masses are added. Thus, $\hat{\bm{x}}_{\epsilon}$ is a probability distribution supported on $\bar{\Pi}_{\epsilon}$. From (\ref{eq:W_epsilon}), for every $W>W_{\epsilon}$, we have,
\begin{align}
    \Delta(W,R)\geq\frac{W}{2}>\Delta(W_{0},\infty).
\end{align}
Thus, replacing policies with $W>W_{\epsilon}$ does not increase the average age. From (\ref{eq:R_epsilon_conditions3}), replacing policies with $R>R_{\epsilon}$ increases the average age by at most $\frac{\epsilon}{2}$. Hence,
\begin{align}\label{eq:xhat_age}
    \sum_{(W,R)\in\bar{\Pi}_{\epsilon}} \hat{x}_{\epsilon}(W,R)\Delta(W,R) \leq \Delta_{\textrm{LP}}+\frac{\epsilon}{2}.
\end{align}
Since the sampling and transmission rates are bounded by $1$, from (\ref{eq:R_epsilon_conditions1}), (\ref{eq:R_epsilon_conditions2}) and (\ref{eq:p_epsilon}), we have
\begin{align}
    \sum_{(W,R)\in\bar{\Pi}_{\epsilon}}\hat{x}_{\epsilon}(W,R)\nu_{s}(W,R) \leq f_{s}+2\xi_{\epsilon},\qquad
    \sum_{(W,R)\in\bar{\Pi}_{\epsilon}}\hat{x}_{\epsilon}(W,R)\nu_{r}(W,R) \leq f_{r}+2\xi_{\epsilon}.
\end{align}
We define
\begin{align}
    \eta_{\epsilon}=\frac{2\xi_{\epsilon}}{\eta_{\min}+2\xi_{\epsilon}},
\end{align}
and construct the distribution
\begin{align}
    \bar{{x}}_{\epsilon}(W,R) = 
    \begin{cases}
        (1-\eta_{\epsilon}) \hat{x}_{\epsilon}(W,R), &(W,R)\in\bar{\Pi}_{\epsilon} \  \textrm{and} \ (W,R) \neq (W_0,\infty), \\ 
        (1-\eta_{\epsilon}) \hat{x}_{\epsilon}(W,R) + \eta_{\epsilon}, & (W,R)=(W_0,\infty). 
    \end{cases}
\end{align}
From the definition of $\eta_{\min}$ in (\ref{eq:176}),
\begin{align}
    \sum_{(W,R)\in\bar{\Pi}_{\epsilon}}\bar{x}_{\epsilon}(W,R)\nu_{s}(W,R) &\leq f_{s},\\
    \sum_{(W,R)\in\bar{\Pi}_{\epsilon}} \bar{x}_{\epsilon}(W,R)\nu_{r}(W,R) &\leq f_{r}.
\end{align}
Thus, $\bar{\bm{x}}_{\epsilon}$ is feasible for the restricted LP. Moreover, from (\ref{eq:xhat_age}),
\begin{align}
    \sum_{(W,R)\in\bar{\Pi}_{\epsilon}} \bar{x}_{\epsilon}(W,R)\Delta(W,R)
     & \leq \Delta_{\textrm{LP}}+\frac{\epsilon}{2} +\eta_{\epsilon}\Delta(W_{0},\infty)\\
     & \leq \Delta_{\textrm{LP}} + \frac{\epsilon}{2}  +  \frac{2\xi_{\epsilon}}{\eta_{\min}}\Delta(W_{0},\infty)\\
     &=\Delta_{\textrm{LP}}+\epsilon.
\end{align}
Therefore,
\begin{align}
    \Delta_{\textrm{LP}}\leq\Delta_{\epsilon}\leq\Delta_{\textrm{LP}}+\epsilon.
\end{align}
From Theorem~\ref{th:6}, $\Delta_{\textrm{LP}}=\Delta^{*}$, which proves (\ref{eq:269}). 

Finally, since $\bar{\Pi}_{\epsilon}$ is finite, the restricted LP admits an optimal basic feasible solution. After adding two slack variables, it has three equality constraints. Therefore, there exists an optimal solution that assigns positive probability to at most three policies.
\end{Proof}

\section{Proof of Theorem~\ref{th:four_approx}}\label{appen:th:four_approx}
\begin{Proof}
For a fixed $R\in\{R_a,R_b\}$, define
\begin{align}\label{eq:Delta_R_mix}
    \Delta_R = \kappa_{R,0}\Delta(W_R,R)+\kappa_{R,1}\Delta(W_R+1,R).
\end{align}
Thus,
\begin{align}\label{eq:Delta_four}
    \Delta_4= \gamma\Delta_{R_a}+(1-\gamma)\Delta_{R_b}.
\end{align}
For $j\in\{0,1\}$, we define
\begin{align}\label{eq:zeta_definition}
    \zeta_{R,j}=\frac{\kappa_{R,j}\ell^{*}}{\mathbb{E}[L(W_R+j,R)]}.
\end{align}
From (\ref{eq:149}), (\ref{eq:208}) and (\ref{eq:R_block_transmission}), we have
\begin{align}
    \frac{\kappa_{R,0}}{\mathbb{E}[L(W_R,R)]}+\frac{\kappa_{R,1}}{\mathbb{E}[L(W_R+1,R)]}=\frac{1}{\ell^{*}}.
\end{align}
Thus,
\begin{align}\label{eq:zeta_sum}
    \zeta_{R,0}+\zeta_{R,1}=1.
\end{align}
Moreover, from (\ref{eq:kappa_R}), we have,
\begin{align}\label{eq:zeta_length}
    \sum_{j=0}^{1} \zeta_{R,j}\mathbb{E}[L(W_R+j,R)]= \ell^{*}.
\end{align}
From (\ref{eq:n158}), (\ref{eq:Delta_R_mix}) and (\ref{eq:zeta_definition}), we have
\begin{align}\label{eq:Delta_R_area}
    \Delta_R=\frac{1}{\ell^{*}}\sum_{j=0}^{1}\zeta_{R,j}\mathbb{E}[Y(W_R+j,R)].
\end{align}
We now obtain an upper bound for $\mathbb{E}[Y(W,R)]$. Whenever $H(W,R)>0$, we have
\begin{align}
    V_R+H(W,R)=W.
\end{align}
Therefore,
\begin{align}\label{eq:226}
    V_RH(W,R) + \frac{H(W,R)(H(W,R) - 1)}{2} \leq  WH(W,R).
\end{align}
The channel realizations during a frame are independent of the age $V_R$ at the beginning of that frame. Thus, $G$ is independent of $V_R$ and $H(W,R)$. As $G$ follows a geometric distribution with success probability $q$, we have
\begin{align}\label{eq:227}
    \mathbb{E}[G]=\frac{1}{q}, \qquad \mathbb{E}\left[\frac{G(G-1)}{2}\right]=\frac{1-q}{q^2}.
\end{align}
From (\ref{eq:549}), we have
\begin{align}\label{eq:228}
    \mathbb{E} [V_R]=\frac{1}{q}- \frac{R(1-q)^R} {1-(1-q)^R}\leq \frac{1}{q}.
\end{align}
Now, from (\ref{eq:n157}), (\ref{eq:226}), (\ref{eq:227}) and (\ref{eq:228}), we have,
\begin{align}\label{eq:Y_upper}
    \mathbb{E}[Y(W,R)]\leq&W\mathbb{E}[H(W,R)]+\frac{\mathbb{E}[H(W,R)]}{q}+\frac{2-q}{q^2}.
\end{align}
Now, note that
\begin{align}\label{eq:230}
    \mathbb{E}[L(W,R)]=&\mathbb{E}[(W-V_R)^{+}]+\frac{1}{q}\geq W-\mathbb{E}[V_R]+\frac{1}{q}\geq W.
\end{align}
From (\ref{eq:210}) and (\ref{eq:230}), it follows that
\begin{align}\label{eq:231}
     W_R\leq\ell^{*}, \qquad W_R+1\leq\ell^{*}+1.
\end{align}
Also, from (\ref{eq:144}), (\ref{eq:zeta_length}) and (\ref{eq:227}), we have
\begin{align}\label{eq:weighted_H}
    \sum_{j=0}^{1} \zeta_{R,j}\mathbb{E}[H(W_R+j,R)] =\ell^{*}-\frac{1}{q}.
\end{align}
Now, from (\ref{eq:zeta_sum}), (\ref{eq:Y_upper}), (\ref{eq:231}) and (\ref{eq:weighted_H}), we get
\begin{align}\label{eq:233}
    \sum_{j=0}^{1}\zeta_{R,j}\mathbb{E}[Y(W_R+j,R)]&\leq(\ell^{*}+1)\left(\ell^{*}-\frac{1}{q}\right)+\frac{1}{q}\left(\ell^{*}-\frac{1}{q}\right)+\frac{2-q}{q^2}\\
    &=(\ell^{*})^2+\ell^{*}-\frac{2}{q}+\frac{1}{q^2}.
\end{align}
From (\ref{eq:Delta_R_area}), we have
\begin{align}\label{eq:319}
    \Delta_{R} \leq \ell^{*} + 1 -\frac{2}{q\ell^{*}} + \frac{1}{q^{2}\ell^{*}}.
\end{align}
Since the same bound holds for both $R=R_{a}$ and $R=R_{b}$, we have
\begin{align}
    \Delta(\pi_{4}) \leq  \ell^{*} + 1 -\frac{2}{q\ell^{*}} + \frac{1}{q^{2}\ell^{*}}.
\end{align}
Now, from (\ref{eq:208}) and (\ref{eq:547}) in Theorem~\ref{th:7}, we get
\begin{align}\label{eq:321}
    2\Delta^{*} \geq 1+ \ell^{*} + \frac{1}{q^{2}\ell^{*}} - \frac{1}{q\ell^{*}}.
\end{align}
Thus, from (\ref{eq:319}), we have
\begin{align}
    \Delta(\pi_{4}) \leq 2 \Delta^{*} - \frac{1}{q \ell^{*}} <2 \Delta^{*}, 
\end{align}
which completes the proof of this theorem.
\end{Proof}

\end{appendices}

\bibliographystyle{unsrt}
\bibliography{references}

@book{puterman2014markov,
  title={Markov Decision Processes: Discrete {S}tochastic {D}ynamic {P}rogramming},
  author={Puterman, M. L.},
  year={2014},
  publisher={John Wiley \& Sons}
}

@article{sennott1989average,
  title={Average cost optimal stationary policies in infinite state {M}arkov decision processes with unbounded costs},
  author={Sennott, L. I.},
  journal={Operations Research},
  volume={37},
  number={4},
  pages={626-633},
  year={1989},
  month={July}}

@book{bertsekas2012dynamic,
  title={Dynamic {P}rogramming and {O}ptimal {C}ontrol: Volume II},
  author={Bertsekas, D.},
  year={2012},
  publisher={Athena Scientific}}

@article{sun2019sampling,
  title={Sampling for data freshness optimization: Non-linear age functions},
  author={Sun, Y. and Cyr, B.},
  journal={Journal of Communications and Networks},
  volume={21},
  number={3},
  pages={204-219},
  year={2019},
 month={June}
}

@book{altman2021constrained,
  title={Constrained Markov decision processes},
  author={Altman, E.},
  year={2021},
  publisher={Routledge}
}

@article{YatesSurvey,
  author = {Yates, R. D. and Sun, Y. and Brown, R. and Kaul, S. K. and Modiano, E. and Ulukus, S.},
  title = {Age of Information: An Introduction and Survey},
  journal = {IEEE Journal on Selected Areas in Communications},
  year = {2021},
  month={May},
  volume = {39},
  number = {5},
  pages = {1183-1210}}

@article{hatami2022demand,
  title={On-demand AoI minimization in resource-constrained cache-enabled IoT networks with energy harvesting sensors},
  author={Hatami, M. and Leinonen, M. and Chen, Z. and Pappas, N. and Codreanu, M.},
  journal={IEEE Transactions on Communications},
  volume={70},
  number={11},
  pages={7446-7463},
  year={2022},
  month={November}
}

@article{tang2020minimizing,
  title={Minimizing age of information with power constraints: Multi-user opportunistic scheduling in multi-state time-varying channels},
  author={Tang, H. and Wang, J. and Song, L. and Song, J.},
  journal={IEEE Journal on Selected Areas in Communications},
  volume={38},
  number={5},
  pages={854-868},
  year={2020},
  month={May}
}

@inproceedings{wang2019minimizing,
  title={Minimizing age of information for real-time monitoring in resource-constrained industrial IoT networks},
  author={Wang, Q. and Chen, H. and Li, Y. and Pang, Z. and Vucetic, B.},
  booktitle={IEEE INDIN},
  year={2019},
  month={July}
}

@article{ceran2019average,
  title={Average age of information with hybrid ARQ under a resource constraint},
  author={Ceran, E. T. and G{\"u}nd{\"u}z, D. and Gy{\"o}rgy, A.},
  journal={IEEE Transactions on Wireless Communications},
  volume={18},
  number={3},
  pages={1900-1913},
  year={2019},
 month={March}
}

@article{chen2023minimizing,
  title={Minimizing age of information in downlink wireless networks with time-varying channels and peak power constraint},
  author={Chen, G. and Chen, Y. and Wang, J. and Song, J.},
  journal={IEEE Transactions on Vehicular Technology},
  volume={72},
  number={7},
  pages={9058-9068},
  year={2023},
  month={July}
}

@article{sun2019sampling2,
  title={Sampling of the Wiener process for remote estimation over a channel with random delay},
  author={Sun, Y. and Polyanskiy, Y. and Uysal, E.},
  journal={IEEE Transactions on Information Theory},
  volume={66},
  number={2},
  pages={1118-1135},
  year={2019},
 month={February}
}

@article{pan2023optimal,
  title={Optimal sampling for data freshness: Unreliable transmissions with random two-way delay},
  author={Pan, J. and Bedewy, A. M. and Sun, Y. and Shroff, N. B.},
  journal={IEEE/ACM Transactions on Networking},
  volume={31},
  number={1},
  pages={408-420},
  year={2023},
  month={February}
}

@inproceedings{banerjee2025equidistant,
  title={Equidistant-Sample or Wait-and-Sample to Minimize Age Under Sampling Constraint?},
  author={Banerjee, S. and Ulukus, S.},
  booktitle={IEEE ICC},
 year={2025},
  month={June}
}

@article{zakeri2022query,
  title={Query-age-optimal scheduling under sampling and transmission constraints},
  author={Zakeri, A. and Moltafet, M. and Leinonen, M. and Codreanu, M.},
  journal={IEEE/ACM Transactions on Networking},
  volume={27},
  number={4},
  pages={1205-1209},
  year={2023},
  month={April}
}

@article{akar2025age,
  title={Age of information in a single-source generate-at-will dual-server status update system},
  author={Akar, N. and Ulukus, S.},
  journal={IEEE Transactions on Communications},
  volume={73},
  number={9},
  pages={7431-7444},
  year={2025},
  month={September}
}

@article{tang2023age,
  title={Age optimal sampling under unknown delay statistics},
  author={Tang, H. and Chen, Y. and Wang, J. and Yang, P. and Tassiulas, L.},
  journal={ IEEE Transactions on Information Theory},
  volume={69},
  number={2},
  pages={1295-1314},
  year={2023},
  month={February}
}

@article{delfani2025optimizing,
  title={Optimizing information freshness in constrained IoT systems: A token-based approach},
  author={Delfani, E. and Pappas, N.},
  journal={IEEE Transactions on Communications},
  volume={73},
  number={8},
  pages={5848-5863},
  year={2024},
  month={August}
}

@article{gong2022sleep,
  author  = {Gong, J. and Zhu, J. and Chen, X. and Ma, X.},
  title   = {Sleep, Sense or Transmit: Energy-Age Tradeoff for Status Update With Two-Threshold Optimal Policy},
  journal = {IEEE Transactions on Wireless Communications},
  volume  = {21},
  number  = {3},
  pages   = {1751-1765},
  year    = {2022},
  month   = {March}
}

@book{shapiro2009lectures,
  title={Lectures on {S}tochastic {P}rogramming: {M}odeling and {T}heory},
  author={Shapiro, A. and Dentcheva, D. and Ruszczy{\'n}ski, A.},
  year={2009},
  publisher={Society for Industrial and Applied Mathematics}
}

\end{document}